\newcommand{\al}{\alpha}

\newcommand{\be}{\beta}

\newcommand{\ga}{\gamma}

\newcommand{\de}{\delta}
\newcommand{\De}{\Delta}
\newcommand{\eps}{\epsilon}
\newcommand{\vareps}{\varepsilon}
\newcommand{\la}{\lambda}

\newcommand{\ka}{\kappa}

\newcommand{\si}{\sigma}

\newcommand{\pa}{\partial}

\newcommand{\been}{\begin{equation}}
\newcommand{\een}{\end{equation}}
\newcommand{\beena}{\begin{eqnarray}}
\newcommand{\eena}{\end{eqnarray}}

\newcommand{\tn}{\textnormal}

\newcommand{\deriv}[2]{\frac{\partial{#1}}{\partial{#2}}}
\newcommand{\derivtwo}[2]{\frac{\partial^{2}{#1}}{\partial{#2}^2}}
\newcommand{\derivtwomix}[3]{\frac{\partial^{2}{#1}}{\partial{#2}\partial{#3}}}
\newcommand{\non}{\nonumber}
\newcommand{\inty}[4]{\int_{#1}^{#2}{#3}\hspace{0.2cm}d{#4}}

\newcommand{\lit}{\hspace{0.2cm}}

\documentclass[a4paper,11pt]{article}

\usepackage{amsmath}
\usepackage{amssymb}
\usepackage{verbatim}
\usepackage{graphicx}
\usepackage{psfrag}
\usepackage{color,xcolor}
\usepackage{amsthm}
\usepackage{bigints}
\usepackage{enumerate}

\DeclareMathOperator{\arccosh}{arccosh}

\author{William J.~Parnell\\
\footnotesize{School of Mathematics, University of Manchester, Oxford Road, Manchester, M13 9PL,UK}}
\title{The Hill and Eshelby tensors for ellipsoidal inhomogeneities in the Newtonian potential problem and linear elastostatics.}

\begin{document}

\maketitle

\numberwithin{equation}{section}
\begin{abstract}
One of the most cited papers in Applied Mechanics is the work of Eshelby from 1957 who showed that a homogeneous isotropic ellipsoidal inhomogeneity embedded in a homogeneous isotropic host would feel uniform strains and stresses when uniform strains or stresses are applied in the far-field. Of specific importance is the uniformity of \textit{Eshelby's tensor} $\mathbf{S}$. Following this paper a vast literature has been generated using and developing Eshelby's result and ideas, leading to some beautiful mathematics and extremely useful results in a wide range of application areas. In 1961 Eshelby conjectured that for anisotropic materials only ellipsoidal inhomogeneities would lead to such uniform interior fields. Although much progress has been made since then, the quest to prove this conjecture is still not complete; numerous important problems remain open. Following a different approach to that considered by Eshelby, a closely related tensor $\mathbf{P}=\mathbf{S}\mathbf{D}^0$ arises, where $\mathbf{D}^0$ is the host medium compliance tensor. The tensor $\mathbf{P}$ is associated with \textit{Hill} and is of course also uniform when ellipsoidal inhomogeneities are embedded in a homogeneous host phase. Two of the most fundamental and useful areas of applications of these tensors are in Newtonian potential problems such as heat conduction, electrostatics, etc.\ and in the vector problems of elastostatics. Knowledge of the Hill and Eshelby tensors permit a number of interesting aspects to be studied associated with inhomogeneity problems and more generally for inhomogeneous media. Micromechanical methods established mainly over the last half-century have enabled bounds on and predictions of the effective properties of composite media. In many cases such predictions can be explicitly written down in terms of the Hill, or equivalently the Eshelby tensor and can be shown to provide excellent predictions in many cases.

Of specific interest is that a number of important limits of the ellipsoidal inhomogeneity can be taken in order to be employed in predictions of the effective properties of e.g.\ layered media, fibre reinforced composites, voids and cracks to name but a few. In the main, results for the Hill and Eshelby tensors associated with these problems are distributed over a wide range of articles and books, using different notation and terminology and so it is often difficult to extract the necessary information for the tensor that one requires. The case of an anisotropic host phase is also frequently non-trivial due to the requirement of the associated Green's tensor. Here this classical problem is revisited and a large number of results for problems that are felt to be of great utility in a wide range of disciplines are derived or recalled. A scaling argument leads to the derivation of the Eshelby tensor for potential problems where the host phase is at most orthotropic, without the requirement of using the anisotropic Green's function. Concentration tensors are derived for a wide variety of problems that can be used directly in the various micromechanical schemes. Both tensor and matrix formulations are considered and contrasted.
\end{abstract}

\section{Introduction} \label{sec:intro}

The canonical \textit{isolated inhomogeneity problem} has been of fundamental importance in a number of materials modelling problems now for well over a century. This problem is the following: a single inhomogeneity, i.e.\ a particle of general shape, with different material properties to that of the surrounding material is embedded inside an \textit{unbounded} (in all directions, i.e.\ free-space) homogeneous host medium. Given some prescribed conditions in the far-field, what form do the fields take within the inhomogeneity? As well as being interesting in its own right, this problem is of utmost importance in homogenization, micromechanics and multiscale modelling.

The first to consider this kind of inhomogeneity problem was Poisson in 1826 \cite{Poi-26} who studied the perturbed field due to an isolated ellipsoid in the context of the Newtonian potential problem. He showed that given a uniform electric polarization (or magnetization), the induced electric (or magnetic) field inside the ellipsoid is also uniform. In 1873 Maxwell \cite{Max-98} derived explicit expressions for this field. Early work in linear elasticity saw a number of studies determine the field inside and around inhomogeneities, including the important case of a cavity (since this was correctly recognized as a defect or flaw). Examples of these works were those associated with the case of spheres \cite{Sou-26}, \cite{Goo-33}, spheroids \cite{Edw-51} and ellipsoids \cite{Sad-47, Sad-49, Rob-51} but all considered specific loadings, usually of the \textit{homogeneous type} in the far field, meaning uniform tractions or displacements that are linear in the independent Cartesian variable say $\mathbf{x}$.

The inhomogeneity problem is now usually associated with the name of Eshelby because in 1957 he showed that for general homogeneous conditions imposed in the far field, the strain set up inside an isotropic homogeneous ellipsoid is uniform \cite{Esh-57}. In 1961 Eshelby \cite{Esh-61} conjectured that ``\textit{...amongst closed surfaces, the ellipsoid alone has this convenient property...."}. Is this true? In the sense of what it is thought that Eshelby meant when he made this conjecture (the so-called \textit{weak} Eshelby conjecture, where the interior field must be uniform for \textit{any} uniform far-field loading), this statement certainly \textit{is} true although this was only proved in 2008, simultaneously by Kang and Milton \cite{Kan-08} and Liu \cite{Liu-08} in the case of isotropic media. There is a slightly different version (the so-called \textit{strong} Eshelby conjecture), where the interior field must be uniform only for a \textit{specific, single} uniform far-field loading. This strong conjecture has still not been proven in the context of three dimensional isotropic linear elasticity, although significant progress has been made in the last decade, see \cite{Kan-09} for a review. Furthermore the results obtained in \cite{Amm-10} go beyond the weak Eshelby conjecture but still do not fully prove the strong conjecture. Interestingly the associated (weak) conjecture for the Newtonian potential problem was proved some time before Eshelby's 1957 elastostatics paper, by Dive in 1931 \cite{Div-31} and Nikliborc \cite{Nik-32} in 1932, see also the discussion in \cite{Kan-08}, \cite{Liu-08}, \cite{Kan-09}. In deriving these results, Dive and Nikliborc proved the converse of Newton's theorem that if $V$ is an ellipsoid of uniform density, the gravitational force in $V$ is zero \cite{Kel-70}. The strong conjecture in the context of the potential problem is true in two dimensions \cite{Ru-96} but is \textit{not} true in dimensions greater than two. A non-ellipsoidal counterexample associated with a \textit{specific} far-field loading (equivalently a specific eigenstress) was found by Liu \cite{Liu-08}.

It is important to note that the proofs of Eshelby's conjectures in elastostatics referred to above correspond to \textit{simply connected}, \textit{isotropic} inhomogeneities with \textit{Lipschitz boundaries}. Eshelby's work was followed up with work by numerous researchers who considered the general anisotropic case \cite{Esh-61}, \cite{Wil-64}, \cite{Wal-67}, \cite{Wal-77}, \cite{Kin-71}, \cite{Lin-73}, \cite{Asa-75}, \cite{Bac-80}, \cite{Wit-89}. In 1974 Cherepanov \cite{Che-74} proved that multiple inhomogeneities of non-ellipsoidal shape can interact in order to render the interior fields uniform; see also Kang and Milton \cite{Kan-08} and Liu \cite{Liu-08} who coined the term \textit{E-inclusions} for such interacting inhomogeneities. Liu and co-workers have also considered the periodic Eshelby problem in two dimensions \cite{Liu-07}, \cite{Liu-09}. Kang and Milton \cite{Kan-08} used their approach to prove Eshelby's weak conjecture in the context of the fully \textit{anisotropic} potential problem. Most notably, it is stressed again that the weak Eshelby conjecture for elasticity has not yet been proved in the context of anisotropic elasticity.

Interest in deriving the Eshelby tensor for non-ellipsoidal inhomogeneities has always been present in order to show that the conjecture holds for specific classes of inhomogeneities. Particular attention has been paid to polygonal and polyhedral inhomogeneities and the associated properties of Eshelby's tensor \cite{Mur-94}, \cite{Rod-96}, \cite{Mur-97}, \cite{Noz-97}, \cite{Mar-98a}, \cite{Mar-98b}, \cite{Lub-98}, \cite{Kaw-01}. The \textit{supersphere} case has been considered recently by \cite{Che-15} building on the work by \cite{Ona-01}, \cite{Ona-02}, \cite{Ona-12}. A general method was developed by Ru \cite{Ru-99}  in order to obtain an analytical solution associated with a two dimensional inhomogeneity of arbitrary cross section and explicit forms of the stress inside hypotrochoidal and rectangular inhomogeneities were derived. Some analytical expressions have recently been derived for two-dimensional problems in the Newtonian potential and plane elastostatics problems where inhomogeneities are either polygonal or their shape can be described by finite Laurent expansions \cite{Zou-10}, \cite{Zou-11}. Furthermore useful properties of the Eshelby tensor have been deduced, including the relationship of the averaged Eshelby tensor for non-ellipsoidal inhomogeneities to their ellipsoidal counterparts \cite{Wan-04}, \cite{Zhe-06}.

More recently the inhomogeneity problem has been studied in the \textit{nonlinear} elasticity context where in two dimensions results associated with Eshelby's conjecture have been proved in two dimensions for so-called \textit{harmonic} materials \cite{Ru-05}, \cite{Kim-07}, \cite{Kim-08}. Although nonlinear problems are generally more difficult that linear elastostatics, the nonlinearity frees up a number of issues that are more constrained in linear problems. The study of nonlinear problems with dilatational eigenstrain was recently carried out in \cite{Yav-13}. Giordano \cite{Gio-08} considered the nonlinearly elastic inhomogeneity problem but where the constitutive behaviour is described via expansions in strain (Landau elasticity).

Here attention is restricted to linear problems for ellipsoidal inhomogeneities and associated limits. A general approach to deriving the Hill tensor and proving many of its properties is to use the integral equation form of the governing equations \cite{Wil-81}. In fact Eshelby approached the problem in quite a different manner, using the concept of \textit{eigenstrain} \cite{Esh-57}. Hill \cite{Hil-65} considered the so-called \textit{polarization} (hence P) of an ellipsoid. The review articles of Walpole \cite{Wal-81} and Willis \cite{Wil-81}, who developed the integral form of the P-tensor have been very influential and the text of Mura \cite{Mur-82} describes the associated Green's tensor and form of Eshelby tensors for elastostatics in detail. The consideration of isolated inhomogeneity problems allows the derivation of so-called \textit{concentration tensors} for dilute micromechanical schemes, where interactions between inhomogeneities are not important \cite{Wu-66}. In the field of micromechanics a number of very ingenious approximations have been made that lead to rather excellent predictions of effective properties in the case where interactions amongst inhomogeneities are important (see e.g.\ \cite{Wil-81}, \cite{Wen-84}, \cite{Mar-00} for broad overviews). Finally it is noted that variational bounds can be conveniently written down in terms of the Hill or Eshelby tensors \cite{Has-63b}, \cite{Wil-77}, \cite{Wil-81}, \cite{Pon-95}, \cite{Cal-14}, \cite{Par-15}.

There is no real preference for the direct integral equation approach leading to the Hill tensor, over the Eshelby eigenstrain approach. It is chiefly down to individual preference although it is important to note that Hill's tensor possesses the major symmetries whereas Eshelby's does not in general. Some find the notion of eigenstrain rather artificial, although in many cases it is a very useful concept as a means for solving harder problems such as the case of multiple inhomogeneities \cite{Mos-75}, \cite{Zho-11}. The simple relation
\begin{align}
\mathbf{S}=\mathbf{P}\mathbf{C}^0 \label{PCS}
\end{align}
between the Hill (P) and Eshelby (S) tensor, where $\mathbf{C}^0$ is the host modulus tensor, means that deriving one immediately yields the other.

The Hill and Eshelby tensors are of great utility in a number of micromechanical methods and what is quite astonishing is that they can be evaluated analytically in a large number of very important cases. However, results are distributed over a large number of articles, reviews and textbooks, and furthermore often in articles that span a wide range of scientific fields due to the wide ranging applicability of the theory. References dealing with derivations of specific results are those of \cite{Wil-81}, \cite{Wal-81}, \cite{Mar-00}, \cite{Qu-06}, \cite{Bur-07}, \cite{Dvo-13} and \cite{Li-08}. The field is still very much alive, pushed forward by both unresolved theoretical issues as well as applications involving not only inhomogeneities but also cracks and dislocations \cite{Mur-96}, \cite{Zho-13} and by the desire to fully resolve the open issues described above. Recent work has focused in more detail on inhomogeneities of general shape and how these can feed into models of inhomogeneous media with distributions of non-canonical inhomogeneities \cite{Bur-07}, \cite{Bur-11}, \cite{Bur-13}, \cite{Zho-11}, \cite{Zou-10}, \cite{Zou-11}. Such studies are important to understand how local stress fields develop in the medium under loading. This is highly dependent upon the inhomogeneity shape.

Here the objective is to gather together important results associated with the Hill and Eshelby tensors for ellipsoidal inhomogeneities in consistent notation, derive a number of important limiting cases such as those associated with cracks and cavities, derive compact results associated with the anisotropic potential problem and finally derive and state associated concentration tensors. This should prove useful to many who frequently require the form of the P- or S-tensors in practice but who struggle to find the appropriate reference.

An important point to note is that using the so-called invariant notation, potential and linear elastostatics problems can be considered simultaneously, only that the latter is a higher order tensor analogue of the former. Here however the applications are made distinct to stress the different results and mechanisms for deriving these expressions. In particular the results from potential theory feed into those from linear elastostatics. As a result index notation shall be used almost entirely throughout.

In much of the literature on micromechanics the terms \textit{inclusion} and \textit{inhomogeneity} are used interchangeably. However in some cases they are used to make an important distinction.  An \textit{inhomogeneity} is defined as a particle of general shape having different material properties to those of the surrounding medium in which it is embedded. On the other hand the terminology \textit{inclusion} is used to represent a general shaped region within some medium that has the \textit{same} properties as the surrounding medium but where this finite inclusion region has been subject to some eigenstrain (e.g.\ thermal strain). This differentiation is used e.g.\ in Mura \cite{Mur-82} and Qu and Cherkaoui \cite{Qu-06}.

In \S \ref{sec:integral} the integral equation formulation of the inhomogeneity problem is stated, yielding integral equations for the potential gradient and strain inside an inhomogeneity. In \S \ref{sec:uniform} it is illustrated that such fields are uniform when the inhomogeneity is ellipsoidal and the general expressions for the associated Hill tensors are stated. The notion of concentration tensors is also discussed. In \S\S \ref{potspecific} and \ref{sec:elasto} specific results are then stated and derived for the cases of the Newtonian potential problem and elastostatics respectively. A closing discussion is given in \S \ref{sec:disc} describing how the results are used in micromechanical methods together with a summary of current areas of associated research. Numerous important details and results are stated in Appendices in order for this review to be comprehensive but also to aid the flow of the reader.

As many pertinent references are given as possible; the focus is specifically on the formulation of the Eshelby, Hill and concentration tensors rather than articles associated with micromechanical methods, of which there are thousands. For the latter the interested reader is referred to the many textbooks that have been written over the last decade, see e.g.\  \cite{Qu-06}, \cite{Bur-07}, \cite{Kan1-08}, \cite{Li-08}, \cite{Dvo-13}.

\section{Integral equation formulation} \label{sec:integral}

Index notation shall be used for tensors throughout, working in Cartesian coordinates and using repeated subscripts to imply summation. The term \textit{unbounded} will be used when referring to free-space, i.e.\ unbounded in all directions. Although a general invariant formulation can be employed to deal with problems in the potential and linear elastostatics context simultaneously \cite{Wil-81}, this approach can obfuscate details that are important when it comes to deriving specific Hill and Eshelby tensors for given anisotropies and inhomogeneity shapes.

\begin{figure} [ht]
\centering
\psfrag{W}{$V_0$}
\psfrag{V}{$V_1$}
\psfrag{d}{$\pa V_1$}
\hspace{-1cm}\includegraphics[width=220pt]{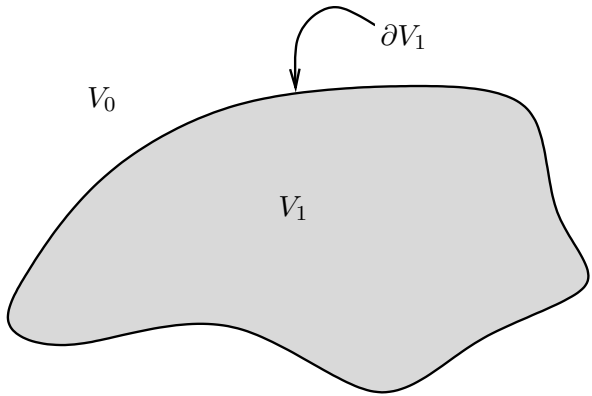}
\caption{An inhomogeneity $V_1$ of general shape and with boundary $\pa V_1$ is embedded perfectly inside the host medium $V_0$. The classical inhomogeneity problem is to determine the fields that arise within the inhomogeneity and host medium given some far-field condition.}
\label{fig:notation}
\end{figure}

Notation is as defined in Fig.\ \ref{fig:notation} for both the potential problem and linear elastostatics. A single isolated inhomogeneity $V_1$, for the time being of general shape and with surface $\pa V_1$ is embedded (perfectly) inside an unbounded homogeneous medium $V$ and we denote the medium exterior to $V_1$ as $V\backslash V_1=V_0$. Both materials are considered generally anisotropic so that their material modulus tensors are
\begin{align}
C_{ij}(\mathbf{x}) &= C_{ij}^1 \chi^1(\mathbf{x}) + C_{ij}^0 (1-\chi^1(\mathbf{x}))
\end{align}
in the context of the potential problem and
\begin{align}
C_{ijk\ell}(\mathbf{x}) &= C_{ijk\ell}^1 \chi^1(\mathbf{x}) + C_{ijk\ell}^0 (1-\chi^1(\mathbf{x}))
\end{align}
in the context of elastostatics. Here the so-called \textit{characteristic function} associated with a domain $V_1$, has been employed, being defined as
\begin{align}
\chi^1(\mathbf{x}) &=
\begin{cases}
1, & \mathbf{x}\in V_1,\\
0, & \mathbf{x}\notin V_1.
\end{cases} \label{5charfunc}
\end{align}
Finally it is noted that the inhomogeneity ($C_{ij}^1$ and $C^1_{ijk\ell}$) and host ($C_{ij}^0$ and $C^0_{ijk\ell}$) modulus tensors are uniform tensors, meaning that each component of the tensor is constant but these constants can be different.

\subsection{The potential problem} \label{subsec:singinctrans}

Since it is often useful to consider a specific physical problem, certainly in terms of language and terminology, the potential problem is described in the context of steady state thermal conductivity. The equation governing the steady state temperature distribution $T(\mathbf{x})$ in the medium described above and depicted in Fig.\ \ref{fig:notation} is
\begin{align}
\deriv{}{x_i}\left(C_{ij}(\mathbf{x})\deriv{T}{x_j}\right) &= 0 \label{4:pdetemp}
\end{align}
where we note that no heat sources are present. The free-space Green's function associated with the \textit{host} phase satisfies
\begin{align}
\deriv{}{x_i}\left(C_{ij}^0\deriv{G}{x_j}(\mathbf{x}-\mathbf{y})\right) + \de(\mathbf{x}-\mathbf{y}) &= 0 \label{4GFscalar}
\end{align}
as well as the far-field condition $\lim_{\mathbf{x}\rightarrow\infty}G(\mathbf{x})=0$. Assuming continuity of temperature and normal flux across $\pa V_1$, the resulting temperature distribution may be straightforwardly derived in integral equation form as
\begin{align}
T(\mathbf{y}) = T^*(\mathbf{y})  - (C_{kj}^1-C_{kj}^0)\int_{V_1}\deriv{T}{x_k}(\mathbf{x})\deriv{G}{x_j}(\mathbf{x}-\mathbf{y})\lit d\mathbf{x} \label{tempinteq}
\end{align}
which holds for all $\mathbf{y}$. Here $T^*(\mathbf{y})$  is the solution to the equivalent problem satisfying \eqref{4:pdetemp} with no inhomogeneity present (or equivalently with $C_{ij}^1 = C_{ij}^0$). Upon taking derivatives of \eqref{tempinteq} with respect to $y_i$ and noting the property $\pa G/\pa x_i= -\pa G/\pa y_i$ it is found that for all $\mathbf{y}\in V$,
\begin{align}
e_i(\mathbf{y}) &= e_i^*(\mathbf{y}) + (C_{kj}^1-C_{kj}^0)\derivtwomix{}{y_i}{y_j}\int_{V_1}e_k(\mathbf{x})G(\mathbf{x}-\mathbf{y})\lit d\mathbf{x}  \label{dTuni}
\end{align}
where the $i$th component of the temperature gradient has been defined as $e_i=\pa T/\pa x_i$.

\subsection{Elastostatics} \label{subsec:singincelastostatics}

The origins of the P-tensor reside in the context of elastostatics rather than in potential problems even though the theory is of course analogous. The P-tensor originated with Hill \cite{Hil-65} who also introduced the compact notation (now commonly referred to as \textit{Hill notation}) for transversely isotropic (TI) fourth order tensors, which we summarize in Appendix \ref{app:apptensTI}. Walpole \cite{Wal-77}, Willis \cite{Wil-77, Wil-80a, Wil-81} and Laws \cite{Law-77} amongst others followed this with influential work associated with inhomogeneities of specific shapes, paying particular attention in many cases to the scenarios of discs, fibres and cracks. A number of P-tensors are also stated in the excellent concise review of micromechanics by Markov \cite{Mar-00} although unfortunately, some typographical errors are present there and we correct those here.

The solution to the isolated inhomogeneity problem in elastostatics proceeds analogously to the potential problem with an expected increase in complexity.  The equations governing the elastic displacement in the medium described above and depicted in Fig.\ \ref{fig:notation} is
\begin{align}
\deriv{}{x_j}\left(C_{ijk\ell}(\mathbf{x})\deriv{u_k}{x_{\ell}}\right) &= 0 \label{4:dispeqn}
\end{align}
where body forces have been neglected. The associated Green's tensor of the host phase satisfies
\begin{align}
\deriv{}{x_j}\left(C_{ijk\ell}^0\deriv{G_{kr}}{x_{\ell}}\right)+\de_{ir}\de(\mathbf{x}-\mathbf{y}) &= 0 \label{4:GFelastic}
\end{align}
as well as the far-field condition $\lim_{\mathbf{x}\rightarrow\infty}G_{ij}(\mathbf{x})=0$, noting that $G_{ij}=G_{ji}$. The resulting displacement field in the medium may be straightforwardly derive in integral equation form as
\begin{align}
u_i(\mathbf{y}) = u_i^*(\mathbf{y}) - (C_{mnk\ell}^1-C_{mnk\ell}^0)\inty{V_1}{}{e_{mn}(\mathbf{x})\deriv{G_{ki}}{x_{\ell}}(\mathbf{x}-\mathbf{y})}{\mathbf{x}}, \label{4:disppert}
\end{align}
which holds for all $\mathbf{y}$. Here $u_i^*(\mathbf{y})$ is the solution to the equivalent problem satisfying \eqref{4:dispeqn} with no inhomogeneity present, or equivalently $C^1_{ijk\ell}=C^0_{ijk\ell}$. As in the potential problem, take derivatives of both sides of \eqref{4:disppert} to form the strain tensor $e_{ij}=(\pa u_i/\pa x_j +\pa u_j/\pa x_i)/2$, using the property $\pa G_{ki}/\pa x_j= -\pa G_{ki}/\pa y_j$ so that we have, for all $\mathbf{y}\in V$,
\begin{align}
e_{ij}(\mathbf{y}) &= e_{ij}^*(\mathbf{y}) + (C_{mnk\ell}^1-C_{mnk\ell}^0)\left[\derivtwomix{}{y_{\ell}}{y_j}\inty{V_1}{}{e_{mn}(\mathbf{x})G_{ki}(\mathbf{x}-\mathbf{y})}{\mathbf{x}}\right]\Bigg|_{(k\ell),(ij)}. \label{4:strainpert}
\end{align}
Here the notation $\big|_{(k\ell),(ij)}$ indicates symmetry with respect to these indices, i.e.\ defining
\begin{align}
Q_{mnki} &= \inty{V_1}{}{e_{mn}(\mathbf{x})G_{ki}(\mathbf{x}-\mathbf{y})}{\mathbf{x}},
\end{align}
we have
\begin{align}
\derivtwomix{Q_{mnki}}{y_{\ell}}{y_j}\Bigg|_{(k\ell),(ij)} &= \frac{1}{4}\left(\derivtwomix{Q_{mnki}}{y_{\ell}}{y_j}+\derivtwomix{Q_{mn\ell i}}{y_k}{y_j}+\derivtwomix{Q_{mnkj}}{y_{\ell}}{y_i}+\derivtwomix{Q_{mn\ell j}}{y_k}{y_i}\right). \label{5:Qsym}
\end{align}

\section{Uniformity of the Hill and Eshelby tensors} \label{sec:uniform}

\subsection{The potential problem}

Impose so-called \textit{homogeneous temperature gradient conditions} (in the language of micromechanics, i.e.\ such conditions would lead to a homogeneous temperature gradient in an inhomogeneous medium) in the far field, i.e. as $|\mathbf{x}|\rightarrow\infty$,
\begin{align}
T &\rightarrow \theta_i x_i,  \label{4temphombc}
\end{align}
where $\theta_i$ is uniform and therefore $T^* = \theta_i x_i$ and $e_i^*=\theta_i$. Referring to \eqref{dTuni}, one then asks, is there an inhomogeneity of any shape that can give rise to a uniform temperature gradient field \textit{inside the inhomogeneity}, i.e.\ for $\mathbf{y}\in V_1$?  If such an inhomogeneity does exist, then \eqref{dTuni} is only consistent for $\mathbf{y}\in V_1$ if the tensor defined as
\begin{align}
P_{ij}(\mathbf{y}) &=  -\derivtwomix{}{y_i}{y_j}\inty{V_1}{}{G(\mathbf{x}-\mathbf{y})}{\mathbf{x}} \label{4:Pijuniform}
\end{align}
is also uniform, i.e.\ is independent of $\mathbf{y}$. The tensor $\mathbf{P}$ with components $P_{ij}$ defined in \eqref{4:Pijuniform} is known as \textit{Hill's Polarization (P) tensor} for the potential problem and it possesses the symmetry $P_{ij}=P_{ji}$. If $\mathbf{P}$ is \textit{not} uniform, it would mean that the assumption of a uniform temperature gradient field \textit{inside} the inhomogeneity was incorrect.

It transpires that when the inhomogeneity region is \textit{ellipsoidal} the P-tensor defined in \eqref{4:Pijuniform} is indeed uniform. This is proved in Appendix \ref{app:ellipsoidpot}, where it is also shown that the general form for the P-tensor can be defined in terms of an integral over the surface of the unit sphere $S^2$.

\vspace{0.2cm}

\begin{center}
\setlength{\fboxsep}{4pt}
\framebox[13cm][c]
{\begin{minipage}{12.5cm}
\begin{center}\textbf{General form of Hill's tensor for the potential problem:\\ Ellipsoid in an unbounded medium}\end{center}
\vspace{0.05cm}
The components of Hill's tensor are defined as
\begin{align}
P^{\tn{ellipsoid}}_{ij} &= \frac{\tn{det}(\mathbf{a})}{4\pi}\int_{S^2}\frac{\Phi_{ij}(\overline{\boldsymbol{\xi}})}{(\overline{\xi}_k a_{k\ell}a_{\ell m}\overline{\xi}_m)^{3/2}}
\hspace{0.1cm}dS(\overline{\boldsymbol{\xi}}) \label{5genPtransport}
\end{align}
where $\overline{\boldsymbol{\xi}}=(\overline{\xi}_1,\overline{\xi}_2,\overline{\xi}_3)$ is a unit vector that points from the origin, i.e.\ the centre of $S^2$, to its surface. Additionally $\Phi_{ij}$ is given by
\begin{align}
\Phi_{ij}(\overline{\boldsymbol{\xi}}) &= \frac{\overline{\xi}_i\overline{\xi}_j}{C^0_{k\ell}
\overline{\xi}_k\overline{\xi}_{\ell}}
\end{align}
and $\mathbf{a}$ is a second order tensor whose components are defined by
\begin{align}
a_{ij}= \sum_{n=1}^3 a_n\de_{in}\de_{jn} \label{analigned}
\end{align}
as long as the semi-axes of the ellipsoid are aligned along the $x_1, x_2$ and $x_3$ axes, so that $\tn{det}(\mathbf{a})=a_1a_2a_3$. In fact it is \textit{always} possible to define $\mathbf{a}$ in this manner by choosing $x_1, x_2$ and $x_3$ to be aligned along the semi-axes of the ellipsoid, as long as one is happy for the principal axes of $C^0_{ij}$ to be defined in different directions to $x_1, x_2$ and $x_3$ should the principal axes of $C^0_{ij}$ and $\mathbf{a}_{ij}$ \textit{not} be aligned.
\end{minipage}}
\end{center}
\vspace{0.2cm}

Clearly since the integral is over the surface of the unit sphere $S^2$, it is sensible to resolve $\overline{\xi}_i$ into spherical coordinates for the purposes of evaluating this integral.
The form of \eqref{5genPtransport} illustrates the important general result that the P-tensor is uniform for an arbitrarily anisotropic ellipsoidal inhomogeneity embedded inside an arbitrarily anisotropic host phase. That Eshelby's (weak) conjecture is true for anisotropic potential problems \cite{Kan-08}, \cite{Liu-08}, means that the ellipsoid is the \textit{only} shaped inhomogeneity for which the interior temperature gradient is uniform under \textit{all} such far-field conditions of the form \eqref{4temphombc}.


To determine the appropriate P-tensor in any circumstance then one can appeal to \eqref{5genPtransport} and carry out the necessary integration. Alternatively, as shall be shown in \S \ref{potspecific}, in many cases it is relatively straightforward to use symmetry arguments and results from potential theory in the isotropic host case together with scalings in some cases of host anisotropy, in order to derive explicit results, often in a more straightforward manner than directly evaluating the general result \eqref{5genPtransport}. In fact in the potential problem context, symmetry arguments and results from potential theory \cite{Kel-70} are often sufficient to derive results for many special cases of \textit{ellipsoids} in host media that are at most \textit{orthotropic}. The general result \eqref{5genPtransport} is thus suitable for more complex anisotropies than orthotropy or for example if the semi-axes of the ellipsoid are \textit{not} aligned with the axes of symmetry of host anisotropy.

We should recall that the P-tensor is independent of the anisotropy of the inhomogeneity and therefore we can retain arbitrary anisotropy for the inhomogeneity domain. The only aspects of the inhomogeneity that influence the P-tensor are its \textit{shape} and, for \textit{anisotropic} host phases, its \textit{orientation} with respect to the axes of anisotropy of the host phase.
It is important to note the following three points:
\begin{itemize}
\item In the \textit{host} region $V_0$ the temperature gradient is generally \textit{not} uniform.
\item For non-homogeneous temperature gradient conditions in the far field, the temperature gradient field inside an ellipsoidal inhomogeneity is generally \textit{not} uniform. However if the prescribed temperature gradient is a polynomial of order $n$, then so is the field inside an ellipsoidal inhomogeneity, see \cite{Asa-75}. This is known as the polynomial conservation property for ellipsoids.
\item Generally for non-ellipsoidal inhomogeneities in unbounded domains and general shaped inhomogeneities in bounded host domains $V$, the temperature gradient inside the inhomogeneities is not uniform, although interacting \textit{E-inclusions} \cite{Liu-08} can lead to uniform interior strains and for specific loadings, non-ellipsoidal inhomogeneities can yield uniform interior strains, e.g.\ the counterexample of the Strong Eshelby conjecture given by Liu \cite{Liu-08}.
\end{itemize}
Regarding the first point, once we know the \textit{interior} field \eqref{5e1eP}, we can use this to determine the \textit{exterior} field by using \eqref{tempinteq} so that for $\mathbf{y}\notin V_1$,
\begin{align}
T(\mathbf{y}) &= \theta_i y_i - (C_{kj}^1-C_{kj}^0)\mathcal{A}_{k\ell}\theta_{\ell}\inty{V_1}{}{\deriv{G}{x_j}(\mathbf{x}-\mathbf{y})}{ \mathbf{x}} \label{4tempexteq}
\end{align}
where $\mathcal{A}_{ij}$ is the temperature gradient concentration tensor linking the interior temperature gradient to that in the far-field, i.e.\ $\theta_j$, see \S \ref{sec:conc2}. The gradient of \eqref{4tempexteq} is \textit{not} uniform since $\mathbf{y}$ now lies outside $V_1$. 


\subsection{Elastostatics} \label{subsec:41elastostatics2}

We used the symmetry relation $C_{ijk\ell}=C_{ij\ell k}$ in deriving \eqref{4:strainpert} as this turns out to be preferable in various contexts. Analogously to the potential problem, let us take \textit{homogeneous displacement gradient conditions} in the far field, i.e.\ as $|\mathbf{x}|\rightarrow \infty$
\begin{align}
u_i &\rightarrow \eps_{ij}x_j, \label{4disphombc}
\end{align}
where $\eps_{ij}$ is uniform and therefore $u^*_i=\eps_{ij}x_j$ and $e^*_{ij}=(\eps_{ij}+\eps_{ji})/2$. Note that $\eps_{ij}$ does not have to be symmetric but if it is then it is simply the strain in the far field.
As in the potential problem case the aim is then determine if there exists an inhomogeneity of any shape that is consistent with the assumption of uniform interior strain. If such an inhomogeneity exists, \eqref{4:strainpert} is only consistent for $\mathbf{y}\in V_1$ if the tensor defined as
\begin{align}
P_{ijk\ell}(\mathbf{y}) &= -\left[\derivtwomix{}{y_j}{y_{\ell}}\inty{V_1}{}{G_{ik}(\mathbf{x}-\mathbf{y})}{\mathbf{x}}\right]\Bigg|_{(ij),(k\ell)}  \label{4:Paniso}
\end{align}
is uniform. The tensor defined here is the P-tensor for elastostatics. It possesses the minor symmetries $P_{ijk\ell}=P_{ij\ell k}=P_{jik\ell}$ by construction. Furthermore, thanks to the symmetry of the free space Green's tensor $G_{ij}=G_{ji}$ it also possesses the major symmetry $P_{ijk\ell}=P_{k\ell ij}$.

It transpires that when the inhomogeneity region is \textit{ellipsoidal} the P-tensor defined in \eqref{4:Paniso} is indeed uniform. This is proved in Appendix \ref{app:ellipsoidpot}, where it is also shown that the general form for the P-tensor can be defined in terms of an integral over the surface of the unit sphere $S^2$.
\vspace{0.2cm}

\begin{center}
\setlength{\fboxsep}{4pt}
\framebox[13cm][c]
{\begin{minipage}{12.5cm}
\begin{center}\textbf{General form of Hill's tensor for linear elastostatics:\\ Ellipsoid in an unbounded medium}\end{center}
\vspace{0.05cm}
The components of Hill's tensor are defined as
\begin{align}
P_{ijk\ell}^{\tn{ellipsoid}} &= \frac{\det(\mathbf{a})}{4\pi}\int_{S^2}\frac{\Phi_{ijk\ell}(\overline{\boldsymbol{\xi}})}
{(\overline{\xi}_m a_{mn}a_{np}\overline{\xi}_p)^{3/2}}dS(\overline{\boldsymbol{\xi}}) \label{5Pijklellipsoid}
\end{align}
where $\overline{\boldsymbol{\xi}}$ and $S^2$ are as defined for the potential problem and $\mathbf{a}$ is defined in \eqref{analigned}. Furthermore
\begin{align}
\Phi_{ijk\ell} &= (\overline{\xi}_j\overline{\xi}_{\ell}N_{ik}(\overline{\boldsymbol{\xi}})))\Big|_{(ij),(k\ell)}
\end{align}
where $N_{ij}$ is defined via
\begin{align}
N_{ik}\tilde{N}_{kj} &=\de_{ij}, & \tilde{N}_{ij}(\overline{\boldsymbol{\xi}}) &= C^0_{ijk\ell}\overline{\xi}_j\overline{\xi}_{\ell}.
\end{align}

\end{minipage}}
\end{center}
\vspace{0.2cm}

That Eshelby's (weak) conjecture is true for isotropic elastostatics problems \cite{Kan-08}, \cite{Liu-08}, means that the ellipsoid is the \textit{only} shaped inhomogeneity for which the interior temperature gradient is uniform under \textit{all} such far-field conditions of the form \eqref{4disphombc}. We stress however that it is not yet clear whether the weak conjecture is true in the context of anisotropic problems.



To determine the P-tensor for an ellipsoid for a given host anisotropy one merely has to evaluate the surface integral in \eqref{5Pijklellipsoid} which can evaluated numerically very efficiently. For host anisotropies more complex than transversely isotropic it is generally recommended that the form \eqref{5Pijklellipsoid} be employed and integrals are evaluated numerically. In what follows here the P-tensor shall be determined in the case of an isotropic host phase by appealing to various symmetries and potential theory. An important result derived by Withers \cite{Wit-89} associated with a transversely isotropic host phase is also stated.


As in the potential problem, the only aspects of the inhomogeneity that influence the P-tensor are its shape and, for anisotropic host phases, its orientation with respect to the axes of anisotropy of the host phase. Note also that the same three points described for the potential problem, preceding equation \eqref{4tempexteq}, also hold here in the elastostatics context. Furthermore, once the field is known inside the inhomogeneity region $V_1$ the exterior field can be determined in terms of the Green's tensor, as
\begin{align}
u_i(\mathbf{y}) &= \bar{\eps}_{ij}y_j - (C_{mnk\ell}^1-C_{mnk\ell}^0)\mathcal{A}_{mnpq}e^*_{pq}\inty{V_1}{}{\deriv{G_{ki}}{x_{\ell}}(\mathbf{x}-\mathbf{y})}{\mathbf{x}} \label{4:disppert2}
\end{align}
where $\mathcal{A}_{ijk\ell}$ are the components of the strain concentration tensor (see \S \ref{sec:conc2}), which links the strain inside the inhomogeneity to that in the far field.

\subsection{The Newtonian potential gradient and strain concentration tensors} \label{sec:conc2}

Defining the volume average
\begin{align}
\overline{f} &= \frac{1}{|V|}\int_V f(\mathbf{x}) \lit d\mathbf{x}
\end{align}
of the function $f$, it is straightforward to show that in the case of the conditions \eqref{4temphombc}, the body averaged temperature gradient is
\begin{align}
\overline{e}_i &= \theta_i = e^*_i. \label{tempgradequiv}
\end{align}
Of immediate interest is the temperature gradient field $e^1_i$ inside an ellipsoidal inhomogeneity $V_1$, which from the theory developed above has been shown to be uniform so that it is equal to its phase average, $e^1_i=\overline{e}^1_i$ where the phase average is defined as
\begin{align}
\overline{f}^1 &= \frac{1}{|V_1|}\int_{V_1} f(\mathbf{x}) \lit d\mathbf{x}.
\end{align}
As such in the case of an isolated ellipsoidal inhomogeneities with homogeneous far-field conditions \eqref{4temphombc}, using \eqref{tempgradequiv}, the expression in \eqref{dTuni} becomes
\begin{align}
e^1_i &= \overline{e}_i - (C^1_{kj}-C^0_{kj})e^1_k P_{ij}. \label{5e1eP}
\end{align}
Using the symmetries $C_{ij}=C_{ji}$ and $P_{ij}=P_{ji}$ and re-arranging, \eqref{5e1eP} can thus be written in the form
\begin{align}
\overline{e}_i &= (\de_{ij}+P_{ik}(C_{kj}^1-C_{kj}^0))e^1_j. \label{5einfe1}
\end{align}
Therefore one can relate the uniform temperature gradient \textit{inside} the inhomogeneity to the average temperature gradient inside the entire body via a second order tensor, which is thus identified as the \textit{temperature gradient concentration tensor} for this problem.
\vspace{0.2cm}

\begin{center}
\setlength{\fboxsep}{4pt}
\framebox[13cm][c]
{\begin{minipage}{12.5cm}
\begin{center}\textbf{Temperature gradient concentration tensor:\\ Ellipsoid in an unbounded medium}\end{center}
\vspace{0.05cm}
For an ellipsoidal inhomogeneity $V_1$ embedded in an otherwise unbounded uniform medium, if homogeneous temperature gradient conditions \eqref{4temphombc} are prescribed in the far field, we have the exact relationship
\begin{align}
\overline{e}_i^1 = e_i^1 &= \mathcal{A}_{ij}\overline{e}_j \label{5transconc2}
\end{align}
where the uniform concentration tensor $\mathcal{A}_{ij}$ is defined by
\begin{align}
\mathcal{A}_{ik}\tilde{\mathcal{A}}_{kj} &= \de_{ij}, & \tilde{\mathcal{A}}_{ij} &= \de_{ij}+P_{ik}(C_{k j}^1-C_{k j}^0) \label{5:conctemp}
\end{align}
and $P_{ij}$ is defined in \eqref{4:Pijuniform}.
\end{minipage}}
\end{center}
\vspace{0.2cm}

Note that $\mathcal{A}_{ij}$ is the concentration tensor associated with an isolated inhomogeneity inside an \textit{unbounded} host medium. The calligraphic notation $\mathcal{A}_{ij}$ has been used to stress the link with (and distinguish from) the \textit{exact} concentration tensor, usually defined as $A_{ij}$, and which links the phase average of the true temperature gradient inside an inhomogeneity to that in the far field in a complex inhomogeneous medium, which may consist of interacting inhomogeneities. For a dilute medium where interaction effects are not important, $A_{ij}=\mathcal{A}_{ij}$.

Moving on to the elastostatics case, it is straightforward to show that in the case of the conditions \eqref{4disphombc}, the body averaged strain is
\begin{align}
\overline{e}_{ij} &= \frac{1}{2}(\eps_{ij}+\eps_{ji}) = e^*_{ij}. \label{strainequiv}
\end{align}
The (uniform) strain $e^1_{ij}$ inside an ellipsoidal inhomogeneity $V_1$ is thus equal to its phase average, $e^1_{ij}=\overline{e}^1_{ij}$. Therefore for an isolated ellipsoidal inhomogeneity with homogeneous far-field conditions \eqref{4disphombc}, using \eqref{strainequiv}, the expression in \eqref{4:strainpert} can be used to determine the expression
\begin{align}
\bar{e}_{ij} = (I_{ijk\ell}+P_{ijmn}(C_{mnk\ell}^1-C_{mnk\ell}^0))\bar{e}_{k\ell}^1 \label{4:strainconciso}
\end{align}
where we remind the reader that $I_{ijk\ell}$ is the fourth order identity tensor defined in \eqref{4thIDapp}. Therefore the uniform strain \textit{inside} the inhomogeneity can be related to the average strain inside the entire body via a fourth order tensor, which is thus identified as the \textit{strain concentration tensor} for this problem.
\vspace{0.1cm}

\begin{center}
\setlength{\fboxsep}{4pt}
\framebox[13cm][c]
{\begin{minipage}{12.5cm}
\begin{center}\textbf{Strain concentration tensor:\\ Ellipsoid in an unbounded medium}\end{center}
\vspace{0.05cm}
For an ellipsoidal inhomogeneity $V_1$ embedded in an otherwise unbounded medium, if homogeneous displacement conditions \eqref{4disphombc} are prescribed in the far field, we have the relationship
\begin{align}
e_{ij}^1 &= \mathcal{A}_{ijk\ell}\overline{e}_{k\ell} \label{5strainrel}
\end{align}
where the uniform concentration tensor $\mathcal{A}_{ijk\ell}$ is defined by
\begin{align}
\mathcal{A}_{ijmn}\tilde{\mathcal{A}}_{mnk\ell} &= I_{ijk\ell}, & \tilde{\mathcal{A}}_{ijk\ell} &= I_{ijk\ell}+P_{ijmn}(C_{mnk\ell}^1-C_{mnk\ell}^0). \label{5strainconcA}
\end{align}
\end{minipage}}
\end{center}
\vspace{0.1cm}

\section{The potential problem: specific cases} \label{potspecific}

\subsection{Isotropic host phase}

Assume that the host phase is isotropic, so that $C_{ij}^0=k_0\de_{ij}$ and therefore the associated free-space Green's function is
\begin{align}
G(\mathbf{x}-\mathbf{y}) &= \frac{1}{4\pi k_0}\frac{1}{|\mathbf{x}-\mathbf{y}|}. \label{5transisoGF}
\end{align}
From \eqref{4:Pijuniform} and \eqref{PCS} therefore
\begin{align}
P_{ij}(\mathbf{x}) &= \frac{1}{k_0}\derivtwomix{\Gamma}{x_i}{x_j}, & S_{ij}(\mathbf{x}) &= \derivtwomix{\Gamma}{x_i}{x_j} \label{Pphi}
\end{align}
where $\Gamma$ is the potential defined by
\begin{align}
\Gamma(\mathbf{x}) &= -\frac{1}{4\pi}\inty{V_1}{}{\frac{1}{|\mathbf{x}-\mathbf{y}|}}{\mathbf{y}}. \label{Npot}
\end{align}
Note that this is the \textit{negative} of the Newtonian potential (see for example Kellogg \cite{Kel-70}) associated with an ellipsoidal domain $V_1$. From potential theory therefore
\begin{align}
\nabla^2\Gamma(\mathbf{x}) = \derivtwomix{\Gamma}{x_k}{x_k}  &= \chi^1(\mathbf{x})
\label{nablaPhi}
\end{align}
and furthermore $\Gamma(\mathbf{x})$ is a quadratic function of the components of $\mathbf{x}$ (see Appendix \ref{app:potential}), illustrating the uniformity of the P-tensor in this case.

As an aside, note that since the host phase is isotropic, the temperature field exterior to the inhomogeneity is determined via \eqref{4tempexteq}, i.e.\
\begin{align}
T(\mathbf{y}) &= \theta_i y_i + (C_{kj}^1-k_0\de_{kj})\mathcal{A}_{k\ell}\theta_{\ell} \frac{1}{k_0}\deriv{\Gamma(\mathbf{y})}{y_j}. \label{tempexteqiso}
\end{align}
This solution tends to $\theta_i y_i$ in the far field, as it should do.

Once $P_{ij}$ is determined for an isotropic host phase the associated concentration tensor for an isolated inhomogeneity may then be found from \eqref{5:conctemp} as is now illustrated in a number of special cases of specific inhomogeneities with given shape and anisotropy.


\subsubsection{Sphere in an isotropic host phase}\label{eganisospheretransport}

When $V_1$ is a sphere, it is clear from \eqref{Npot} that $\Gamma(\mathbf{x})$ must be spherically symmetric and hence $\frac{\pa^2\Gamma}{\pa x_i\pa x_j}$ must be isotropic (and uniform), i.e.\
\begin{align}
\derivtwomix{\Gamma}{x_i}{x_j} &= \ga\de_{ij} \label{nablaPhi2}
\end{align}
for some constant $\ga$. Note that the form \eqref{nablaPhi2} is a result of the spherical shape and \textit{not} any assumption regarding isotropy of the inhomogeneity as such an assumption has not been made. Performing a contraction in \eqref{nablaPhi2} and using \eqref{nablaPhi} with $\mathbf{x}\in V_1$ yields  $\ga=\frac{1}{3}$. Therefore from \eqref{Pphi} and \eqref{PCS}
\begin{align}
P_{ij} &= \frac{1}{3k_0}\de_{ij}, & S_{ij} &= \frac{1}{3}\de_{ij}. \label{PtensisoT}
\end{align}
For practical purposes, especially for use in micromechanical methods for bounds and estimates of effective material properties, it is useful to write down the associated concentration tensors.

\subsubsection*{Isotropic sphere}

If the spherical inhomogeneity is isotropic with conductivity $C_{ij}^1=k_1\de_{ij}$, one can show straightforwardly using \eqref{5:conctemp} and properties of second order tensors (see Appendix \ref{app:tensorstructure2}) that
\begin{align}
\mathcal{A}_{ij} &= \frac{3k_0}{k_1+2k_0}\de_{ij}. \label{Concspheretemp}
\end{align}

\subsubsection*{Anisotropic sphere}

Consider a transversely isotropic sphere where the plane of isotropy is the $x_1 x_2$ plane. The conductivity tensor therefore takes the form $C_{ij}^1 = k_1(\Theta_{ij}+\upsilon\de_{i3}\de_{j3})$ where $\Theta_{ij}$ is defined according to
\begin{align}
\de_{ij}=\Theta_{ij}+\de_{i3}\de_{j3} \label{5Thetaij}
\end{align}
and $\upsilon$ indicates the degree of anisotropy, with $\upsilon=1$ giving isotropy. Using the result derived in \eqref{PtensisoT} and \eqref{5Thetaij} together with properties from Appendix \ref{app:tensorstructure2}, the concentration tensor $\mathcal{A}_{ij}$ can be written down in the form
\begin{align}
\mathcal{A}_{ij} &= \frac{3k_0}{k_1+2k_0}\Theta_{ij} + \frac{3k_0}{\upsilon k_1+2k_0}\de_{i3}\de_{j3}. \label{Concanisosph}
\end{align}
Setting $\upsilon=1$ recovers the isotropic result \eqref{Concspheretemp}.

Averaging over all orientations of the anisotropy of the inhomogeneity will yield an isotropic concentration tensor of the form
\begin{align}
\underline{\mathcal{A}}_{ij} &= \ga\de_{ij}, \label{5anisosphava}
\end{align}
where the underline denotes averaging over orientations. By performing this orientation averaging (see Appendix \ref{app:avor2}) on \eqref{Concanisosph} it is straightforwardly shown that
\begin{align}
\ga &= \frac{2k_0}{k_1+2k_0} + \frac{k_0}{\upsilon k_1+2k_0}. \label{5anisosphav}
\end{align}

\subsubsection{Circular cylinder in an isotropic host phase}\label{egcirccyl}

When $V_1$ is a circular cylinder with axis of symmetry in the $x_3$ direction, it is clear that $\Gamma(\mathbf{x})$ should be independent of $x_3$ and isotropic in the $x_1x_2$ plane so that
\begin{align}
\derivtwomix{\Gamma}{x_i}{x_j} &= \ga\Theta_{ij} \label{5Phicyl}
\end{align}
for some constant $\ga$ where $\Theta_{ij}$ was defined in \eqref{5Thetaij}. Performing a contraction in \eqref{5Phicyl} and using \eqref{nablaPhi} the result $\ga=\frac{1}{2}$ is obtained. From \eqref{Pphi} therefore
\begin{align}
P_{ij} &= \frac{1}{2k_0}\Theta_{ij}, & S_{ij} &= \frac{1}{2}\Theta_{ij}. \label{5PTij}
\end{align}
If the cylinder is \textit{isotropic} with conductivity tensor
\begin{align}
C^1_{ij} &= k_1\de_{ij},
\end{align}
it is straightforward to show that
\begin{align}
\mathcal{A}_{ij} &= \frac{2k_0}{k_1+k_0}\Theta_{ij} + \de_{i3}\de_{j3}. \label{5cylA}
\end{align}
Alternatively, suppose that the cylinder is \textit{transversely isotropic} with conductivity tensor
\begin{align}
C_{ij}^1 &= k_1(\Theta_{ij}+\upsilon\de_{i3}\de_{j3}). \label{anisoCij1}
\end{align}
Interestingly one can show that in this case the concentration tensor is \textit{identical} to the isotropic case, i.e.\ that in \eqref{5cylA}: the parameter $\upsilon$ does not appear in the concentration tensor. Of course if the axis of symmetry of transverse isotropy is \textit{not} aligned with the cylinder axis then this concentration tensor \textit{would} then depend on $\upsilon$.

If the (uniform) orientation average of \eqref{5cylA} is taken, the associated concentration tensor is derived:
\begin{align}
\underline{\mathcal{A}}_{ij} &= \frac{k_1+5k_0}{3(k_1+k_0)}\de_{ij}. \label{5Acylav}
\end{align}
This last result is often used when very long, thin needle-like inhomogeneities are uniformly distributed and oriented throughout some host medium.

\subsubsection{Ellipsoid in an isotropic host phase} \label{chap5ellipsoidtransport}

Consider now the general case of an ellipsoidal inhomogeneity and as usual denote the semi-axes of the ellipsoid as $a_j$, $j=1,2,3$. It is straightforward to show, using the theory of the potential, as in Appendix \ref{app:potential} that for an ellipsoid in an \textit{isotropic} host phase, the function $\Gamma(\mathbf{x})$ is quadratic in the components of $\mathbf{x}$ and can be written in the closed form
\begin{align}
\Gamma(\mathbf{x}) 
&= \left(\frac{x_1^2}{a_1^2}+\frac{x_2^2}{a_2^2}+\frac{x_3^2}{a_3^2}-1\right)\Upsilon - \sum_{j=1}^3 \frac{x_j^2}{a_j}\deriv{\Upsilon}{a_j} \label{5:phiups}
\end{align}
where
\begin{align}
\Upsilon &= \frac{1}{4}a_1a_2a_3\int_0^{\infty}\frac{dt}{\sqrt{(a_1^2+t)(a_2^2+t)(a_3^2+t)}}.
\end{align}
In Appendix \ref{app:potential} it is then shown that
%
\begin{align}
\derivtwomix{\Gamma}{x_i}{x_j} &= \sum_{n=1}^3 \mathcal{E}(\vareps_n;\vareps_1,\vareps_2)\de_{in}\de_{jn} \label{Ptransportellipsoid}
\end{align}
where with $\vareps_n=a_3/a_n$,
\begin{align}
\mathcal{E}(x;\vareps_1,\vareps_2) &= \frac{x^2}{2}\int_0^{\infty}\frac{ds}{(1+sx^2)\sqrt{(1+s\vareps_1^2)(1+s\vareps_2^2)(1+s)}}. \label{5ellipsoidfunction}
\end{align}
Therefore
\begin{align}
P_{ij} &= \frac{1}{k_0}\sum_{n=1}^3 \mathcal{E}(\vareps_n;\vareps_1,\vareps_2)\de_{in}\de_{jn}, & S_{ij} &= k_0P_{ij}. \label{5Pspheroid}
\end{align}
Finally note that using \eqref{nablaPhi} it is easily shown that $\ga_1+\ga_2+\ga_3 = 1$.


\subsubsection{Spheroid in an isotropic host phase}\label{5spheroideg}

Denote the semi-axes of the spheroid as $a_1=a_2=a\neq a_3$ and use this in \eqref{appUpsilonphi} which becomes
\begin{align}
\Upsilon &= \frac{1}{2}a_3^2\int_0^{\pi/2}\frac{\cos\psi}{\vareps^2+(1-\vareps^2)\sin^2\psi}\hspace{0.1cm}d\psi
\end{align}
where $\vareps=a_3/a$. Make the substitution $\be=\sin\psi$ to find
\begin{align}
\Upsilon = \frac{1}{2}a_3^2\int_0^1 \frac{d\be}{\vareps^2+(1-\vareps^2)\be^2}
&= \frac{a_3^2}{2}\begin{cases}
 \frac{\arccosh(\vareps)}{\vareps\sqrt{\vareps^2-1}}, & \vareps>1, \\
 \frac{\arccos(\vareps)}{\vareps\sqrt{1-\vareps^2}}, & \vareps<1, \\
 1, & \vareps=1,
\end{cases}
\end{align}
noting that $\vareps=1$ is the case of a sphere.

Therefore from \eqref{5:phiups} it is clear that
\begin{align}
\Gamma(\mathbf{x}) &= \frac{1}{2}(x_1^2+x_2^2)\mathcal{T}(\vareps)
+\frac{1}{2}x_3^2\mathcal{S}(\vareps)-\Upsilon \label{5:GAMMAUPS}
\end{align}
where
\begin{align}
\mathcal{S}(\vareps) &= \frac{2}{a_3^2}\Upsilon - \frac{2}{a_3}\deriv{\Upsilon}{a_3} \non\\
&= \frac{1}{1-\vareps^2}-\frac{\vareps}{1-\vareps^2}\begin{cases}
\frac{1}{\sqrt{\vareps^2-1}}\arccosh(\vareps), & \vareps>1, \\
\frac{1}{\sqrt{1-\vareps^2}}\arccos(\vareps), & \vareps<1
\end{cases} \label{5spheroidfunction}
\end{align}
and $\mathcal{T}(\vareps)=\frac{1}{2}(1-\mathcal{S}(\vareps))$. The function $\mathcal{S}(\vareps)$ has taken many forms in the literature but it is felt that this is a most clear, consistent and concise formulation. Note that $\mathcal{S}(\vareps)\rightarrow \frac{1}{3}$ as $\vareps\rightarrow 1$ for the spherical case (see further details below). 

Using \eqref{5:GAMMAUPS} in \eqref{Pphi}, the resulting Hill and Eshelby tensors take the form
\begin{align}
P_{ij} &= \frac{1}{k_0}\left(\ga\Theta_{ij} + \ga_3 \de_{i3}\de_{j3}\right), & S_{ij} &= k_0 S_{ij}
\end{align}
where $\ga_3=\mathcal{S}(\vareps)$ and $\ga=\mathcal{T}(\vareps)$.


Expressions for the concentration tensors associated with the spheroidal inhomogeneity case can now be determined straightforwardly. For an isotropic spheroid,
\begin{align}
\mathcal{A}_{ij} &= \dfrac{k_0}{k_0+(k_1-k_0)\ga}\Theta_{ij} + \dfrac{k_0}{k_0+(k_1-k_0)\ga_3}\de_{i3}\de_{j3}. \label{5Aijspheroid}
\end{align}
Averaging uniformly over orientations of the axes of the spheroid yields
\begin{align}
\underline{\mathcal{A}}_{ij} &= \frac{1}{3}\left(\dfrac{2k_0}{k_0+(k_1-k_0)\ga}+\dfrac{k_0}{k_0+(k_1-k_0)\ga_3}\right)\de_{ij}. \label{5Aijspheroidaveraged}
\end{align}

One can take limits in the case of the spheroidal inhomogeneity in order to derive the following results, some of which confirm cases considered above. When $V_1$ is
\hspace{-1cm}
\begin{enumerate}[(i)]
\item a sphere, i.e.\ $\vareps\rightarrow 1$, it is deduced that $\ga=\ga_3 = \frac{1}{3}$,
\item a cylinder, i.e.\ $\vareps\rightarrow\infty$, it is deduced that $\ga_3= 0$,  $\ga=\frac{1}{2}$,
\item a disk or layer, i.e.\ $\vareps\rightarrow 0$, it is deduced that $\ga_3=1$,  $\ga=0$.
\end{enumerate}
When used in \eqref{5Aijspheroid} (i) and (ii) confirm the results derived for the concentration tensors for isotropic spheres and cylinders derived in \S\S \ref{eganisospheretransport} and \ref{egcirccyl} respectively. One has to be rather careful in taking these limits and for (i) use L'Hopital's rule appropriately, noting that as $\vareps\rightarrow 1$
\begin{align}
\mathcal{S}(\vareps) = \frac{1}{3} -\frac{4}{15}(\vareps-1) + \frac{6}{35}(\vareps-1)^2+O((\vareps-1)^3). \label{5:Seps1lim}
\end{align}
In (ii) one has to use the fact that $\arccosh x\sim \log x$ as $x\rightarrow\infty$.  The result for layers in (iii) can also be obtained via straightforward symmetry arguments.

\subsubsection{Limiting case of an elliptical cylinder} \label{chap5:ellipcyl}

One can use the formulation for general ellipsoids above in order to obtain a result for an elliptical cylinder, unbounded in the $x_3$ direction with semi-axes $a_1$ and $a_2$ lying along the $x_1$ and $x_2$ axes respectively. Taking
the limit $a_3\rightarrow \infty$ in \eqref{5:gammana}, one can show that
\begin{align}
\derivtwomix{\Gamma}{x_i}{x_j} &= \sum_{n=1}^2 \ga_n \de_{in}\de_{jn}
\end{align}
where
\begin{align}
\ga_1 &= \frac{a_1a_2}{2}\int_0^{\infty}\frac{ds}{(a_1^2+s)^{\frac{3}{2}}(a_2^2+s)^{\frac{1}{2}}}, & \ga_2 &= \frac{a_1a_2}{2}\int_0^{\infty}\frac{ds}{(a_1^2+s)^{\frac{1}{2}}(a_2^2+s)^{\frac{3}{2}}},
\end{align}
noting the fact that $x_3$ dependence is eliminated as should be expected. The integrals can be determined explicitly, noting the indefinite forms
\begin{align}
\int \frac{ds}{(a_1^2+s)^{\frac{3}{2}}(a_2^2+s)^{\frac{1}{2}}} &= \frac{2}{(a_1^2-a_2^2)}\left(\frac{a_2^2+s}{a_1^2+s}\right)^{\frac{1}{2}}, \\
\int \frac{ds}{(a_1^2+s)^{\frac{1}{2}}(a_2^2+s)^{\frac{3}{2}}} &= \frac{2}{(a_2^2-a_1^2)}\left(\frac{a_1^2+s}{a_2^2+s}\right)^{\frac{1}{2}}
\end{align}
and therefore
\begin{align}
\ga_1 &= \frac{a_2}{a_1+a_2}, & \ga_2 &= \frac{a_1}{a_1+a_2}.
\end{align}
The Hill and Eshelby tensors therefore take the form
\begin{align}
P_{ij} &= \frac{1}{k_0(a_1+a_2)} (a_2\de_{i1}\de_{j1} + a_1\de_{i2}\de_{j2}), & S_{ij} &= k_0P_{ij}.
\end{align}
As regards the concentration tensor for an \textit{isotropic} cylinder with $C^1_{ij}=k_1\de_{ij}$, this is determined in the form
\begin{align}
\mathcal{A}_{ij} &= \frac{k_0(1+\eps)}{k_0+k_1 \eps}\de_{i1}\de_{j1} + \frac{k_0(1+\eps)}{k_1+k_0 \eps}\de_{i2}\de_{j2}
 + \de_{i3}\de_{j3} \label{5ellipticcylAij}
\end{align}
where $\eps=a_2/a_1$ is the aspect ratio of the ellipse.

\subsubsection{Limiting cases of a cavity, penny-shaped crack and ribbon-crack} \label{chap5:cavcrackpot}

It does not really make sense to define a \textit{temperature gradient concentration tensor} in the context of cracks or cavities because clearly there is no interior field. However it turns out that this concept \textit{is} useful and can be interpreted as linking the far-field to the field on the surface of such inhomogeneities \cite{Hor-93} with an appropriate definition of ``cavity temperature gradient''. As such here the results above are used in order to derive associated concentration tensors for cracks and cavities.

Consider a spheroidal inhomogeneity and the limit $k_1\rightarrow 0$ in \eqref{5Aijspheroid}. This yields
\begin{align}
\mathcal{A}_{ij} &= \dfrac{1}{1-\ga}\Theta_{ij} + \dfrac{1}{1-\ga_3}\de_{i3}\de_{j3}. \label{5Aijspheroidcavity}
\end{align}
This is the concentration tensor for potential problems involving spheroidal \textit{cavities}.

Next consider the so-called ``penny-shaped crack'' limit. We require the asymptotic form of $\ga(\vareps)$ and $\ga_3(\vareps)$ as $\vareps\rightarrow 0$. These are easily shown to be
\begin{align}
\ga_3(\vareps) &= 1-\frac{\pi}{2}\vareps + 2\vareps^2 + O(\vareps^3), & \ga(\vareps)  &= \frac{\pi}{4}\vareps -\vareps^2 + O(\vareps^3). \label{5:gaga3crack}
\end{align}
As such one can derive the form
\begin{align}
\mathcal{A}_{ij} 
 &= \Theta_{ij} + \left(\frac{2}{\pi\vareps} + \frac{1}{2}\right)\de_{i3}\de_{j3} +O(\vareps) \label{5Aijspheroidcavity3}
\end{align}
where expansions have been taken for $\vareps\ll 1$ and terms up to $O(1)$ have been retained since higher order terms will clearly vanish as $\vareps\rightarrow 0$.

The coefficient of $\de_{i3}\de_{j3}$ in \eqref{5Aijspheroidcavity3} involves an apparently singular limit as $\vareps\rightarrow 0$. That this is not a problem arises from the fact that this expression is used in formulae for effective properties of cracked media where this term is always multiplied by a volume-fraction term in such micromechanical methods, (or rather a ``crack-density'') that is proportional to $\vareps$ \cite{Hoe-79}, \cite{Hoe-83}. Note that taking the limits in the opposite order, i.e. $\vareps\rightarrow 0$ and \textit{then} $k_1\rightarrow 0$ yields an inconsistent result, giving rise to singular effective material behaviour in the crack limit which cannot be correct.


%


Finally, consider a different limit, the so-called ``ribbon-crack limit''. Take $k_1=0$ in the elliptical cylinder result \eqref{5ellipticcylAij} to find
\begin{align}
\mathcal{A}_{ij} &= (1+\eps)\de_{i1}\de_{j1} + \frac{(1+\eps)}{\eps}\de_{i2}\de_{j2}
 + \de_{i3}\de_{j3}. \label{5ellipticcylAij2}
\end{align}
Therefore as $\eps\rightarrow 0$
\begin{align}
\mathcal{A}_{ij} &= \frac{1}{\eps}\de_{i2}\de_{j2}
 + \de_{ij} +O(\eps). \label{5ellipticcylAij3}
\end{align}
As in the penny-shaped crack result above, the concentration tensor for the ribbon-crack is singular.

\subsection{Anisotropic host phase} \label{5:sec:anisohost}

The general form \eqref{5genPtransport} for the P-tensor associated with arbitrary host anisotropy requires the necessary surface integral to be evaluated. In the case of transversely isotropic and orthotropic media however, where principal axes are aligned with the semi-axes of the ellipsoid, the problem can be simplified significantly by employing a scaling of the Cartesian variables in order to reduce the isolated ellipsoidal inhomogeneity problem in an anisotropic medium to the case of an ellipsoidal inhomogeneity (with different semi-axes) in an \textit{isotropic} medium. Therefore the results derived above for the isotropic host phase case can be used in the scaled domain and then map back to the physical domain to obtain the appropriate physical Hill and Eshelby tensors.

As usual consider the case of an ellipsoid with semi-axes $a_j, j=1,2,3$ but now embedded in an \textit{orthotropic} host medium (with principal axes aligned along $x_j$, i.e.\ with the semi-axes of the ellipsoid) so that
\begin{align}
C^0_{ij} &= k_0\left(\de_{i1}\de_{j1}+\upsilon_2\de_{i2}\de_{j2}+\upsilon_3\de_{i3}\de_{j3}\right), \label{5:Cpotortho}
\end{align}
where $\upsilon_2=1$ (or $\upsilon_3=1$) for transverse isotropy. The governing partial differential equation is
\begin{align}
\deriv{}{x_i}\left(C^0_{ij}\deriv{T}{x_j}\right) &= 0.
\end{align}

Now employ the simple rescaling
\begin{align}
x_j &= \sqrt{\upsilon_j}\hat{x}_j, & j &=1,2,3
\end{align}
where $\upsilon_1=1$ is introduced for notational convenience (the conductivity along the $x_1$ axis is thus $k_0$), so that the semi-axes of the ellipsoid in the mapped domain become $\hat{a}_j=a_j/\sqrt{\upsilon_j}, j=1,2,3$ and denote the scaled ellipsoid as $\hat{V}_1$. The governing equation then becomes that governing isotropic media so that
\begin{align}
P_{ij} &= \frac{1}{k_0}\sum_{n=1}^3 \frac{\pa^2\Gamma(\mathbf{x})}{\pa x_n^2}\de_{in}\de_{jn} = \frac{1}{k_0}\sum_{n=1}^3 \frac{1}{\upsilon_n}\frac{\pa^2\hat{\Gamma}(\hat{\mathbf{x}})}{\pa \hat{x}_n^2}\de_{in}\de_{jn}
\end{align}
where $\hat{\Gamma}(\hat{\mathbf{x}})$ is defined in terms of the \textit{isotropic} (due to scaling) Green's tensor as defined in \eqref{5transisoGF} but now integrated over the scaled ellipsoid $\hat{V}_1$, i.e.\
\begin{align}
\hat{\Gamma}(\hat{\mathbf{x}}) &= -\frac{1}{4\pi}\int_{\hat{V}_1}\frac{1}{|\hat{\mathbf{y}}-\hat{\mathbf{x}}|}d\hat{\mathbf{y}}.
\end{align}
As a consequence for a general ellipsoid, the result \eqref{appgenPhirep} can be used but with $x_j$ replaced by $\hat{x}_j$ and $a_j$ replaced by $\hat{a}_j, j=1,2,3$. Therefore with reference to \eqref{5ellipsoidfunction}
\begin{align}
\derivtwomix{\hat{\Gamma}}{\hat{x}_i}{\hat{x}_j} &= \sum_{n=1}^3 \mathcal{E}(\hat{\vareps}_n;\hat{\vareps}_1,\hat{\vareps}_2)  \de_{in}\de_{jn}
\end{align}
where $\hat{\vareps}_n=\hat{a}_3/\hat{a}_n$. The P and S-tensors for an ellipsoid embedded inside an orthotropic host medium can then be written as
\begin{align}
P_{ij} &= \frac{1}{k_0}\sum_{n=1}^3 \ga_n\de_{in}\de_{jn}, & S_{ij} &= \sum_{n=1}^3 \upsilon_n\ga_n\de_{in}\de_{jn} \label{5Pgenaniso2}
\end{align}
where $\ga_n = \frac{1}{\upsilon_n}\mathcal{E}(\hat{\vareps}_n;\hat{\vareps}_1,\hat{\vareps}_2)$.

Note that the above scaling approach is considerably simpler than carrying out the necessary integrals in the corresponding general expression \eqref{5genPtransport} for the P-tensor. Now consider some specific cases of anisotropy of the host phase. First consider an inhomogeneity embedded in a \textit{transversely isotropic} (TI) host phase with conductivity tensor
\begin{align}
C_{ij}^0 &= k_0(\Theta_{ij}+\upsilon\de_{i3}\de_{j3}). \label{5hostTI}
\end{align}
Hence the (orthotropic) P-tensor for an ellipsoid with semi-axes aligned with the principal directions of anisotropy is found by setting $\upsilon_1=\upsilon_2=1$ and $\upsilon_3=\upsilon$ in \eqref{5Pgenaniso2} above. Simplifications arise for a spheroid of course as shall now be illustrated.

\subsubsection{Spheroid in a transversely isotropic host phase} \label{5exspheroidinTI}

Consider a spheroid in a transversely isotropic medium where the major/minor axis of the spheroid is aligned with the axis of transverse isotropy of the host phase. Denote the semi-axes of the spheroid as $a=a_1=a_2\neq a_3$ and the axis of transverse isotropy as $x_3$. We use the scaling argument above to see immediately that the P and S-tensors are given by
\begin{align}
P_{ij} &= \frac{1}{k_0}\left(\ga\Theta_{ij}+\ga_3\de_{i3}\de_{j3}\right), & S_{ij} &= k_0P_{ij} \label{5transportPTI}
\end{align}
where with reference to \eqref{5spheroidfunction}
\begin{align}
\ga_3(\vareps) &= \frac{1}{\upsilon}\mathcal{S}\left(\frac{\vareps}{\sqrt{\upsilon}}\right), \label{I1expressiontilde}
\end{align}
$\vareps=a_3/a$ and $\ga=\frac{1}{2}(1-\upsilon\ga_3)$, the latter being derived by using
\begin{align}
\derivtwo{\hat{\Gamma}}{\hat{x}_1}+\derivtwo{\hat{\Gamma}}{\hat{x}_2}+\derivtwo{\hat{\Gamma}}{\hat{x}_3} &= 1
\end{align}
for $\hat{\mathbf{x}}\in V_1$. As is evident, given the calculations already made associated with isotropy, this is a much simpler mechanism for obtaining results for an anisotropic medium than using the general form for the P-tensor and carrying out the necessary subsequent surface integral.

Assuming the spheroid itself is isotropic with conductivity tensor $C^1_{ij} = k_1\de_{ij}$, and using \eqref{5hostTI} together with the form of P-tensor in \eqref{5transportPTI}, the concentration tensor defined in \eqref{5:conctemp} can be straightforwardly determined as
\begin{align}
\mathcal{A}_{ij} &= \frac{k_0}{k_0+(k_1-k_0)\ga}\Theta_{ij} + \frac{k_0}{k_0+(k_1-\upsilon k_0)\ga_3}\de_{i3}\de_{j3} \label{I1expressiontildea}
\end{align}
with $\ga$ and $\ga_3$ as defined above. Alternatively, supposing that the spheroid is now transversely isotropic with the same axis of symmetry as the host, i.e.\ $C^1_{ij} = k_1(\Theta_{ij}+\zeta\de_{i3}\de_{j3})$, one finds that
\begin{align}
\mathcal{A}_{ij} &= \frac{k_0}{k_0+(k_1-k_0)\ga}\Theta_{ij} + \frac{k_0}{k_0+(\zeta k_1-\upsilon k_0)\ga_3}\de_{i3}\de_{j3}.
\end{align}

\subsubsection{Circular cylinder in a transversely isotropic host phase}

The circular cylinder limit can be taken in the spheroid case considered in \S \ref{5exspheroidinTI} where the cross-section of the cylinder sits in the plane of isotropy of the host medium. It is then anticipated that the P and S-tensors will be TI. It has been discussed above that $\mathcal{S}(x)\rightarrow 0$ as $x\rightarrow 0$ and therefore as with the isotropic host case from \eqref{I1expressiontilde} $\ga_3\rightarrow 0$. As such $\ga=\frac{1}{2}(1-\nu\ga_3)= \frac{1}{2}$ and then
\begin{align}
P_{ij} &= \frac{1}{2k_0}\Theta_{ij}, & S_{ij} &= \frac{1}{2}\Theta_{ij}
\end{align}
so that in fact this tensor is unchanged from the case of an isotropic host phase as in \eqref{5PTij}. The concentration tensor for an isotropic cylinder can be straightforwardly determined as
\begin{align}
\mathcal{A}_{ij} &= \frac{2k_0}{k_1+k_0}\Theta_{ij}+\de_{i3}\de_{j3}. \label{5TIcylTI}
\end{align}
The concentration tensor associated with a transversely isotropic cylinder is also given by that in \eqref{5TIcylTI}.


An interesting non-standard example is the case of a spheroid embedded inside a transversely isotropic host phase where the axes of symmetry and semi-axes are \textit{not} coincident. In this case the general (surface integral) form of the P and S-tensors must be used with the semi-axes aligned with the $\mathbf{x}$ axes but with all components of the modulus tensor being generally non-zero.

\subsubsection{Ellipsoid in an orthotropic host phase}

Consider an ellipsoid with semi-axes $a_j, j=1,2,3$ that are aligned with the axes of anisotropy of the host medium with orthotropic conductivity tensor as defined in \eqref{5:Cpotortho}. Analogous scaling arguments can be used as above in order to scale this problem into an ellipsoid in an \textit{isotropic} host and then scale back to the physical domain, as described above to show that
\begin{align}
P_{ij} &= \frac{1}{k_0}\sum_{n=1}^3\ga_n\de_{in}\de_{jn}, & S_{ij} &= \sum_{n=1}^3\ga_n\de_{in}\de_{jn}
\end{align}
where
\begin{align}
\ga_n &= \frac{1}{\upsilon_n}\mathcal{E}\left(\hat{\vareps}_n;\hat{\vareps}_1,\hat{\vareps}_2\right)
\end{align}
where $\upsilon_1=1$ and noting that $\hat{\vareps}_n = \sqrt{\frac{\upsilon_n}{\upsilon_3}}\vareps_n$ where $\vareps_n=\frac{a_3}{a_n}$.

For reference, P-tensors for a variety of problems are summarized in table \ref{tab:Ptensortransport}.

\begin{center}
\begin{table}
\begin{tabular}{|c|c|c|}
\hline
Host anisotropy  & Inclusion shape & P-tensor \\
\hline
Isotropic & Ellipsoid  & Use potential theory:\\
 & $a_1\neq a_2\neq a_3$ & $P_{ij} = \dfrac{1}{k_0}\sum_{n=1}^3\mathcal{E}(\vareps_n;\vareps_1,\vareps_2)\de_{in}\de_{jn}$\\
  & $\vareps_n=a_3/a_n$ & \\
\hline
 & Spheroid & Use potential theory:\\
  & $a_1=a_2=a\neq a_3$ & $P_{ij} = \dfrac{1}{k_0}\left(\ga\Theta_{ij}+\ga_3\de_{i3}\de_{j3}\right)$\\
  &  $\vareps=a_3/a$  & $\ga=\dfrac{1}{2}(1-\ga_3), \ga_3 = \mathcal{S}(\vareps)$\\
\hline
 & Sphere & Use symmetry: \\
 & & $P_{ij} = \dfrac{1}{3k_0}\de_{ij}$ \\
\hline
Transversely  & Ellipsoid & Use scalings and potential theory:\\
isotropic &  $a_1\neq a_2\neq a_3$ &  $P_{ij} = \dfrac{1}{k_0}\sum_{n=1}^3\frac{1}{\upsilon_n}\mathcal{E}(\hat{\vareps}_n;\hat{\vareps}_1,\hat{\vareps}_2)\de_{in}\de_{jn}$\\
$\upsilon_1=\upsilon_2=1\neq \upsilon_3=\upsilon$ &  $\vareps_n=a_3/a_n$ & $\hat{\vareps}_n = \hat{a}_3/\hat{a}_n$ and $\hat{a}_n = a_n/\sqrt{\upsilon_n}$.\\
 \hline
 & Spheroid, $a_1=a_2=a\neq a_3$ &                Use scalings and potential theory:       \\
 &  and $a_3$ is aligned with    &  $P_{ij} = \dfrac{1}{k_0}\left(\ga\Theta_{ij}+\ga_3\de_{i3}\de_{j3}\right)$\\
 & axis $x_3$ of transverse isotropy & $\ga=\dfrac{1}{2}(1-\upsilon\ga_3), \ga_3 = \frac{1}{\upsilon}\mathcal{S}\left(\frac{\vareps}{\sqrt{\upsilon}}\right)$\\
 \hline
 & Spheroid, $a_1=a_2=a\neq a_3$ &              Use scalings and potential theory:       \\
 &  and $a$ is aligned with &    $P_{ij} = \dfrac{1}{k_0}\sum_{n=1}^3\frac{1}{\upsilon_n}\mathcal{E}(\hat{\vareps}_n;\hat{\vareps}_1,\hat{\vareps}_2)\de_{in}\de_{jn}$\\
  & axis $x_3$ of transverse isotropy & $\hat{\vareps}_n = \hat{a}_3/\hat{a}_n$ and $\hat{a}_n = a_n/\sqrt{\upsilon_n}$.\\
 \hline
 & Sphere & Special case of spheroid result above: \\
 &        &  $P_{ij} = \dfrac{1}{k_0}\left(\ga\Theta_{ij}+\ga_3\de_{i3}\de_{j3}\right)$\\
 &        & $\ga=\dfrac{1}{2}(1-\upsilon\ga_3), \ga_3 = \frac{1}{\upsilon}\mathcal{S}\left(\frac{1}{\sqrt{\upsilon}}\right)$\\
 \hline
Orthotropic & Ellipsoid  & Use scalings and potential theory:\\
$\upsilon_1=1\neq\upsilon_2\neq \upsilon_3$ & $a_1\neq a_2\neq a_3$ & $P_{ij} = \dfrac{1}{k_0}\sum_{n=1}^3\frac{1}{\upsilon_n}\mathcal{E}(\hat{\vareps}_n;\hat{\vareps}_1,\hat{\vareps}_2)\de_{in}\de_{jn}$\\
 & & $\hat{\vareps}_n = \hat{a}_3/\hat{a}_n$ and $\hat{a}_n = a_n/\sqrt{\upsilon_n}$.\\
\hline
Worse than orthotropic & & Use general integral form:\\
or semi-axes of ellipsoids & & $P_{ij}=\dfrac{\tn{det}(\mathbf{a})}{4\pi}\bigint_{S^2}\dfrac{\Phi_{ij}\lit dS(\overline{\boldsymbol{\xi}})}{(\overline{\xi}_k a_{k\ell}a_{\ell m}\overline{\xi}_m)^{3/2}}$ \\
not aligned with axes  & & \\
of anisotropy. & & $\Phi_{ij}= \overline{\xi}_i\overline{\xi}_j/(C_{k\ell}^0\overline{\xi}_k\overline{\xi}_{\ell})$\\
\hline
\end{tabular}
\caption{Table summarizing results for the P-tensor associated with ellipsoidal inhomogeneities for potential problems.}
\label{tab:Ptensortransport}
\end{table}
\end{center}

\newpage
\section{Elastostatics: specific cases} \label{sec:elasto}

\subsection{Isotropic host phase}

For the case of an isotropic host phase case the elastic modulus tensor is defined as
\begin{align}
C_{ijk\ell}^0 &= 3\ka_0 I_{ijk\ell}^1+2\mu_0 I_{ijk\ell}^2 \label{5hostiso}
\end{align}
in terms of the isotropic fourth order basis tensors \eqref{appnot:I14} and \eqref{appnot:I24}. Here $\ka_0$ and $\mu_0$ are the bulk and shear moduli of the host, noting the relation to Poisson's ratio $\nu_0$
\begin{align}
\ka_0 &= \frac{2\mu_0(1+\nu_0)}{3(1-2\nu_0)}. \label{5:elcorel}
\end{align}
The appropriate isotropic Green's tensor is
\begin{align}
G_{ij}(\mathbf{x}-\mathbf{y}) &= \frac{1}{4\pi\mu_0}\frac{\de_{ij}}{|\mathbf{x}-\mathbf{y}|}-\frac{1}{16\pi\mu_0(1-\nu_0)}\derivtwomix{|\mathbf{x}-\mathbf{y}|}{x_i}{x_j} \label{4GFisoelast}
\end{align}
and the expression for the P-tensor in \eqref{4:Paniso} therefore becomes
\begin{multline}
P_{ijk\ell} = \dfrac{1}{4\mu_0}\left(\derivtwomix{\Gamma}{x_j}{x_{\ell}}\de_{ik}+\derivtwomix{\Gamma}{x_j}{x_k}\de_{i\ell}
+\derivtwomix{\Gamma}{x_i}{x_{\ell}}\de_{jk}+\derivtwomix{\Gamma}{x_i}{x_k}\de_{j\ell}\right)\\
+\dfrac{1}{4\mu_0(1-\nu_0)}\frac{\pa^4 \Psi}{\pa x_i\pa x_j \pa x_k \pa x_{\ell}}. \label{4:Piso}
\end{multline}
The potential $\Gamma$ is that already encountered and defined in \eqref{Npot} and $\Psi$ is defined by
\begin{align}
\Psi(\mathbf{x}) &= \frac{1}{4\pi}\inty{V_1}{}{|\mathbf{x}-\mathbf{y}|}{\mathbf{y}}, \label{5:Psipot}
\end{align}
which satisfies (see for example Kellogg \cite{Kel-70})
\begin{align}
\nabla^4\Psi &= -2\nabla^2\Gamma = -2\chi^1(\mathbf{x}).
\label{4:Psicases}
\end{align}

Once $P_{ijk\ell}$ is determined, the components of the Eshelby tensor can be calculated from \eqref{PCS} and the associated concentration tensor can be found from \eqref{5strainconcA}. Recall that no assumptions have been made regarding the anisotropy of the inhomogeneity. This is not required in order for the P-tensor to be determined. The only aspects of the inhomogeneity that influence the P-tensor are its \textit{shape} and, for \textit{anisotropic} host phases, its \textit{orientation} with respect to the axes of anisotropy of the host phase.


\subsubsection{Sphere in an isotropic host phase}
\label{elastisosphereexample}

Assume that the host phase is isotropic with elastic modulus tensor given in \eqref{5hostiso} and consider the case where $V_1$ is a \textit{sphere}. The (uniform) tensors $\pa^2\Gamma/\pa x_i\pa x_j$ and $\pa^4\Psi/\pa x_i\pa x_j\pa x_k\pa x_{\ell}$ will be spherically symmetric, i.e.\ isotropic and must possess full symmetry with respect to the interchange of any index. As with the potential problem therefore,
\begin{align}
\derivtwomix{\Gamma}{x_i}{x_j} &= \frac{1}{3}\de_{ij} \label{4:Phi1}
\end{align}
and for $\Psi$ the general isotropic form, with the additional constraint regarding symmetry with respect to interchange of any indices, is
\begin{align}
\frac{\pa^4\Psi}{\pa x_i\pa x_j\pa x_k\pa x_{\ell}} &= \psi(\de_{ij}\de_{k\ell}+\de_{ik}\de_{j\ell}+\de_{i\ell}\de_{jk}) \label{4:Psi1}
\end{align}
where $\psi$ is a constant to be determined. Performing the contractions $j=i$, $\ell=k$ in \eqref{4:Psi1} and utilizing \eqref{4:Psicases}, $\psi=-2/15$. Using \eqref{4:Phi1} and \eqref{4:Psi1} in \eqref{4:Piso}, the components of the P-tensor are
\begin{align}
P_{ijk\ell} &= \frac{1}{6\mu_0}\left(\de_{ik}\de_{j\ell}+\de_{i\ell}\de_{jk}\right)
               -\frac{1}{30\mu_0(1-\nu_0)}\left(\de_{ij}\de_{k\ell}+\de_{ik}\de_{j\ell}+\de_{i\ell}\de_{jk}\right). \label{Pisosphere22}
\end{align}
After simplification and writing in terms of the tensors $I_{ijk\ell}^1$ and $I_{ijk\ell}^2$ this becomes
\begin{align}
P_{ijk\ell} &= p_1 I_{ijk\ell}^1 + p_2 I_{ijk\ell}^2, \label{5:Pisosphere}
\end{align}
where upon using the relationship \eqref{5:elcorel} the components of the P-tensor are determined as
\begin{align}
p_1 &= \frac{1-2\nu_0}{6\mu_0(1-\nu_0)} = \frac{1}{3\ka_0+4\mu_0}, &
p_2 &= \frac{4-5\nu_0}{15\mu_0(1-\nu_0)} = \frac{3(\ka_0+2\mu_0)}{5\mu_0(3\ka_0+4\mu_0)}. \label{5p2}
\end{align}
From this form the Eshelby tensor is easily determined via \eqref{PCS} as
\begin{align}
S_{ijk\ell} &= s_1 I_{ijk\ell}^1 + s_2 I_{ijk\ell}^2, \label{5:Sisosphere}
\end{align}
where
\begin{align}
s_1 &= \frac{1+\nu_0}{3(1-\nu_0)} = \frac{3\ka_0}{3\ka_0+4\mu_0}, &
s_2 &= \frac{2(4-5\nu_0)}{15(1-\nu_0)} = \frac{6(\ka_0+2\mu_0)}{5(3\ka_0+4\mu_0)}. \label{5s2}
\end{align}


\subsubsection*{Isotropic sphere}

The P-tensor derived in \eqref{5:Pisosphere} holds for a spherical inhomogeneity of arbitrary anisotropy, embedded inside an isotropic host phase. If the inhomogeneity is also assumed isotropic with elastic modulus tensor of the form
\begin{align}
C_{ijk\ell}^1 &= 3\ka_1 I_{ijk\ell}^1+2\mu_1 I_{ijk\ell}^2, \label{5incmod}
\end{align}
one can use \eqref{4thIDapp} and the contraction properties of the tensors $I_{ijk\ell}^1, I_{ijk\ell}^2$ as defined in Appendix \ref{app:tensorstructure4} in order to write
\begin{align}
\tilde{\mathcal{A}}_{ijk\ell} &= I_{ijk\ell}^1+I_{ijk\ell}^2 + P_{ijmn}(C_{mnk\ell}^1-C_{mnk\ell}^0)\non\\
                           &= \left(1+3(\ka_1-\ka_0)p_1\right)I_{ijk\ell}^1 + \left(1+2(\mu_1-\mu_0)p_2\right)I_{ijk\ell}^2. \label{5:Aelasticsphereiso}
\end{align}
Finally the inversion properties of such a tensor are employed (as described in Appendix \ref{app:apptensTI}) together with \eqref{5p2} in order to obtain the strain concentration tensor:
\begin{align}
\mathcal{A}_{ijk\ell} &= \left(\frac{3\ka_0+4\mu_0}{3\ka_1+4\mu_0}\right)I_{ijk\ell}^1 + \frac{5\mu_0(3\ka_0+4\mu_0)}{3\ka_0(3\mu_0+2\mu_1)+4\mu_0(2\mu_0+3\mu_1)}I_{ijk\ell}^2. \label{5finalAisosphere1}
\end{align}
Taking $\ka_1, \mu_1\rightarrow 0$ in \eqref{5finalAisosphere1} the concentration tensor for a spherical cavity is obtained purely in terms of the host Poisson ratio, i.e.\
\begin{align}
\mathcal{A}_{ijk\ell} &= \frac{3(1-\nu_0)}{2(1-2\nu_0)}I^1_{ijk\ell} + \frac{15(1-\nu_0)}{7-5\nu_0}I^2_{ijk\ell}, \label{5:spherevoidvv}
\end{align}


\subsubsection*{Cubic sphere}

Assume now that the sphere has cubic symmetry with elastic modulus tensor
\begin{align}
C_{ijk\ell}^1 &= 3\ka_1 I_{ijk\ell}^1+2\mu_1 I_{ijk\ell}^2 + \eta_1\de_{ijk\ell}
\end{align}
where
\begin{align}
\de_{ijk\ell} &=
\begin{cases}
1, & i=j=k=\ell,\\
0, & \tn{otherwise}.
\end{cases}
\end{align}
Instead of \eqref{5:Aelasticsphereiso}, it is found that
\begin{align}
\tilde{\mathcal{A}}_{ijk\ell} = \left(1+3(\ka_1-\ka_0)p_1\right)I_{ijk\ell}^1 + \left(1+2(\mu_1-\mu_0)p_2\right)I_{ijk\ell}^2 + \eta_1 P_{ijmn}\de_{mnk\ell}. \label{5tildeAaniso}
\end{align}
Next, using the form of the P-tensor in \eqref{5:Pisosphere} and the relationships \eqref{2prods2}, the expression \eqref{5tildeAaniso} becomes
\begin{align}
\tilde{\mathcal{A}}_{ijk\ell} = \tilde{\al}_1 I_{ijk\ell}^1 + \tilde{\al}_2 I_{ijk\ell}^2 + \tilde{\al}_3\de_{ijk\ell},
\end{align}
where
\begin{align}
\tilde{\al}_1 &= 1+3(\ka_1-\ka_0)p_1+\eta_1(p_1-p_2), &
\tilde{\al}_2 &= 1+2(\mu_1-\mu_0)p_2, &
\tilde{\al}_3 &= \eta_1p_2.
\end{align}
Finally, appealing to the theory regarding cubic tensors in Appendix \ref{app:app4cubic} the concentration tensor is
\begin{align}
\mathcal{A}_{ijk\ell} &=\al_1 I_{ijk\ell}^1 + \al_2 I_{ijk\ell}^2 + \al_3\de_{ijk\ell}
\end{align}
where
\begin{align}
\al_1 &= \frac{\tilde{\al}_2^2+\tilde{\al}_1\tilde{\al}_3}{\tilde{\al}_2^2(\tilde{\al}_1+\tilde{\al}_3)}, & \al_2 &= \frac{1}{\tilde{\al}_2}, & \al_3 &= -\frac{\tilde{\al}_3}{\tilde{\al}_2^2}.
\end{align}
Note that when $\eta_1=0$ the case of an isotropic sphere is recovered as in \eqref{5:Aelasticsphereiso}.

\subsubsection{Circular cylinder in an isotropic host phase} \label{examplecirccyl}

Now assume that $V_1$ is a circular cylinder with axis of symmetry in the $x_3$ direction.  As for the potential problem, $\Gamma(\mathbf{x})$ must be independent of $x_3$ and isotropic in the $x_1x_2$ plane. Hence, as described in \eqref{5Phicyl}
\begin{align}
\derivtwomix{\Gamma}{x_i}{x_j} &= \frac{1}{2}\Theta_{ij}.
\end{align}
Furthermore $\Psi$ must also be independent of $x_3$ and be isotropic in the $x_1x_2$ plane and hence write
\begin{align}
\frac{\pa\Psi}{\pa x_i\pa x_j\pa x_k \pa x_{\ell}} &= \psi(\Theta_{ij}\Theta_{k\ell}+\Theta_{ik}\Theta_{j\ell}+\Theta_{i\ell}\Theta_{jk}), \label{4:Psicases2}
\end{align}
for some constant $\psi$, where the fact that this tensor should be fully symmetric with respect to interchange of any of its indices has been used.
Performing the contractions $j=i$, $\ell=k$ and using \eqref{4:Psicases} leads to the conclusion that $\psi=-\frac{1}{4}$. From \eqref{4:Piso} therefore
\begin{multline}
P_{ijk\ell} = \dfrac{1}{8\mu_0}\left(\Theta_{j\ell}\de_{ik}+\Theta_{jk}\de_{i\ell}+\Theta_{i\ell}\de_{jk}+\Theta_{ik}\de_{j\ell}\right)\\
-\dfrac{1}{16\mu_0(1-\nu_0)}(\Theta_{ij}\Theta_{k\ell}+\Theta_{ik}\Theta_{j\ell}+\Theta_{i\ell}\Theta_{jk}). \label{5:Pcyl}
\end{multline}
By writing $\de_{ij}=\Theta_{ij}+\de_{i3}\de_{j3}$ in the first term of \eqref{5:Pcyl} and then recognizing the appropriate Hill TI basis tensors (see Appendix \ref{app:apptensTI}) that arise as a result of the various contraction terms, the P-tensor can be written in the form
\begin{align}
P_{ijk\ell} &= \frac{1}{8\mu_0}\left[\frac{2(1-2\nu_0)}{1-\nu_0}\mathcal{H}_{ijk\ell}^1 + \frac{3-4\nu_0}{1-\nu_0}\mathcal{H}_{ijk\ell}^5+2\mathcal{H}_{ijk\ell}^6\right] \\
         &= \frac{1}{4\mu_0}\left[\frac{2\mu_0}{\la_0+2\mu_0}\mathcal{H}_{ijk\ell}^1 + \frac{\la_0+3\mu_0}{\la_0+2\mu_0}\mathcal{H}_{ijk\ell}^5+\mathcal{H}_{ijk\ell}^6\right]. \label{5Pcyliso}
\end{align}

In order to derive the Eshelby tensor one contracts $\mathbf{P}$ with the host modulus tensor $\mathbf{C}^0$. In order to do this write the isotropic basis tensors in terms of the Hill basis tensors using the expressions \eqref{5ITI}-\eqref{5I2TI} and then use the contractions summarized in Table \ref{tab:Pbasistab} in order to determine that
\begin{align}
S_{ijk\ell} &= \frac{1}{4(1-\nu_0)}\left(2\mathcal{H}^1_{ijk\ell} + 2\nu_0 \mathcal{H}^2_{ijk\ell} +(3-4\nu_0)\mathcal{H}^5_{ijk\ell}+2(1-\nu_0)\mathcal{H}^6_{ijk\ell}\right).
\end{align}
It may well be the case that a different basis set should be used if it transpires that the cylinder itself is more anisotropic than TI, for use in concentration tensors for example. However for computation, the matrix formulation, as described in Appendix \ref{app:matrix4tensor} can be of great utility when anisotropic basis tensors are becoming rather cumbersome.

Suppose now that the cylinder is isotropic with elastic modulus tensor as defined in \eqref{5incmod}. Once again using \eqref{5ITI}-\eqref{5I2TI}, constructing $\tilde{A}_{ijk\ell}$ is then just a matter of exploiting the contractions in Table \ref{tab:Pbasistab}.
It transpires that $\tilde{\mathcal{A}}_{ijk\ell}$ takes the form
\begin{align}
\tilde{\mathcal{A}}_{ijk\ell} &= \sum_{n=1}^6 \tilde{\al}_n\mathcal{H}^n_{ijk\ell} \label{ex5Ainvcc}
\end{align}
where
\begin{align}
\tilde{\al}_1 &= \frac{\la_1+\mu_1+\mu_0}{(\la_0+2\mu_0)}, & \tilde{\al}_2 &= \frac{(\la_1-\la_0)}{2(\la_0+2\mu_0)}, &
\tilde{\al}_3 &= 0, \\
\tilde{\al}_4 &= 1, &
\tilde{\al}_5 &= \frac{\mu_0+\mu_1(3-4\nu_0)}{4\mu_0(1-\nu_0)}, &
\tilde{\al}_6 &= \frac{(\mu_1+\mu_0)}{2\mu_0}.
\end{align}
Using the inversion expression for transversely isotropic tensors stated in Appendix \ref{app:apptensTI}, the appropriate concentration tensor is thus derived as
\begin{align}
\mathcal{A}_{ijk\ell} &= \sum_{n=1}^6 \al_n \mathcal{H}_{ijk\ell}^n  \label{cylstrainconc}
\end{align}
where
\begin{align}
\al_1 &= \frac{\la_0+2\mu_0}{(\la_1+\mu_1+\mu_0)}, & \al_2 &= \frac{(\la_0-\la_1)}{2(\la_1+\mu_1+\mu_0)}, &
\al_3 &= 0, \\
\al_4 &= 1, &
\al_5 &= \frac{4\mu_0(1-\nu_0)}{\mu_0+\mu_1(3-4\nu_0)}, &
\al_6 &= \frac{2\mu_0}{\mu_1+\mu_0}.
\end{align}
Since $\al_2\neq\al_3$ it is seen that $\mathcal{A}_{1133}\neq\mathcal{A}_{3311}$. The circular cylindrical cavity limit can be obtained by setting $\mu_1=0$. Using the expression $\la=2\mu\nu/(1-2\nu)$,
\begin{align}
\al_1 &= \frac{2(1-\nu_0)}{(1-2\nu_0)}, & \al_2 &= \frac{\nu_0}{1-2\nu_0}, &
\al_3 &= 0, \\
\al_4 &= 1, &
\al_5 &= 4(1-\nu_0), &
\al_6 &= 2.
\end{align}
Finally note that the fourth order tensor \eqref{cylstrainconc} can be represented in matrix form (see Appendix \ref{app:matrix4tensor}) as
\begin{align}
[\mathcal{A}] &=
\left( \begin{array}{cccccc}
a_{11} & a_{12} & a_{13} & 0 & 0 & 0  \\
a_{12} & a_{11} & a_{13} & 0 & 0 & 0  \\
a_{31} & a_{31} & a_{33} & 0 & 0 & 0  \\
0 & 0 & 0 & a_{33} & 0 & 0 \\
0 & 0 & 0 & 0 & a_{33}  & 0 \\
0 & 0 & 0 & 0 & 0 & a_{66}
\end{array} \right) \label{5:Amatrix}
\end{align}
where
\begin{align}
a_{11} &= \frac{(1-\nu_0)(3-4\nu_0)}{(1-2\nu_0)}, &  a_{12} &= \frac{(\nu_0-1)(1-4\nu_0)}{(1-2\nu_0)}, & a_{13} &= \frac{\nu_0}{1-2\nu_0},\\
a_{31} &= 0, & a_{33} &= 1, & a_{66} &= 2(1-\nu_0).
\end{align}

Clearly it is possible to write down explicit expressions for the concentration tensor when the circular cylinder is anisotropic. This merely complicates the tensorial (or matrix) operations after the derivation of the P-tensor in \eqref{5:Pcyl}. Given this P-tensor, perhaps the most important aspect is then to choose the tensor basis set correctly, given the anisotropy of the inhomogeneity. For example, when the cylinder itself is transversely isotropic (a common occurrence in applications) it is considered sensible to use a TI tensor basis set. For practical purposes and especially for the sake of computation, using the matrix formulation of tensors is advantageous in cases where the tensor basis sets become rather cumbersome. The following procedure is used, in the usual notation, referring to Appendix \ref{app:matrix4tensor}, and defining the $6\times 6$ matrix $[P]$ associated with the P-tensor, define
\begin{align}
[\tilde{\mathcal{A}}] &= [I] + [P][W][C^1-C^0] \label{5:APmatform1}
\end{align}
where $[W]$ is defined in \eqref{app:Wmat} and therefore
\begin{align}
[\mathcal{A}] &= [W]^{-1}[\tilde{\mathcal{A}}]^{-1}[W]^{-1}. \label{5:APmatform2}
\end{align}

\subsubsection{Spheroid in an isotropic host phase} \label{5:example:spheroidelasto}

When $V_1$ is a spheroid, the potential theory outlined in Appendix \ref{app:potential} is once again of use. It is clear that the P-tensor must be transversely isotropic and therefore will take the form
\begin{align}
P_{ijk\ell} &= \sum_{n=1}^6 p_n \mathcal{H}_{ijk\ell}^n.
\end{align}
The separate contributions to the P-tensor shall therefore first be decomposed into this form. Firstly, from the potential case described in Example \ref{5spheroideg}
\begin{align}
\derivtwomix{\Gamma}{x_i}{x_j} &= \ga\Theta_{ij} + \ga_3\de_{i3}\de_{j3}
\end{align}
where $\ga_3 = \mathcal{S}(\vareps)$ and $\ga=\frac{1}{2}(1-\ga_3)$. Representing all terms in the Hill basis one can show that
\begin{multline}
\frac{1}{4}\left(\derivtwomix{\Gamma}{x_j}{x_{\ell}}\de_{ik}+\derivtwomix{\Gamma}{x_j}{x_k}\de_{i{\ell}} \non
+\derivtwomix{\Gamma}{x_i}{x_{\ell}}\de_{jk}+\derivtwomix{\Gamma}{x_i}{x_k}\de_{j{\ell}}\right) \non\\
= \ga(\mathcal{H}^1_{ijk{\ell}}+\mathcal{H}^5_{ijk{\ell}}) + \ga_3\mathcal{H}^4_{ijk{\ell}}+ \frac{1}{2}(\ga+\ga_3)\mathcal{H}_{ijk{\ell}}^6. \label{5:spheroidtensorP}
\end{multline}
This is seen by using $\Theta_{ik}=\de_{ik}-\de_{i3}\de_{k3}$ and writing for example,
\begin{align}
\derivtwomix{\Gamma}{x_j}{x_{\ell}}\de_{ik} &= \derivtwomix{\Gamma}{x_j}{x_{\ell}}(\Theta_{ik}+\de_{i3}\de_{k3}) \non\\
 &= (\ga\Theta_{j\ell} + \ga_3\de_{j3}\de_{\ell 3})(\Theta_{ik}+\de_{i3}\de_{k3}) \non\\
 &= \ga(\Theta_{j\ell}\Theta_{ik} + \Theta_{j\ell}\de_{i3}\de_{k3}) + \ga_3(\Theta_{ik}\de_{j3}\de_{\ell 3}+\de_{i3}\de_{j3}\de_{k3}\de_{\ell 3}).
\end{align}
Doing this for each term on the left hand side of \eqref{5:spheroidtensorP}, combining and using the definitions of the TI basis tensors in Appendix \ref{app:apptensTI} leads to the form on the right hand side of \eqref{5:spheroidtensorP}.

Further, after much algebraic manipulation using the simplifications of the integrals in Appendix \ref{app:potential} in the case of spheroids one can show that
\begin{align}
\frac{1}{4}\frac{\pa^4\Psi}{\pa x_i\pa x_j\pa x_k\pa x_{\ell}} &= \sum_{n=1}^6 \psi_n \mathcal{H}_{ijk{\ell}}^n
\end{align}
where, upon using $\ga_3=1-2\ga$,
\begin{align}
\psi_1 &= \frac{\vareps^2(4\ga-1)-\ga}{4(1-\vareps^2)},  & & \psi_2 = \psi_3 = \frac{\vareps^2(1-2\ga)-\ga}{4(1-\vareps^2)}, \label{5:psi1spheroid} \\
\psi_4 &= \frac{3 \ga-1}{2(1-\vareps^2)}, & \psi_5 &= \frac{1}{2}\psi_1, & \psi_6 &= 2\psi_2. \label{5:psi6spheroid}
\end{align}
Therefore 
\begin{align}
p_1 &= \frac{1}{\mu_0}\left(\ga+\frac{1}{(1-\nu_0)}\psi_1\right), & p_2 &= p_3 = \frac{\psi_2}{\mu_0(1-\nu_0)}, \label{5:spheroid:p1}\\
p_4 &= \frac{1}{\mu_0}\left(1-2\ga+\frac{1}{(1-\nu_0)}\psi_4\right), &
p_5 &= \frac{1}{\mu_0}\left(\ga+\frac{1}{2(1-\nu_0)}\psi_1\right), \\
p_6 &= \frac{1}{\mu_0}\left(\frac{1}{2}(1-\ga)+\frac{2}{(1-\nu_0)}\psi_2\right).  \label{5:spheroid:p6}
\end{align}
A good check is to ascertain that the result for a sphere in an isotropic host phase is recovered by taking $\vareps\rightarrow 1$ and using \eqref{5:Seps1lim}. This is easily done and yields the result derived in \S \ref{elastisosphereexample} associated with a sphere. Another useful limit is to take $\vareps\rightarrow 0$. Results already derived can be employed, e.g.\ \eqref{5:gaga3crack} which when used in \eqref{5:psi1spheroid}-\eqref{5:psi6spheroid} yield
\begin{align}
\psi_1 &= -\frac{\pi}{16}\vareps+O(\vareps^3), & \psi_2 = \psi_3 &= -\frac{\pi}{16}\vareps + \frac{\vareps^2}{2} + O(\vareps^3),
\end{align}
\begin{align}
\psi_4 &= -\frac{1}{2}+\frac{3\pi}{8}\vareps - 2\vareps^2 + O(\vareps^3), &
\psi_5 &= -\frac{\pi}{32}\vareps + O(\vareps^3), & \psi_6 &= -\frac{\pi}{8}\vareps + \vareps^2 + O(\vareps^3).
\end{align}
For the components of the P-tensor this then gives
\begin{align}
p_1 &= \frac{1}{\mu_0}\left(\frac{\pi(3-4\nu_0)}{16(1-\nu_0)}\vareps -\vareps^2\right)+O(\vareps^3), \label{p1eps0}\\
p_2 = p_3 &=  \frac{1}{\mu_0}\left(-\frac{\pi}{16(1-\nu_0)}\vareps + \frac{1}{2(1-\nu_0)}\vareps^2\right) + O(\vareps^3), \\
p_4 &= \frac{1}{\mu_0}\left(\frac{1-2\nu_0}{2(1-\nu_0)} - \frac{\pi(1-4\nu_0)}{8(1-\nu_0)}\vareps - \frac{2\nu_0}{1-\nu_0}\vareps^2\right) + O(\vareps^3), \\
p_5 &= \frac{1}{\mu_0}\left(\frac{\pi(7-8\nu_0)}{32(1-\nu_0)}\vareps - \vareps^2\right) + O(\vareps^3),  \\
p_6 &= \frac{1}{\mu_0}\left(\frac{1}{2} - \frac{\pi(2-\nu_0)}{8(1-\nu_0)}\vareps + \frac{3-\nu_0}{2(1-\nu_0)}\vareps^2\right) + O(\vareps^3) \label{p6eps0}
\end{align}

The Eshelby tensor with respect to a TI basis, in the form
\begin{align}
S_{ijk\ell} &= \sum_{n=1}^6 s_n \mathcal{H}^n_{ijk\ell}
\end{align}
for either the spheroid with components of P-tensor \eqref{5:spheroid:p1}-\eqref{5:spheroid:p6} or the $\vareps\rightarrow 0$ limit of the spheroid with components of the P-tensor \eqref{p1eps0}-\eqref{p6eps0} has components that are related directly to the components of the P-tensor via the expressions
\begin{align}
s_1 &= 2\mu_0\left(\frac{p_1+2 p_2 \nu_0}{1-2\nu_0}\right), & s_2 &= 2\mu_0\left(\frac{p_1\nu_0+(1-\nu_0)p_2}{1-2\nu_0}\right), \\
s_3 &= 2\mu_0\left(\frac{p_3 +\nu_0 p_4}{1-2\nu_0}\right), &
s_4 &= 2\mu_0\left(\frac{2p_3\nu_0+(1-\nu_0)p_4}{1-2\nu_0}\right), \\
s_5 &= 2\mu_0p_5, & s_6 &= 2\mu_0p_6.
\end{align}
Finally, it is straightforward, but rather tedious, to show, using the P-tensor derived in the previous example, that the strain concentration tensor associated with an isotropic spheroid embedded in an isotropic host phase is
\begin{align}
\mathcal{A}_{ijk\ell} &= \sum_{n=1}^6 \al_n \mathcal{H}_{ijk\ell}^n  \label{5:spheroidstrainconc}
\end{align}
where with $\Delta=\frac{1}{2}q_1q_4-q_2q_3$,
\begin{align}
\al_1 &=  \frac{q_4}{2\Delta}, & \al_2 &= - \frac{q_2}{2\Delta}, & \al_3 &= -\frac{q_3}{2\Delta}, \label{5:spheroidal1}\\
\al_4 &= \frac{q_1}{2\Delta}, & \al_5 &= \frac{1}{q_5}, & \al_6 &= \frac{1}{q_6} \label{5:spheroidal6}
\end{align}
and where
\begin{align}
q_1 &= 1 + 2p_1\left(\la_d+\mu_d\right) + 2p_2\la_d, &
q_2 &=     p_1\la_d + p_2\left(\la_d+2\mu_d\right), \\
q_3 &=     2p_3\left(\la_d+\mu_d\right) + p_4\la_d, &
q_4 &= 1 + 2p_3\la_d + p_4\left(\la_d+2\mu_d\right), \\
q_5 &= 1 + 2p_5\mu_d, &
q_6 &= 1 + 2p_6\mu_d,
\end{align}
with
\begin{align}
\la_d&=\la_1-\la_0, & \mu_d &= \mu_1-\mu_0. \label{5:ladmudc}
\end{align}
The spheroidal cavity result is simply \eqref{5:spheroidstrainconc}-\eqref{5:ladmudc} with $\la_1=\mu_1=0$ and so every occurrence of $\la_d$ and $\mu_d$ is simply replaced with $-\la_0$ and $-\mu_0$ respectively. The result can be obtained in terms of $\nu_0$ alone by using the expression $\la_0=2\mu_0\nu_0/(1-2\nu_0)$.

The average of the concentration tensor over uniform orientations of spheroids can be obtained using the result in \eqref{app:4thordertensorisoav} in order to derive an expression of the form
\begin{align}
\underline{\mathcal{A}}_{ijk\ell} &= \al_1 I^1_{ijk\ell} + \al_2 I^2_{ijk\ell}.
\end{align}

\subsubsection{Elastic layer} \label{5:example:layer}

The result for the spheroid can be used in order to determine the P-tensor and concentration tensor for an elastic layer, taking $\vareps=0$ in \eqref{p1eps0}-\eqref{p6eps0},
\begin{align}
P_{ijk\ell}
 &= \frac{1}{2\mu_0}\left(\frac{1-2\nu_0}{1-\nu_0}\mathcal{H}^4_{ijk\ell} + \mathcal{H}^6_{ijk\ell}\right),
\end{align}
Eshelby's tensor easily follows as
\begin{align}
S_{ijk\ell} &= \frac{\nu_0}{1-\nu_0}\mathcal{H}^3_{ijk\ell} + \mathcal{H}^4_{ijk\ell} + \mathcal{H}^6_{ijk\ell}.
\end{align}

Using the P-tensor the concentration tensor for an isotropic layer is straightforwardly determined as
\begin{align}
\mathcal{A}_{ijk\ell} = \mathcal{H}^1_{ijk\ell} + \left(\frac{\la_0-\la_1}{\la_1+2\mu_1}\right)\mathcal{H}^3_{ijk\ell}
+ \left(\frac{\la_0+2\mu_0}{\la_1+2\mu_1}\right)\mathcal{H}^4_{ijk\ell} + \mathcal{H}^5_{ijk\ell} + \frac{\mu_0}{\mu_1}\mathcal{H}^6_{ijk\ell}.
\end{align}

%

\subsubsection{Limiting case of a penny-shaped crack} \label{chap5:cavcrackelast}

For a penny shaped crack, terms up to $O(\vareps)$ are retained in \eqref{p1eps0}-\eqref{p6eps0} to obtain
\begin{multline}
P_{ijk\ell} = \frac{1}{\mu_0}\Bigg[\frac{\pi(3-4\nu_0)}{16(1-\nu_0)}\vareps\mathcal{H}^1_{ijk\ell} -\frac{\pi}{16(1-\nu_0)}\vareps(\mathcal{H}^2_{ijk\ell}+\mathcal{H}^3_{ijk\ell}) \\
+\left(\frac{1-2\nu_0}{2(1-\nu_0)} - \frac{\pi(1-4\nu_0)}{8(1-\nu_0)}\vareps\right)\mathcal{H}^4_{ijk\ell}\\
+\frac{\pi(7-8\nu_0)}{32(1-\nu_0)}\vareps\mathcal{H}^5_{ijk\ell} +
\left(\frac{1}{2} - \frac{\pi(2-\nu_0)}{8(1-\nu_0)}\vareps\right)\mathcal{H}^6_{ijk\ell}
\Bigg].
\end{multline}
Using this to determine the concentration tensor with $\mu_1=0$ as in the cavity limit, it is straightforwardly shown that
\begin{multline}
\mathcal{A}_{ijk\ell} = (1-\nu_0)\mathcal{H}^1_{ijk\ell} -\frac{1}{2}(1-\nu_0)\mathcal{H}^2_{ijk\ell} + \\
\left(\frac{4\nu_0(1-\nu_0)}{\pi(1-2\nu_0)}\frac{1}{\vareps}-\frac{1}{2}(1-\nu_0)(1+2\nu_0)\right)\mathcal{H}^3_{ijk\ell}\\
+\left(\frac{4(1-\nu_0)^2}{\pi(1-2\nu_0)}\frac{1}{\vareps}+\frac{1}{2}(1+2\nu_0)(1-\nu_0)\right)\mathcal{H}^4_{ijk\ell}+ \mathcal{H}^5_{ijk\ell}\\
+\left(\frac{4(1-\nu_0)}{\pi(2-\nu_0)}\frac{1}{\vareps}+\frac{16(3-\nu_0)(1-\nu_0)}{\pi^2(2-\nu_0)^2}\right)\mathcal{H}^6_{ijk\ell} \label{5:elastcrackor}
\end{multline}
where terms of $O(\vareps)$ have been neglected in \eqref{5:elastcrackor}. This expression has singular behaviour as $\vareps\rightarrow 0$ akin to the potential problem result \eqref{5Aijspheroidcavity3} and when deriving effective properties for distributions of cracks, this singular nature is necessary to yield the correct effective behaviour \cite{Hoe-79}. In fact although $O(1)$ coefficients have been retained in in \eqref{5:elastcrackor} only the singular terms are required in order to determine effective properties. The expression \eqref{5:elastcrackor} corrects the typographical errors given on p.\ 104 of \cite{Mar-00}.

A common requirement is the determination of the effective properties of a medium comprising penny shaped cracks that are uniformly distributed \textit{and} uniformly oriented inside some host material. Using \eqref{app:4thordertensorisoav} the associated concentration tensor is shown to be
\begin{align}
\underline{\mathcal{A}}_{ijk\ell} &= \frac{4(1-\nu_0^2)}{3\pi\vareps(1-2\nu_0)}I^1_{ijk\ell} + \frac{8(1-\nu_0)(5-\nu_0)}{15\pi\vareps(2-\nu_0)}I^2_{ijk\ell} + O(1). \label{5:avpennycrack}
\end{align}

We shall now consider the case of an ellipsoid in an isotropic medium. In order to deal with this generally in a tensor setting, ideally an orthotropic tensor basis should be used. Although it is possible to write down such a basis, details are rather lengthy and in fact for practical computation, it is perhaps most sensible to write down the nine independent components of the P-tensor and use matrix computations in the manner described after \S \ref{examplecirccyl} above.

\subsubsection{Ellipsoid in an isotropic host phase} \label{5ex:elastellipsoid}

The nine independent components of the P-tensor for an ellipsoid can be defined in terms of the function $\mathcal{E}(\vareps_n;\vareps_1,\vareps_2)$ and the semi-axes ratios $\vareps_n$.

The nine independent components of the Eshelby tensor for an ellipsoid in an isotropic medium are usually stated in terms of the four components $S_{1111}, S_{1122}, S_{1133}$ and $S_{1212}$ together with cyclic properties of the indices, in terms of $I_{mn}$ and $I_m$ as defined in \eqref{app:Imn}-\eqref{app:Imn2}. In turn these lead to expressions in terms of the fundamental integral $\mathcal{E}(\vareps_n;\vareps_1,\vareps_2)$ via \eqref{app:InmathcalE} and \eqref{app:Imn1}-\eqref{app:Imn3}. As such use \eqref{4:Piso} with \eqref{Ptransportellipsoid} and \eqref{app:Psiresultellipsoid} and employ the properties \eqref{app:Imn1}-\eqref{app:Imn3} to derive the following compact forms
\begin{align}
P_{1111} &= \frac{3}{16\pi\mu_0(1-\nu_0)}I_{11} + \frac{1-4\nu_0}{16\pi\mu_0(1-\nu_0)}I_1, \label{PellipsoidA}\\
P_{1122} &= \frac{1}{16\pi\mu_0(1-\nu_0)}(I_{21}-I_1), \\
P_{1133} &= \frac{1}{16\pi\mu_0(1-\nu_0)}(I_{31}-I_1),  \\
P_{1212} &= \frac{1}{32\pi\mu_0(1-\nu_0)}(I_{12}+I_{21})+\frac{(1-2\nu_0)}{32\pi\mu_0(1-\nu_0)}(I_1+I_2). \label{PellipsoidD}
\end{align}
All other non-zero components are obtained by cyclic permutation of the indices in the above equations. Those components that cannot be obtained via cyclic permutation are zero, e.g.\ $P_{1112}=P_{1223}=P_{1323}=0$.

For representations and calculations of the concentration tensor it is convenient to use the matrix representation of the tensors. This in discussed in the next \S by considering the elliptical cylinder and ribbon crack limits. First however the components of the Eshelby tensor are stated, using \eqref{PCS} and noting the slightly modified notation for $I_{mn}$ in \eqref{app:Imn} (i.e.\ the factor of $a_m^2$) as compared with the standard definition, e.g.\ Mura \cite{Mur-82}. The components are expressed as
\begin{align}
S_{1111} &= \frac{3}{8\pi(1-\nu_0)}I_{11} + \frac{1-2\nu_0}{8\pi(1-\nu_0)}I_1, \label{SellipsoidA}\\
S_{1122} &= \frac{1}{8\pi(1-\nu_0)}\frac{\vareps_1^2}{\vareps_2^2}I_{12} - \frac{1-2\nu_0}{8\pi(1-\nu_0)}I_1, \\
S_{1133} &= \frac{1}{8\pi(1-\nu_0)}\frac{\vareps_1^2}{\vareps_3^2}I_{13} - \frac{1-2\nu_0}{8\pi(1-\nu_0)}I_1, \\
S_{1212} &= \frac{1+\vareps_1^2/\vareps_2^2}{16\pi(1-\nu_0)}I_{12}+\frac{1-2\nu_0}{16\pi(1-\nu_0)}(I_1+I_2) \label{SellipsoidD}
\end{align}
and permutation rules follow as for the P-tensor.

\subsubsection{Elliptical cylinder and ribbon-crack limit} \label{5:ribbon}

In \S \ref{chap5:ellipcyl} it was shown that in the limit as $a_3\rightarrow\infty$, $\mathcal{E}(1;\vareps_1,\vareps_2) \rightarrow 0$ and
\begin{align}
\mathcal{E}(\vareps_1;\vareps_1,\vareps_2) &\rightarrow \frac{a_2}{a_1+a_2} = \frac{\eps}{1+\eps}, &
\mathcal{E}(\vareps_2;\vareps_1,\vareps_2) &\rightarrow \frac{a_1}{a_1+a_2} = \frac{1}{1+\eps},
\end{align}
where $\eps=a_2/a_1$. 
These are used in the expressions for $I_{mn}$ and $I_n$ in Appendix \ref{app:potential} and substituted into \eqref{PellipsoidA}-\eqref{PellipsoidD} to determine
the associated P-tensor components. Since the P-tensor is still orthotropic there are nine independent components:
\begin{align}
P_{1111} &= \eps\left(\frac{4(1+\eps)(1-\nu_0)-(1+2\eps)}{4\mu_0(1-\nu_0)(1+\eps)^2}\right), \\
P_{2222} &= \frac{4(1+\eps)(1-\nu_0)-(2+\eps)}{4\mu_0(1-\nu_0)(1+\eps)^2}, \\
P_{3333} &= 0,
\end{align}
\vspace{-1cm}
\begin{align}
P_{1122} &= \frac{-\eps}{4\mu_0(1-\nu_0)(1+\eps)^2} & P_{1133} &= 0, \\
P_{2233} &= 0,  &    P_{1313} &= \frac{\eps}{4\mu_0(1+\eps)}, \\
P_{2323} &= \frac{1}{4\mu_0(1+\eps)},  & P_{1212} &= \frac{(1-\nu_0)(1+\eps)^2-\eps}{4\mu_0(1-\nu_0)(1+\eps)^2}.
\end{align}
The Eshelby tensor components follow as
\begin{align}
S_{1111} &= \eps\left(\frac{1+2(1+\eps)(1-\nu_0)}{2(1-\nu_0)(1+\eps)^2}\right), & S_{2222} &= \frac{\eps+2(1-\nu_0)(1+\eps)}{2(1-\nu_0)(1+\eps)^2}, \\
S_{3333} &= 0, & S_{1133} &= \frac{\nu_0\eps}{(1-\nu_0)(1+\eps)},  \\
S_{3311} &= 0, & S_{1122} &= \frac{-\eps+2\eps(1+\eps)\nu_0}{2(1-\nu_0)(1+\eps)^2} \\
S_{2211} &= \frac{-\eps+2\nu_0(1+\eps)}{2(1-\nu_0)(1+\eps)^2} & S_{2233} &=  \frac{\nu_0}{(1-\nu_0)(1+\eps)},  \\
S_{3322} &= 0, &  S_{1313} &= \frac{\eps}{2(1+\eps)}, \\
S_{2323} &= \frac{1}{2(1+\eps)},   &    S_{1212} &= \frac{-\eps+(1-\nu_0)(1+\eps)^2}{2(1-\nu_0)(1+\eps)^2}
\end{align}
noting that Eshelby's tensor does not possess the major symmetry, unlike Hill's tensor.

Suppose that the host and elliptical cylinder are both isotropic. In order to determine the concentration tensor orthotropic tensors are required. Although it is possible to use an orthotropic basis set, it is perhaps most convenient to work with the matrix formulation of the tensors and derive the concentration tensor using a symbolic mathematical package such as Mathematica. In doing this the matrix formulation $[\mathcal{A}]$ of the tensor $\boldsymbol{\mathcal{A}}$ is employed as noted in \eqref{5:APmatform1}-\eqref{5:APmatform2}. The components of the matrix are so long that to list these here would not be beneficial but two very useful limits shall be written down. The elliptical cylindrical \textit{cavity} limit is obtained by taking $\mu_1\rightarrow 0$ which yields a matrix form of the tensor (referring to e.g.\ \eqref{5:Amatrix}) as
\begin{align}
[\mathcal{A}] &=
\left( \begin{array}{cccccc}
a_{11} & a_{12} & a_{13} & 0 & 0 & 0  \\
a_{12} & a_{22} & a_{23} & 0 & 0 & 0  \\
a_{13} & a_{23} & a_{33} & 0 & 0 & 0  \\
0 & 0 & 0 & a_{44} & 0 & 0 \\
0 & 0 & 0 & 0 & a_{55}  & 0 \\
0 & 0 & 0 & 0 & 0 & a_{66}
\end{array} \right) \label{5:Amatrix2}
\end{align}
where
\begin{align}
a_{11} &= \frac{(1-\nu_0)(1+2\eps-2(1+\eps)\nu_0)}{(1-2\nu_0)}, &
a_{12} &= \frac{(1-\nu_0)(-1+2(1+\eps)\nu_0)}{(1-2\nu_0)}, \\
a_{13} &= \frac{(2\eps-1+2(1-\eps)\nu_0)\nu_0}{(1-2\nu_0)}, & a_{21} &= \frac{(1-\nu_0)(-\eps+2(1+\eps)\nu_0)}{\eps(1-2\nu_0)}, \\
a_{22} &= \frac{(1-\nu_0)(2+\eps-2(1+\eps)\nu_0)}{\eps(1-2\nu_0)}, &
a_{23} &= \frac{(2-\eps+2(\eps-1)\nu_0)\nu_0}{\eps(1-2\nu_0)}, \\
a_{33} &= 1, & a_{44} &= \frac{1+\eps}{2\eps}, \\
a_{55} &= \frac{1+\eps}{2}, & a_{66} &= \frac{(1+\eps)^2(1-\nu_0)}{2\eps}
\end{align}
and taking the limit as $\eps\rightarrow 0$ yields the ribbon-crack limit, retaining terms up to $O(1)$ in $\eps$,
\begin{align}
a_{11} &= 1-\nu_0, & a_{12} &= -(1-\nu_0), \\
a_{13} &= -\nu_0, &
a_{21} &= \frac{2(1-\nu_0)\nu_0}{(1-2\nu_0)\eps}-(1-\nu_0), \\
a_{22} &= \frac{2(1-\nu_0)^2}{(1-2\nu_0)\eps}+1-\nu_0, &
a_{23} &= \frac{2(1-\nu_0)\nu_0}{(1-2\nu_0)\eps}-\nu_0, \\
a_{33} &= 1, & a_{44} &= \frac{1}{2\eps}+\frac{1}{2}, \\
a_{55} &= \frac{1}{2}, & a_{66} &= \frac{1}{2\eps}(1-\nu_0)+(1-\nu_0).
\end{align}

\subsubsection{Flat ellipsoid}

Consider the case when $a_1> a_2\gg a_3$. It is straightforward to take this limit in \eqref{I1FE}-\eqref{thetak} in order to obtain
\begin{align}
I_1 &= 4\pi\vareps_2\frac{(F(k)-E(k)}{((\vareps_2/\vareps_1)^2-1)}, \label{I1flat}\\
I_2 &= 4\pi\vareps_2E(k) - I_1, \\
I_3 &= 4\pi - 4\pi\vareps_2 E(k) \label{I3flat}
\end{align}
where with reference to \eqref{FE} and \eqref{thetak}, $F(k)$ and $E(k)$ are introduced as the complete Elliptic integrals of the first and second kind, respectively
\begin{align}
E(k) &= \int_0^{\pi/2}\frac{dx}{(1-k^2\sin^2 x)^{1/2}}, & F(k) &= \int_0^{\pi/2}(1-k^2\sin^2 x)^{1/2}\hspace{0.1cm} dx
\end{align}
and $k=1-\frac{\vareps_1^2}{\vareps_2^2}$. From \eqref{I1flat}-\eqref{I3flat}, the $I_{mn}$ can be straightforwardly determined via \eqref{app:Imn1}-\eqref{app:Imn3} and thus the components of the P-tensor from \eqref{PellipsoidA}-\eqref{PellipsoidD} and \eqref{SellipsoidA}-\eqref{SellipsoidD}.

\subsubsection{Spheroid limit check}

One can straightforwardly take the spheroidal inhomogeneity limit $a_1=a_2=a\neq a_3$ in the ellipsoidal result above. In particular it is noted that in the limit as $a_1\rightarrow a_2=a$, referring to \S \ref{5spheroideg},
\begin{align}
I_1 = I_2 = 2\pi(1-\mathcal{S}(\eps)), \\
I_3 = 4\pi\mathcal{S}(\eps),
\end{align}
where $\vareps=a_3/a$. This then gives
\begin{align}
I_{11} = I_{22} = I_{12} = I_{21} &= \pi - \frac{I_1-I_3}{4(\vareps^2-1)}, \\
I_{13} = I_{23} &= \frac{I_1-I_3}{\vareps^2-1}, \\
I_{33} &= \frac{4\pi}{3} - \frac{2}{3}\vareps^2 I_{13}, \\
I_{31} = I_{32} &= \vareps^2 I_{13}.
\end{align}
These can then be used in \eqref{PellipsoidA}-\eqref{PellipsoidD} together with the cyclic properties to derive the components of the P-tensor for a spheroid. It is straightforward to check that this gives rise to the coefficients $p_1-p_6$ as defined for a TI tensor in \eqref{5:spheroid:p1}-\eqref{5:spheroid:p6}.

\subsection{Anisotropic host phase}

In the potential problem case, scaling coordinate systems assisted in the derivation of results associated with anisotropic media. Although such methods can sometimes lead to modest simplifications in elasticity, the general theory does not lead to any significant advances, certainly for the problems that are of greatest interest in micromechanics. In particular such methods do not lead to significant simplifications for generally transversely isotropic media which is a material symmetry of great importance. Therefore to derive the P-tensors associated with inhomogeneities in anisotropic host phases, it is best to work with the integral form of the P-tensor as defined in \eqref{5Pijklellipsoid} for an ellipsoid.

Few explicit results are available in general however since the Green's tensor cannot generally be determined analytically. One of the few that can however is that associated with TI media. Withers derived the associated Eshelby tensor for an ellipsoid \cite{Wit-89} using the form of the Green's function determined by Pan and Chou \cite{Pan-76}. Let us here state his result in the case of a spheroid in a TI medium where the semi-major or minor axis of the spheroid is aligned with the axis of transverse isotropy. This result shall then be checked by employing the general integral form \eqref{5Pijklellipsoid}. Only in the last decade have articles started to appear that compute effective properties via micromechanical methods, see e.g.\ \cite{Sev-05}, \cite{Gir-07}. It is also important to note specific results for the Eshelby and Hill tensors associated with \textit{cracks} in anisotropic media. See e.g.\ Gruescu et al. \cite{Gru-05} and Barth\'el\'emy \cite{Bar-09}.

\subsubsection{Spheroid in a transversely isotropic host phase} \label{5egelasticspheroidTI}

Consider a spheroid with semi-axes $a=a_1=a_2\neq a_3$ embedded in a transversely isotropic host phase where the $x_1x_2$ plane is the plane of isotropy. The elastic modulus tensor of the host is
\begin{align}
C_{ijk\ell}^0 &= \sum_{n=1}^6 c^0_n \mathcal{H}_{ijk\ell}^n \label{5hostTIel}
\end{align}
where
\begin{align}
c_1^0 &= 2K_0, & c^0_2 &= \ell_0, & c^0_3 &= \ell_0, \\
c_4^0 &= n_0, & c^0_5 &= 2 m_0, & c^0_6 &= 2 g_0.
\end{align}
Here $K_0$ and $m_0$ are the in-plane bulk and shear moduli and $g_0$ is the antiplane shear modulus (often $p_0$ is used for the anti-plane modulus but this is not employed here in order to avoid confusion associated with components $p_j$ of the Hill tensor).

\subsection*{Derivation from Withers' Eshelby tensor}

Withers derived the Eshelby tensor for a \textit{spheroid} in a transversely isotropic host medium. In order to state this result it is useful to first define the parameters
$$\begin{array}{l}
v_1 = \left(\displaystyle\frac{(\hat{\ell}_0-\ell_0)(\hat{\ell}_0+\ell_0+2g_0)}{4n_0g_0}\right)^{1/2}+\left(\displaystyle\frac{(\hat{\ell}_0+\ell_0)(\hat{\ell}_0-\ell_0-2g_0)}{4n_0g_0}\right)^{1/2},\\
v_2 = \left(\displaystyle\frac{(\hat{\ell}_0-\ell_0)(\hat{\ell}_0+\ell_0+2g_0)}{4n_0g_0}\right)^{1/2}-\left(\displaystyle\frac{(\hat{\ell}_0+\ell_0)(\hat{\ell}_0-\ell_0-2g_0)}{4n_0g_0}\right)^{1/2},\\
v_3 = \left(\displaystyle\frac{m_0}{g_0}\right)^{1/2},
\end{array}$$
where $\hat{\ell}_0=(n_0(K_0+m_0))^{1/2}$. We note that for elastic
materials $v_3\in\mathbb{R}$ but $v_1,v_2\in\mathbb{C}$ in general
with $v_2=\overline{v_1}$, where an overbar denotes the complex
conjugate\footnote{This latter point does not appear to have been
recognized in the original papers on this subject, e.g.\
\cite{Wit-89}. An example of a transversely isotropic material for
which $v_2=\overline{v_1}\in\mathbb{C}$ is zinc with (all in GPa)
$K=80, \ell=33, n=50, m=63, g=40$, for which $v_1=1.1284+0.6465 i$
to 4dp.} For $v_i\in\mathbb{R}$ define
\begin{align}
v_iI_3(v_i) &= 4\pi\mathcal{S}(v_i\vareps), &
I_1(v_1) &= \frac{4\pi}{v_i}-2 I_3(v_i).
\end{align}
When $v_i\in\mathbb{C}$, either of the cases in $\mathcal{S}(v_i\vareps)$ are valid since they are merely an analytic
continuation of the function (of $v_i$) into the complex $v_i$-plane. Note that in the case of isotropy, $\ell_0=\la_0=\ell_0'$,
$g_0=m_0=\mu_0$, $\hat{\ell}_0=n_0=K_0+m_0=\la_0+2\mu_0$ and thus $v_1=v_2=v_3=1$. The notation $I_i=I_i(1)$ is therefore appropriate for the isotropic case, as already introduced.

For a TI host phase defined by elastic properties \eqref{5hostTIel} Withers determined the components $S_{ijk\ell}$ in the form
\begin{align}
S_{1111} &= \sum_{i=1}^2 \left[2g_0(1+M_i)v_i^2-m_0\right] L_iv_i I_1(v_i) + \frac{1}{2}D m_0 I_1(v_3), \label{S1}\\
S_{1122} &= \sum_{i=1}^2 \left[2g_0(1+M_i)v_i^2 - 3m_0\right]L_i v_iI_1(v_i)-\frac{1}{2}D m_0 I_1(v_3),\\
S_{3333} &= 2\sum_{i=1}^2[\ell_0-n_0 M_iv_i^2] v_i^3 M_i L_i  I_3(v_i),\\
S_{1133} &= 2\sum_{i=1}^2[\ell_0-n_0 M_i v_i^2]v_i L_i I_1(v_i),\\
S_{3311} &= 2\sum_{i=1}^2 [g_0 v_i^2(1+M_i)-m_0] M_iL_iv_i^3 I_3(v_i),\\
S_{1313} &= \frac{1}{2}g_0\sum_{i=1}^2
L_iv_i^3(1+M_i)(I_3(v_i)-2M_i I_1(v_i))+\frac{1}{4}D g_0
I_3(v_3)v_3^2 \label{S1313}
\end{align}
where
\begin{align}
D &= \frac{1}{4\pi g_0 v_3}, & M_i &=
\frac{(K_0+m_0)/v_i^2-g_0}{\ell_0+g_0}, & L_i &=
(-1)^i\frac{g_0-n_0v_i^2}{8\pi n_0g_0(v_1^2-v_2^2)v_i^2}. \non
\end{align}
Note that slightly different notation has been used here from that in \cite{Wit-89} and in particular the notation $I_3(v_i)$ has been used whereas \cite{Wit-89} used $I_2(v_i)$ for this term in the corresponding equations. This is done here to preserve the symmetry with the isotropic case above so that as $v_i\rightarrow 1$, $I_3(v_i)\rightarrow I_3$.

Via straightforward contraction with the TI compliance tensor $D_0$, i.e.\ the inverse of \eqref{5hostTIel}, \eqref{PCS} then yields
\begin{align}
P_{1111}+P_{1122} &= \frac{n_0(S_{1111}+S_{1122})-2\ell_0 S_{1133}}{2\Delta}, \\
P_{1111}-P_{1122} &= \frac{S_{1111}-S_{1122}}{2m_0}, \\
P_{3333} &= \frac{K_0 S_{3333}-\ell_0 S_{3311}}{\Delta}, \\
P_{1133} &= \frac{n_0 S_{3311}-\ell_0S_{3333}}{2\Delta}, \\
P_{1313} &= \frac{S_{1313}}{2g_0}, \\
P_{1212} &= \frac{S_{1212}}{2m_0}
\end{align}
and where $\Delta=K_0 n_0-\ell_0^2$. Of course this calculation could also be done with the help of matrices rather than tensor forms. Since the P-tensor is TI however it is rather straightforward to write down the TI tensor basis forms
\begin{align}
P_{ijk\ell} &= \sum_{n=1}^6 p_n \mathcal{H}^n_{ijk\ell}, & S_{ijk\ell} &= \sum_{n=1}^6 s_n \mathcal{H}^n_{ijk\ell}, \label{5elasticPTI}
\end{align}
where
\begin{align}
p_1 &= 2P_{1111}-P_{1212}, & p_2 &= p_3 = P_{1133}, \\
p_4 &= P_{3333}, & p_5 &= P_{1212}, & p_6 &= 2P_{1313}.
\end{align}
and similarly for the Eshelby tensor with $p_n\rightarrow s_n$ and $P_{ijk\ell}\rightarrow S_{ijk\ell}$.

\subsection*{Derivation from the direct integral form}

As noted above, the P-tensor will itself be transversely isotropic of the form \eqref{5elasticPTI}. Using the direct integral formulation of the P-tensor \eqref{5Pijklellipsoid}, let the unit vector $\overline{\boldsymbol{\xi}}$ pointing to the surface of the unit sphere be parametrized by the two angles $\varphi\in[0,2\pi)$ and $\vartheta\in[0,\pi)$, i.e.\
\begin{align}
\overline{\xi}_1 &= \cos\varphi\sin\vartheta, & \overline{\xi}_2 &= \sin\varphi\sin\vartheta, & \overline{\xi}_3 &= \cos\vartheta.
\end{align}
As such, together with \eqref{5hostTIel}, \eqref{5Pijklellipsoid} becomes
\begin{align}
P_{ijk\ell} &= \frac{\vareps}{4\pi}\int_0^{2\pi}\int_0^{\pi}\frac{\Phi_{ijkl}}{(1+(\vareps^2-1)\cos^2\vartheta)^{3/2}}\sin\vartheta d\varphi d\vartheta \label{5:eshdirectform}
\end{align}
where
\begin{align}
\Phi_{ijk\ell} &= \left(\overline{\xi}_j\overline{\xi}_{\ell}N_{ik}\right)\Big|_{(ij),(k\ell)}
\end{align}
with $N_{ij}$ defined via $N_{ik}\tilde{N}_{kj} = \de_{ij}$. The components of $\mathbf{\tilde{N}}$ are defined as $\tilde{N}_{ij}=C_{ijk\ell}^0\overline{\xi}_j\overline{\xi}_{\ell}$. It is straightforward to implement this in a variety of mathematical packages or programming languages.

Here let us plot the five independent components of the P-tensor, illustrating that the two approaches above agree. Let us take the elastic properties to be transversely isotropic and choose the material PZT-7A\footnote{This material is \textit{Lead Zirconate Titanate}, a material frequently used in piezoelectric composites}. Although realistically this material would normally be chosen as the reinforcing phase in a composite, it is appropriate to illustrate the calculations for a real material. Its properties are (all stated in GPa)
\begin{align}
K &= 121.2, & m &= 35.8, & \ell &= 73, & n&= 175, & g &= 47.2.
\end{align}

\begin{figure} [ht]
\centering
\psfrag{x}{$\vareps$}
\psfrag{A}{$P_{1111}$}
\psfrag{B}{\textcolor{blue}{$P_{3333}$}}
\psfrag{C}{\textcolor{red}{$P_{1313}$}}
\hspace{-1cm}\includegraphics[width=280pt]{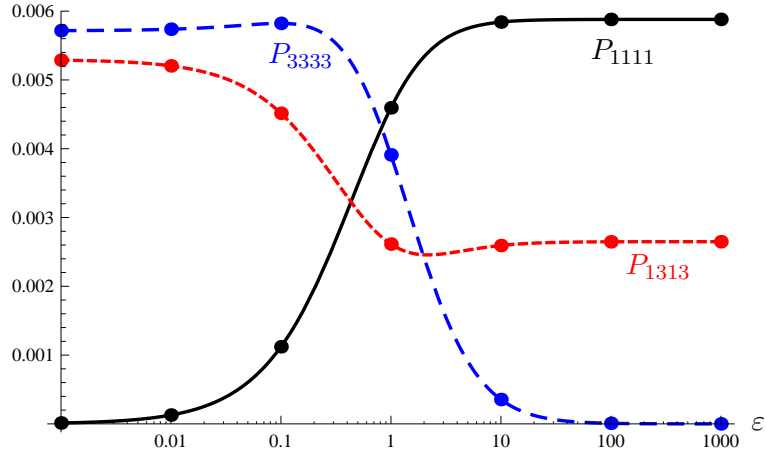}
\caption{Plot of the components $P_{1111}$ (solid black), $P_{3333}$ (dashed blue) and $P_{1313}$ (dotted red) associated with a spheroidal inhomogeneity (varying the aspect ratio $\vareps$) embedded in the transversely isotropic host phase PZT-7A, using the form of the P-tensor derived from the explicit form of the Eshelby tensor for this problem. These explicit results are confirmed by calculating the components at discrete values of the aspect ratio by evaluating the direct integral form of the P-tensor given in \eqref{5:eshdirectform}. Note that the limiting values as $\vareps\rightarrow\infty$ correspond to the circular cylinder result \eqref{5Pcyliso} and when $\vareps\rightarrow 0$ the layer limit is obtained.}
\label{fig:Patensoraniso}
\end{figure}

\begin{figure} [ht]
\centering
\psfrag{x}{$\vareps$}
\psfrag{B}{$P_{1122}$}
\psfrag{A}{\textcolor{red}{$P_{1133}$}}
\hspace{-1cm}\includegraphics[width=280pt]{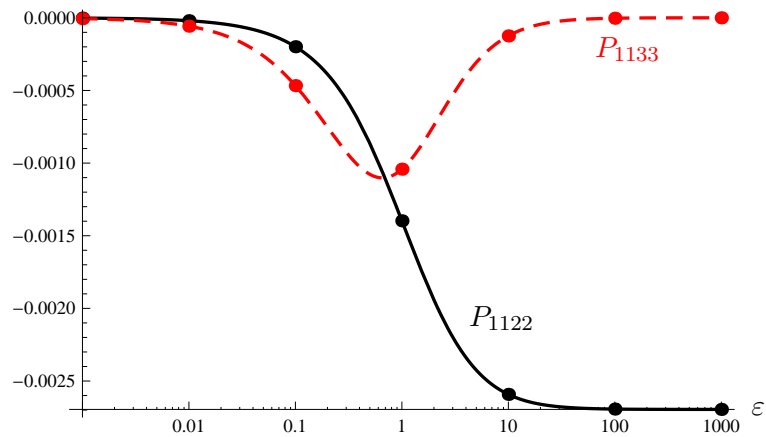}
\caption{As with Figure \ref{fig:Patensoraniso} but for the components $P_{1122}$ (solid black) and $P_{1133}$ (dashed red) of the P-tensor.}
\label{fig:Pbtensoraniso}
\end{figure}

In figures \ref{fig:Patensoraniso}-\ref{fig:Pbtensoraniso} the components of the P-tensor are plotted, using the explicit form arising from the Eshelby tensor and using the direct evaluation of the integral in order to confirm the results.

%

\subsection*{Concentration tensor}

Since the P-tensor is known one can now go on to deduce the associated concentration tensor. First assume that the spheroid is isotropic with elastic modulus tensor
\begin{align}
C^1_{ijkl} &= 3\ka_1 I_{ijkl}^1 + 2\mu_1 I_{ijkl}^2. \label{5TIspheroid}
\end{align}
Since the concentration tensor will be transversely isotropic, it is convenient to write $C^1_{ijkl}$ with respect to the transversely isotropic tensor basis, i.e.
\begin{align}
C^1_{ijkl} &= \sum_{n=1}^6 c^1_n \mathcal{H}_{ijkl}^n \label{5inclTIel}
\end{align}
where the $c^1_n$ coefficients are defined in terms of the two independent elastic moduli $\ka_1$ and $\mu_1$:
\begin{align}
c^1_1 &= 2\ka_1+\frac{2}{3}\mu_1, & c^1_2 &= \ka_1-\frac{2}{3}\mu_1, & c^1_3 &= \ka_1-\frac{2}{3}\mu_1, \\
c^1_4 &= \ka_1+\frac{4}{3}\mu_1, & c^1_5 &= 2\mu_1, & c^1_6 &= 2\mu_1.
\end{align}
Let us employ \eqref{5hostTIel}, together with the form of P-tensor defined in \eqref{5elasticPTI}. We shall also exploit the properties of the TI basis tensors $\mathcal{H}_{ijkl}^n$ described in Appendix \ref{app:apptensTI} (and in particular the contraction properties in table \ref{tab:Pbasistab}), together with the expressions written down in \eqref{5ITI}-\eqref{5I2TI}. The inverse of the concentration tensor defined in \eqref{5strainconcA} can then be determined in the form
\begin{align}
\tilde{\mathcal{A}}_{ijkl} &= \sum_{n=1}^6 \tilde{\alpha}_n \mathcal{H}_{ijkl}^n, \label{5TIATI}
\end{align}
where upon defining $c_n=c^1_n-c^0_n$,
\begin{align}
\tilde{\alpha}_1 &= 1+p_1 c_1+2 p_2 c_3, &
\tilde{\alpha}_2 &= p_1 c_2+p_2 c_4, &
\tilde{\alpha}_3 &= p_3 c_1+p_4 c_3, \label{5TIAc1}\\
\tilde{\alpha}_4 &= 1+2 p_3 c_2 + p_4 c_4, &
\tilde{\alpha}_5 &= 1+p_5 c_5, &
\tilde{\alpha}_6 &= 1+p_6c_6. \label{5TIAc3}
\end{align}
The tensor $\tilde{\mathcal{A}}_{ijkl}$ is then inverted, following the procedure in Appendix \ref{app:apptensTI}, to yield the concentration tensor
\begin{align}
\mathcal{A}_{ijkl} &= \sum_{n=1}^6 \alpha_n \mathcal{H}_{ijkl}^n \label{5AelasticTI1}
\end{align}
where
\begin{align}
\alpha_1 &= \frac{\tilde{\alpha}_1}{2\Delta}, &
\alpha_2 &= -\frac{\tilde{\alpha}_2}{2\Delta}, &
\alpha_3 &= -\frac{\tilde{\alpha}_3}{2\Delta},\\
\alpha_4 &= \frac{\tilde{\alpha}_4}{2\Delta}, & \alpha_5 &= \frac{1}{\tilde{\alpha_5}}, &
\alpha_6 &= \frac{1}{\tilde{\alpha}_6}. \label{5AelasticTI2}
\end{align}

Alternatively suppose that the inhomogeneity is transversely isotropic with the same symmetry axis as the host, i.e.\ it possesses the elastic modulus tensor of the form \eqref{5inclTIel} but where now the constants $c^1_n$ are defined in terms of the $5$ independent components of this tensor. Then the concentration tensor is again defined by \eqref{5AelasticTI1} but of course now with the $c^1_n$ associated with the \textit{transversely isotropic} cylinder. This indicates the merit of using the above notation since one can still use \eqref{5AelasticTI1}-\eqref{5AelasticTI2} in this case, merely modifying the $c^1_n$ to account for the transverse isotropy of the cylinder.

As usual, the matrix form of these fourth order tensors can be employed for computational efficiency when the problems lack simple symmetries.

\subsubsection{Circular cylinder in a transversely isotropic host}

Suppose now that the inhomogeneity is a circular cylinder with its cross-section residing in the plane of isotropy of the TI host phase. One can arrive at the corresponding P-tensor in two ways. The first is to take the limit $\vareps=a_3/a\rightarrow\infty$ in the prolate spheroid case in \S \ref{5egelasticspheroidTI}. The second way is to recognize that since the anisotropy of the host will not affect the in-plane components of the P-tensor, the tensor will simply be the same as that for an isotropic host as derived in \S \ref{examplecirccyl} but the elastic properties are modified via $\la_0+\mu_0\rightarrow K_0$ and $\mu_0\rightarrow m_0$ for in-plane components and $\mu_0\rightarrow g_0$ for the anti-plane component. Therefore, from \eqref{5Pcyliso}
\begin{align}
p_1 &= \frac{1}{2(K_0+m_0)}, & p_2 &= 0, & p_3&=0, \\
p_4 &= 0, & p_5 &= \frac{K_0+2m_0}{4m_0(K_0+m_0)}, & p_6 &= \frac{1}{4g_0}.   \label{5pncyl2cc}
\end{align}
The concentration tensor may then be derived by using these coefficients in \eqref{5TIAc1}-\eqref{5TIAc3} and the expressions that follow. 



\section{Discussion} \label{sec:disc}

\subsection{Association with micromechanics}

One of the primary reasons for deriving the Hill or Eshelby tensors and associated concentration tensors is to understand a multitude of aspects of the behaviour of inhomogeneous media, including their macroscopic constitutive response and so-called effective properties. Following a relatively straightforward argument regarding volume averaging, the effective modulus tensor $\mathbf{C}^*$ of an $n+1$ phase medium with a distinguishable host phase (phase $0$) can be stated as \cite{Mar-00}, \cite{Wil-81}
\begin{align}
\mathbf{C}^* &= \mathbf{C}^0 +  \sum_{r=1}^n \phi_r(\mathbf{C}^r-\mathbf{C}^0)\mathbf{A}^r \label{Cstar}
\end{align}
where $\phi_r$ is the volume fraction of phase $r$ and $\mathbf{A}^r$ is the \textit{exact} concentration tensor associated with embedded phase $r$. This is in contrast to the concentration tensor $\boldsymbol{\mathcal{A}}$ introduced in \S \ref{sec:conc2} which is the concentration tensor associated with an \textit{isolated} inhomogeneity, i.e.\ the presence of other inhomogeneities is not accounted for in $\boldsymbol{\mathcal{A}}$. As such if the inhomogeneity phases are distributed \textit{dilutely} then one can merely use the approximation $\mathbf{A}\approx\boldsymbol{\mathcal{A}}$ in \eqref{Cstar}. Most micromechanical methods use a more sophisticated approximation that can account, in an approximate manner at least, for interactions. One of the most commonly employed methods is the so-called classical self consistent method \cite{Kan1-08}. Interaction is \textit{approximated} in this most simple self-consistent scheme by taking the host medium in the determination of $\boldsymbol{\mathcal{A}}$ to be the unknown \textit{effective medium}. In general then \eqref{Cstar} gives rise to a nonlinear system of equations for the determination of effective properties. In many cases these are not even algebraic equations. Furthermore for the self consistent method, one has to make an assumption in advance of the symmetry properties of the effective tensor. For example in the case of aligned spheroids $\mathbf{C}^*$ will be TI. 

The textbooks referred to at the end of \S \ref{sec:intro} provide an excellent introduction to the numerous micromechanical methods, many of which are based on the form of effective modulus tensor defined in \eqref{Cstar}. A similar form can be deduced for media where there is no distinguishable host phase (e.g.\ polycrystals) and also for media where multiphysics effects are important as described in the next section.

\subsection{Beyond the potential problem and elastostatics} \label{sec:conc}

A large number of explicit, compact results associated with the Hill and Eshelby tensors for ellipsoidal inhomogeneities, as well as their associated concentration tensors have been collected, stated and in some cases derived. The intention is that this will be of great utility to a large number of researchers for implementation in micromechanical and bounding schemes. A thorough discussion of both matrix and tensor (where possible due to space limitations) formulations has been carried out. Typographical errors in past articles and reviews have been corrected and a common notation has been employed.

Although the general integral forms \eqref{5genPtransport} and \eqref{5Pijklellipsoid} are useful they should generally be avoided where explicit forms are available. Gavazzi and Lagoudas \cite{Gav-90} described a numerical implementation for elasticity. It should be noted that recently Masson \cite{Mas-08} derived a new form of the P-tensor in terms of a \textit{single} integral, although the integrand is inevitably more complex than that in the surface integral in \eqref{5Pijklellipsoid}.

In the literature many of the cases described above are considered as approximations to more complicated shaped inhomogeneities. In terms of the derivation of overall effective properties this is extremely useful, certainly as a first approximation, since it avoids complex computational simulations. However it must be stressed that more advanced analysis is required if detailed micromechanical information such as stress concentration calculations close to inhomogeneities of a complex shape is required \cite{Bur-11}. For finite domains, provided the host phase is in some sense \textit{much larger} than the inhomogeneity, if the inhomogeneity is ellipsoidal then the temperature gradient field inside the ellipsoid is well approximated as being uniform. The inhomogeneity problem associated with bounded domains is described in the book by Li and Wang \cite{Li-08} which summarizes the work in \cite{Li-07a}, \cite{Li-07b}.

Still remaining in the context of the potential problem and elastostatics, an important extension of the inhomogeneity problem is that of the \textit{coated} inhomogeneity. This problem is popular, not least because it arises as a micromechanics problem in the generalized self-consistent method (GSCM) \cite{Chr-79p}. The so-called \textit{double inclusion} problem dates back many decades and was solved approximately by Hori and Nemat-Nasser \cite{Hor-93} although the approximations involved lead to some rather counter-intuitive predictions when used in the GSCM \cite{Hu-00}. Exact solutions in the case of concentric spheroids of ellipsoids have been derived by Hatta and Taya \cite{Hat-86} in the thermal context and Jiang et al.\ \cite{Jia-03} in two-dimensional elasticity. The case of inhomogeneities with radially dependent material properties has been considered in Chapter 3 of \cite{Kan1-08} amongst others. The coated inclusion is also of interest due to its association with the \textit{neutral inclusion} problem \cite{Mil-01}. Associated with the coated inhomogeneity is the scenario when the interface of an inhomogeneity with the host phase is \textit{imperfect} \cite{Gao-95}, \cite{LeQ-10}. This imperfection can itself be used as the basis for a neutral inclusion \cite{Ru-98}, \cite{Big-02}.


It is important to note that when the inhomogeneity becomes very small, i.e.\ the case of a \textit{nano-inhomogeneity}, then surface energies become non-negligible. This problem has been considered by Sharmi and Ganti \cite{Sha-04} and Duan et al.\ \cite{Dua-05} for example. Including surface energies is important in order to incorporate size-dependent effects in effective properties. These are absent in classical micromechanical methods that use the standard Eshelby or Hill tensors.


Eshelby's problem has also been considered in the context of micro-continuum elasticity models, which themselves were introduced in order to bridge the gap between continuum and atomistic/molecular models \cite{Eri-99}. Micropolar (Cosserat) theory has been considered by Cheng and He \cite{Che-95}, \cite{Che-97} and Ma and Hu \cite{Ma-06}. Micro-stretch theory has been developed by Ma and Hu \cite{Ma-07}. Strain gradient constitutive behaviour was studied by Gao and Ma \cite{Gao-09}, \cite{Gao-10}.


The dynamic problem was considered for spheres and cylinders by Mikata and Nemat-Nasser \cite{Mik-90}, \cite{Mik-91} and more generally in \cite{Che-99},  \cite{Mic-03}. Rate dependence of the Hill and Eshelby tensors has been considered by Suvarov and Dvorak \cite{Suv-02} and viscoelastic properties have been studied by Wang and Weng \cite{Wan-92} by using transform techniques and correspondence principles. Nguyen et al.\ \cite{Ngu-11} studied cracked viscoelastic solids using the appropriate Eshelby tensor. Extensions to plasticity were considered by e.g.\ \cite{Ju-01}, \cite{Lee-01}, \cite{Fri-12}.


The Newtonian potential and elastostatics problems are canonical problems that can assist with the development of coupled (multiphysics) problems. Dunn and Taya \cite{Dun-93}, Dunn and Wienecke \cite{Dun-97} and Mikata \cite{Mik-00}, \cite{Mik-01} considered the case of piezoelectricity and the prediction of the electroelastic moduli. Li and Dunn \cite{Li-98} and Zhang and Soh \cite{Zha-05} considered full coupling and the resulting effective moduli associated with piezoelectromagnetic media. The theory associated with poroelastic and thermoelastic behaviour was developed by Berryman \cite{Ber-97} and extended to the anisotropic case by Levin and Alvarez-Tostado \cite{Lev-03}.

Upon closing it should be noted that it is very fortuitous that such elegant and concise uniformity results hold for ellipsoidal inhomogeneities.  These results allow a large number of expressions to be derived analytically and as such the results have been utilized a great deal. Having said that there is much work to be done. As has been noted, the Eshelby conjecture is still not fully resolved \cite{Amm-10}, analysis for general shaped inhomogenities continues \cite{Bur-13}, specifically in the context of stress analysis and resulting effective properties and although computational methods are powerful, they are still only able to solve elasticity problems for inhomogeneous media with an order of 1000 inhomogeneities in ``reasonable'' times. For use in Monte-Carlo schemes this is therefore still computationally expensive. Nonlinear problems in the context of finite elasticity still require attention \cite{Yav-13} and this applies to coupled problems as well.


\vspace{1cm}
\normalsize

\textit{\textbf{Acknowledgements}}:

The author is grateful to EPSRC for funding his research fellowship (EP/L018039/1).

\bibliography{bigrefs2}

\begin{thebibliography}{100}

\bibitem{Amm-10}
{\sc H.~Ammari, Y.~Capdeboscq, H.~Kang, H.~Lee, G.W. Milton, and H.~Zribi}.
\newblock Progress on the strong {E}shelby's conjecture and extremal structures
  for the elastic moment tensor.
\newblock {\em J.\ Math.\ Pures.\ Appl.}, {\bf 94}:93--106, 2010.

\bibitem{Asa-75}
{\sc R.J. Asaro and D.M. Barnett}.
\newblock The non-uniform transformation strain problem for an anisotropic
  ellipsoidal inclusion.
\newblock {\em J. Mech. Phys. Solids}, {\bf 23}:77--83, 1975.

\bibitem{Bac-80}
{\sc D.J. Bacon, D.M. Barnett, and R.O. Scattergood}.
\newblock Anisotropic continuum theory of lattice defects.
\newblock {\em Prog.\ Mater.\ Sci.}, {\bf 23}:51--262, 1980.

\bibitem{Bar-09}
{\sc J.-F. Barth\'el\'emy}.
\newblock Compliance and hill polarization tensor of a crack in an anisotropic
  matrix.
\newblock {\em Int. J. Solids Struct.}, {\bf 46}:4064--4072, 2009.

\bibitem{Ber-97}
{\sc J.G. Berryman}.
\newblock Generalization of {E}shelby's formula for a single ellipsoidal
  elastic inclusion to poroelasticity and thermoelasticity.
\newblock {\em Phys. Rev. Lett.}, {\bf 79}:1142--1145, 1997.

\bibitem{Big-02}
{\sc D.~Bigoni and A.B. Movchan}.
\newblock Statics and dynamics of structural interfaces in elasticity.
\newblock {\em Int. J. Solids Struct.}, {\bf 39}:4843--4865, 2002.

\bibitem{Bur-07}
{\sc V.~Buryachenko}.
\newblock {\em Micromechanics of heterogeneous materials}.
\newblock Springer Science, New York, 2007.

\bibitem{Bur-11}
{\sc V.\ Buryachenko and M.~Brun}.
\newblock {FEA} in elasticity of random structure composites reinforced by
  heterogeneities of non canonical shape.
\newblock {\em Int. J. Solids Struct.}, {\bf 48}:719--728, 2011.

\bibitem{Bur-13}
{\sc V.\ Buryachenko and M.~Brun}.
\newblock Iteration method in linear elasticity of random structure composites
  containing heterogeneities of non canonical shape.
\newblock {\em Int. J. Solids Struct.}, {\bf 50}:1130--1140, 2013.

\bibitem{Cal-14}
{\sc C.\ Calvo-Jurado and W.J. Parnell}.
\newblock Hashin-{S}htrikman bounds on the effective thermal conductivity of a
  transversely isotropic two-phase composite material.
\newblock {\em J.\ Math.\ Chemistry}, {\bf 53}:828--843, 2014.

\bibitem{Che-15}
{\sc F.~Chen, A.~Giraud, I.~Sevostianov, and G.~Dragan}.
\newblock Numerical evaluation of the {E}shelby tensor for a concave
  superspherical inclusion.
\newblock {\em Int. J. Engng. Sc.}, {\bf 93}:51--58, 2015.

\bibitem{Che-95}
{\sc Z.-Q. Cheng and L.-H. He}.
\newblock Micropolar elastic fields due to a spherical inclusion.
\newblock {\em Int. J. Engng. Sc.}, {\bf 33}:389--397, 1995.

\bibitem{Che-97}
{\sc Z.-Q. Cheng and L.-H. He}.
\newblock Micropolar elastic fields due to a circular cylindrical inclusions.
\newblock {\em Int. J. Engng. Sc.}, {\bf 35}:659--668, 1997.

\bibitem{Che-99}
{\sc Z.Q. Cheng and R.C. Batra}.
\newblock Exact eshelby tensor for a dynamic circular cylindrical inclusion.
\newblock {\em J. Appl. Mech. ASME}, {\bf 66}:563--565, 1999.

\bibitem{Che-74}
{\sc G.P. Cherepanov}.
\newblock Inverse problems of the plate theory of elasticity.
\newblock {\em J.\ Appl.\ Math.\ Mech.}, {\bf 38}:963--979, 1974.

\bibitem{Chr-79p}
{\sc R.M. Christensen and K.H. Lo}.
\newblock Solutions for effective shear properties in three phase sphere and
  cylinder models.
\newblock {\em J. Mech. Phys. Solids}, {\bf 27}:315--330, 1979.

\bibitem{Div-31}
{\sc P.~Dive}.
\newblock Attraction des ellipsoides homog\'{e}nes et r\'{e}ciproques d'un
  th\'{e}or\`{e}me de {N}ewton.
\newblock {\em Bull.\ Soc.\ Math.\ France}, {\bf 59}:128--140, 1931.

\bibitem{Dua-05}
{\sc H.L. Duan, J.~Wang, Z.P. Huang, and B.L. Karihaloo}.
\newblock Eshelby formalism for nano-inhomogeneities.
\newblock {\em Proc. R. Soc. A}, {\bf 461}:3335--3353, 2005.

\bibitem{Dun-93}
{\sc M.L.\ Dunn and M.~Taya}.
\newblock Micromechanics predictions of the effective electroelastic moduli of
  piezoelectric composites.
\newblock {\em Int. J. Solids Struct.}, {\bf 30}:161--175, 1993.

\bibitem{Dun-97}
{\sc M.L.\ Dunn and H.A. Wienecke}.
\newblock Inclusions and inhomogeneities in transversely isotropic
  piezoelectric solids.
\newblock {\em Int. J. Solids Struct.}, {\bf 34}:3571--3582, 1997.

\bibitem{Dvo-13}
{\sc G.~Dvorak}.
\newblock {\em Micromechanics of composite materials}.
\newblock Springer, 2013.

\bibitem{Edw-51}
{\sc R.H. Edwards}.
\newblock Stress concentrations around spheroidal inclusions and cavities.
\newblock {\em J.\ Appl.\ Mech.\ ASME}, {\bf 18}:19--30, 1951.

\bibitem{Eri-99}
{\sc A.C. Eringen}.
\newblock {\em Microcontinuum Field Theories I: Foundations and Solids}.
\newblock Springer-Verlag, New York, 1999.

\bibitem{Esh-57}
{\sc J.D. Eshelby}.
\newblock The determination of the elastic field of an ellipsoidal inclusion,
  and related problems.
\newblock {\em Proc. R. Soc. A}, {\bf 241}:376--396, 1957.

\bibitem{Esh-61}
{\sc J.D. Eshelby}.
\newblock Elastic inclusions and inhomogeneities.
\newblock In {\sc I.N. Sneddon and R.~Hill}, editors, {\em Progress in Solid
  Mechanics}, ~{\bf 2}, pages 87--140. The Netherlands, North-Holland
  Publishing Company, 1961.

\bibitem{Fri-12}
{\sc F.~Fritzen, S.~Forest, T.~B{\"o}hlke, D.~Kondo, and T.~Kanit}.
\newblock Computational homogenization of elasto-plastic porous metals.
\newblock {\em Int.\ J.\ Plasticity}, {\bf 29}:102--119, 2012.

\bibitem{Gao-09}
{\sc X.-L. Gao and H.M. Ma}.
\newblock Green's function and {E}shelby's tensor based on a simplified strain
  gradient elasticity theory.
\newblock {\em Acta Mech.}, {\bf 207}:163--181, 2009.

\bibitem{Gao-10}
{\sc X.-L. Gao and H.M. Ma}.
\newblock Strain gradient solution for {E}shelby's ellipsoidal inclusion
  problem.
\newblock {\em Proc. R. Soc. A}, {\bf 466}:20090631, 2010.

\bibitem{Gao-95}
{\sc Z.~Gao}.
\newblock A circular inclusion with imperfect interface: {E}shelby's tensor and
  related problems.
\newblock {\em J. Appl. Mech. ASME}, {\bf 62}:860--866, 1995.

\bibitem{Gav-90}
{\sc A.C. Gavazzi and D.C. Lagoudas}.
\newblock On the numerical evaluation of {E}shelby's tensor and its application
  to elastoplastic fibrous composites.
\newblock {\em Computational Mechanics}, {\bf 7}:13--19, 1990.

\bibitem{Gio-08}
{\sc S.~Giordano, P.L. Palla, and L.~Colombo}.
\newblock Nonlinear elastic {L}andau coefficients in heterogeneous materials.
\newblock {\em Eur.\ Phys.\ Letters}, {\bf 83}:66003, 2008.

\bibitem{Gir-07}
{\sc A.~Giraud, Q.~V. Huynh, D.~Hoxha, and D.~Kondo}.
\newblock Application of results on {E}shelby tensor to the determination of
  effective poroelastic properties of anisotropic rocks-like composites.
\newblock {\em Int. J. Solids Struct.}, {\bf 44}:3756--3772, 2007.

\bibitem{Goo-33}
{\sc J.N. Goodier}.
\newblock Concentration of stress around spherical and cylindrical inclusions
  and flaws.
\newblock {\em J. Appl. Mech. ASME}, {\bf 55}:39--44, 1933.

\bibitem{Gru-05}
{\sc C.~Gruescu, V.~Montchiet, and D.~Kondo}.
\newblock Eshelby tensor for a crack in an orthotropic elastic medium.
\newblock {\em C.\ R.\ M\'echanique}, {\bf 333}:467--473, 2005.

\bibitem{Has-63b}
{\sc Z.~Hashin}.
\newblock Theory of mechanical behaviour of heterogeneous solids.
\newblock {\em Appl. Mech. Rev.}, {\bf 17}:1--0, 1963.

\bibitem{Hat-86}
{\sc H.~Hatta and M.~Taya}.
\newblock Thermal conductivity of coated filler composites.
\newblock {\em J. Appl. Phys.}, {\bf 59}:1851--1860, 1986.

\bibitem{Hil-65}
{\sc R.~Hill}.
\newblock A self-consistent mechanics of composite materials.
\newblock {\em J. Mech. Phys. Solids}, {\bf 13}:213--222, 1965.

\bibitem{Hoe-79}
{\sc A.~Hoenig}.
\newblock Elastic moduli of a non-randomly cracked body.
\newblock {\em Int.\ J.\ Solids Structures}, {\bf 15}:137--154, 1979.

\bibitem{Hoe-83}
{\sc A.~Hoenig}.
\newblock Thermal conductivities of a cracked solid.
\newblock {\em J.\ Comp.\ Materials}, {\bf 17}:231--237, 1983.

\bibitem{Hor-93}
{\sc M.~Hori and S.~Nemat-Nasser}.
\newblock Double-inclusion model and overall moduli of multi-phase composites.
\newblock {\em Mech. Mater.}, {\bf 14}:189--206, 1993.

\bibitem{Hu-00}
{\sc G.K. Hu and G.J. Weng}.
\newblock The connections between the double-inclusion model and the
  {P}onte-{C}astaneda--{W}illis, {M}ori--{T}anaka, and {K}uster--{T}{\"o}ksoz
  models.
\newblock {\em Mech. Mater.}, {\bf 32}:495--503, 2000.

\bibitem{Jia-03}
{\sc C.P. Jiang, Z.H. Tong, and Y.K. Cheung}.
\newblock A generalized self-consistent method accounting for fiber section
  shape.
\newblock {\em Int. J. Solids Struct.}, {\bf 40}:2589--2609, 2003.

\bibitem{Ju-01}
{\sc J.W. Ju and L.Z. Sun}.
\newblock Effective elastoplastic behavior of metal matrix composites
  containing randomly located aligned spheroidal inhomogeneities. {P}art {I}
  micromechanics based formulation.
\newblock {\em Int. J. Solids Struct.}, {\bf 38}:183--201, 2001.

\bibitem{Kan1-08}
{\sc S.K. Kanaun and V.M. Levin}.
\newblock {\em Self-Consistent methods for composites. {V}olume 1 - {S}tatic
  problems}.
\newblock Springer, Dordrecht, 2008.

\bibitem{Kan-09}
{\sc H.~Kang}.
\newblock Conjectures of {P}\'{o}lya-{S}zeg{\"o} and {E}shelby, and the
  {N}ewtonian potential problem: {A} review.
\newblock {\em Mech. Mater.}, {\bf 41}:405--410, 2009.

\bibitem{Kan-08}
{\sc H.~Kang and G.W. Milton}.
\newblock Solutions to the {P}\'{o}lya-{S}zeg{\"o} conjecture and the weak
  {E}shelby conjecture.
\newblock {\em Arch.\ Rational Mech.\ Anal.}, {\bf 188}:93--116, 2008.

\bibitem{Kaw-01}
{\sc M.~Kawashita and H.~Nozaki}.
\newblock Esehlby tensor of a polygonal inclusion and its special properties.
\newblock {\em J.\ Elasticity Phys.\ Science Solids}, {\bf 64}:71--84, 2001.

\bibitem{Kel-70}
{\sc O.D. Kellogg}.
\newblock {\em Foundations of Potential Theory}.
\newblock Frederick Ungar Publishing Company, 1970.

\bibitem{Kim-07}
{\sc C.I. Kim and P.~Schiavone}.
\newblock Designing an inhomogeneity with uniform interior stress in finite
  plane elastostatics.
\newblock {\em Int.\ J.\ Non-Linear Mech.}, {\bf 197}:285--299, 2007.

\bibitem{Kim-08}
{\sc C.I. Kim, M.~Vasudevan, and P.~Schiavone}.
\newblock Eshelby's conjecture in finite plane elastostatics.
\newblock {\em Q.\ J.\ Mech.\ Appl.\ Math.}, {\bf 61}:63--73, 2008.

\bibitem{Kin-71}
{\sc N.~Kinoshita and T.~Mura}.
\newblock Elastic fields of inclusions in anisotropic media.
\newblock {\em Phys.\ Stat.\ Sol.\ (a)}, {\bf 5}:759--768, 1971.

\bibitem{Law-77}
{\sc N.~Laws}.
\newblock The determination of stress and strain concentrations at an
  ellipsoidal inclusion in an anisotropic material.
\newblock {\em J. Elasticity}, {\bf 7}(1):91--97, 1977.

\bibitem{LeQ-10}
{\sc H.~Le~Quang, G.~Bonnet, and Q.-C. He}.
\newblock Size-dependent {E}shelby tensor fields and effective conductivity of
  composites made of anisotropic phases with highly conducting interfaces.
\newblock {\em Phys. Rev. B}, {\bf 81}:064203, 2010.

\bibitem{Lee-01}
{\sc H.K. Lee and S.~Simunovic}.
\newblock A damage constitutive model of progressive debonding in aligned
  discontinuous fiber composites.
\newblock {\em Int. J. Solids Struct.}, {\bf 38}:875--895, 2001.

\bibitem{Lev-03}
{\sc V.M. Levin and J.M. Alvarez-Tostado}.
\newblock Eshelby's formula for an ellipsoid elastic inclusion in anisotropic
  poroelasticity and thermoelasticity.
\newblock {\em Int.\ J.\ Fracture}, {\bf 119}:77--82, 2003.

\bibitem{Li-98}
{\sc J.Y. Li and M.L. Dunn}.
\newblock Anisotropic coupled-field inclusion and inhomogeneity problems.
\newblock {\em Phil.\ Mag.\ A}, {\bf 77}:1341--1350, 1998.

\bibitem{Li-07a}
{\sc S.~Li, R.A. Sauer, and G.~Wang}.
\newblock The {E}shelby tensors in a finite spherical domain -- {P}art {I}:
  {T}heoretical formulations.
\newblock {\em J. Appl. Mech. ASME}, {\bf 74}:770--783, 2007.

\bibitem{Li-08}
{\sc S.~Li and G.~Wang}.
\newblock {\em Introduction to {M}icromechanics and {N}anomechanics}.
\newblock World Scientific, 2008.

\bibitem{Li-07b}
{\sc S.~Li, G.~Wang, and R.A. Sauer}.
\newblock The {E}shelby tensors in a finite spherical domain -- {P}art {II}:
  {A}pplications to homogenization.
\newblock {\em J. Appl. Mech. ASME}, {\bf 74}:784--797, 2007.

\bibitem{Lin-73}
{\sc S.C. Lin and T.~Mura}.
\newblock Elastic fields of inclusions in anistropic media (ii).
\newblock {\em Phys.\ Stat.\ Sol.\ (a)}, {\bf 15}:281--285, 1973.

\bibitem{Liu-09}
{\sc L.~Liu}.
\newblock Solutions to the periodic eshelby inclusion problem in two
  dimensions.
\newblock {\em Math. Mech. Solids}, {\bf 15}:557--590, 2009.

\bibitem{Liu-07}
{\sc L.~Liu, R.D. James, and P.H. Leo}.
\newblock Periodic inclusion--matrix microstructures with constant field
  inclusions.
\newblock {\em Met.\ Mat.\ Trans.\ A}, {\bf 38}:781--787, 2007.

\bibitem{Liu-08}
{\sc L.P. Liu}.
\newblock Solutions to the {E}shelby conjectures.
\newblock {\em Proc.\ Roy.\ Soc.\ A}, {\bf 464}:573--594, 2008.

\bibitem{Lub-98}
{\sc V.A. Lubarda and X.~Markenscoff}.
\newblock On the absence of {E}shelby property for non-ellipsoidal inclusions.
\newblock {\em Int. J. Solids Struct.}, {\bf 35}:3405--3411, 1998.

\bibitem{Ma-06}
{\sc H.~Ma and G.~Hu}.
\newblock Eshely tensors for an ellipsoidal inclusion in a micropolar material.
\newblock {\em Int. J. Engng. Sc.}, {\bf 44}:595--605, 2006.

\bibitem{Ma-07}
{\sc H.~Ma and G.~Hu}.
\newblock Eshelby tensors for an ellipsoidal inclusion in a microstretch
  material.
\newblock {\em Int. J. Engng. Sc.}, {\bf 44}:3094--3061, 2007.

\bibitem{Mar-98b}
{\sc X.~Markenscoff}.
\newblock Inclusions with constant eigenstress.
\newblock {\em J. Mech. Phys. Solids}, {\bf 46}:2297--2301, 1998.

\bibitem{Mar-98a}
{\sc X.~Markenscoff}.
\newblock On the shape of the {E}shelby inclusions.
\newblock {\em J. Elasticity}, {\bf 49}:163--166, 1998.

\bibitem{Mar-00}
{\sc K.~Markov}.
\newblock Elementary micromechanics of heterogeneous media.
\newblock In {\sc K.~Markov and L.~Preziosi}, editors, {\em Heterogeneous
  media. Micromechanics modeling. {M}ethods and {S}imulations.}, chapter~1,
  pages 1--162. Boston, Birkh{\"a}user, 2000.

\bibitem{Mas-08}
{\sc R.~Masson}.
\newblock New explicit expressions of the {H}ill polarization tensor for
  general anisotropic elastic solids.
\newblock {\em Int. J. Solids Struct.}, {\bf 45}(3):757--769, 2008.

\bibitem{Max-98}
{\sc J.C. Maxwell}.
\newblock {\em A Treatise on Electricity and Magnetism. Vols 1 and 2}.
\newblock Oxford University Press, Oxford, 1998.

\bibitem{Mic-03}
{\sc T.M. Michelitsch, H.~Gao, and V.M. Levin}.
\newblock Dynamic {E}shelby tensor and potentials for ellipsoidal inclusions.
\newblock {\em Proc. R. Soc. A}, {\bf 459}:863--890, 2003.

\bibitem{Mik-00}
{\sc Y.~Mikata}.
\newblock Determination of piezoelectric {E}shelby tensor in transversely
  isotropic piezoelectric solids.
\newblock {\em Int. J. Engng. Sc.}, {\bf 38}:605--641, 2000.

\bibitem{Mik-01}
{\sc Y.~Mikata}.
\newblock Explicit determination of piezoelectric {E}shelby tensors for a
  spheroidal inclusion.
\newblock {\em Int. J. Solids Struct.}, {\bf 38}:7045--7063, 2001.

\bibitem{Mik-90}
{\sc Y.~Mikata and S.~Nemat-Nasser}.
\newblock Elastic field due to a dynamically transforming spherical inclusion.
\newblock {\em Int. J. Solids Struct.}, {\bf 38}:7045--7063, 1990.

\bibitem{Mik-91}
{\sc Y.~Mikata and S.~Nemat-Nasser}.
\newblock Interaction of a harmonic wave with a dynamically transforming
  inhomogeneity.
\newblock {\em J. Appl. Phys.}, {\bf 70}:2071--2078, 1991.

\bibitem{Mil-01}
{\sc G.W. Milton and S.K. Serkov}.
\newblock Coated inclusions in conductivity and anti-plane elasticity.
\newblock {\em Proc. R. Soc. A}, {\bf 457}:1973--1999, 2001.

\bibitem{Moe-08}
{\sc M.~Moekher}.
\newblock Fourth-order {C}artesian tensors: {O}ld and new facts, notions and
  applications.
\newblock {\em Q. J. Mech. Appl. Math.}, {\bf 61}:181--203, 2008.

\bibitem{Mos-75}
{\sc Z.A. Moschovidis and T.~Mura}.
\newblock Two-ellipsoidal inhomogeneities by the equivalent inclusion method.
\newblock {\em J.\ Appl.\ Mech.\ ASME}, {\bf 42}:847--852, 1975.

\bibitem{Mur-82}
{\sc T.~Mura}.
\newblock {\em Micromechanics of Defects in Solids}.
\newblock Kluwer, The Hague, 1982.

\bibitem{Mur-97}
{\sc T.~Mura}.
\newblock The determination of the elastic field of a polygonal star shaped
  inclusion.
\newblock {\em Mech.\ Res.\ Comm.}, {\bf 24}:473--482, 1997.

\bibitem{Mur-96}
{\sc T.~Mura, H.M. Shodja, and Y.~Hirose}.
\newblock Inclusion problems.
\newblock {\em Appl.\ Mech.\ Rev.}, {\bf 49}:S118--S127, 1996.

\bibitem{Mur-94}
{\sc T.~Mura, H.M. Shojda, T.Y. Lin, and A.~Makkawy}.
\newblock The determination of the elastic field of a pentagonal star shaped
  inclusion.
\newblock {\em Bull.\ Tech.\ Univ.\ Istanbul}, {\bf 47}:267--280, 1994.

\bibitem{Ngu-11}
{\sc S.T. Nguyen, L.~Dormieux, Y.~Le~Pape, and J.~Sanahuja}.
\newblock A {B}urger model for the effective behavious of a microcracked
  viscoelastic solid.
\newblock {\em Int.\ J.\ Damage Mech.}, page 1056789510395554, 2011.

\bibitem{Nik-32}
{\sc W.~Nikliborc}.
\newblock Eine bemerkung {\"u}ber die {V}olumpotentiale.
\newblock {\em Math.\ Zeit.}, {\bf 35}:625--631, 1932.

\bibitem{Noz-97}
{\sc H~Nozaki and M.~Taya}.
\newblock Elastic fields in a polygon-shaped inclusion with uniform
  eigenstrains.
\newblock {\em J. Appl. Mech. ASME}, {\bf 64}:495--502, 1997.

\bibitem{Ona-01}
{\sc S.~Onaka}.
\newblock Averaged eshelby tensor and elastic strain energy of a superspherical
  inclusion with uniform eigenstrains.
\newblock {\em Phil.\ Mag.\ Letters}, {\bf 81}:265--272, 2001.

\bibitem{Ona-02}
{\sc S.~Onaka}.
\newblock Elastic states of doughnut-like inclusions with uniform eigenstrains
  treated by averaged eshelby tensors.
\newblock {\em Phil.\ Mag.\ Letters}, {\bf 82}:1--7, 2002.

\bibitem{Ona-12}
{\sc S.~Onaka}.
\newblock Superspheres: Intermediate shapes between spheres and polyhedra.
\newblock {\em Symmetry}, {\bf 4}:336--343, 2012.

\bibitem{Pan-76}
{\sc Y.-C. Pan and T.-W. Chou}.
\newblock Point force solution for an infinite transversely isotropic solid.
\newblock {\em J.\ Appl.\ Mech.\ ASME}, {\bf 43}:608--612, 1976.

\bibitem{Par-15}
{\sc W.J.\ Parnell and C.~Calvo-Jurado}.
\newblock On the computation of the {H}ashin-{S}htrikman bounds for
  transversely isotropic two-phase linear elastic fibre-reinforced composites.
\newblock \textit{J.\ Eng.\ Mathematics}, In press, 2015.

\bibitem{Poi-26}
{\sc S.D. Poisson}.
\newblock Second m\'{e}moire sur la th\'{e}orie de magnetisme.
\newblock {\em M\'{e}m. Acad. R. Sci. Inst. France}, {\bf 5}:488--533, 1826.

\bibitem{Pon-95}
{\sc P.\ Ponte~Casta{\~n}eda and J.R. Willis}.
\newblock The effect of spatial distribution on the effective behaviour of
  composite materials and cracked media.
\newblock {\em J. Mech. Phys. Solids}, {\bf 43}:1919--1951, 1995.

\bibitem{Qu-06}
{\sc J.~Qu and M.~Cherkaoui}.
\newblock {\em Fundamentals of {M}icromechanics of {S}olids}.
\newblock Wiley, 2006.

\bibitem{Rob-51}
{\sc K.~Robinson}.
\newblock Elastic energy of an ellipsoidal inclusion in an infinite solid.
\newblock {\em J. Appl. Phys.}, {\bf 22}:1045--1054, 1951.

\bibitem{Rod-96}
{\sc G.J. Rodin}.
\newblock Eshelby's inclusion problem for polygons and polyhedra.
\newblock {\em J. Mech. Phys. Solids}, {\bf 44}:1977--1995, 1996.

\bibitem{Ru-98}
{\sc C.Q. Ru}.
\newblock Interface design of neutral elastic inclusions.
\newblock {\em Int. J. Solids Struct.}, {\bf 35}:559--572, 1998.

\bibitem{Ru-99}
{\sc C.Q. Ru}.
\newblock Analytic solution for eshelby's problem of an inclusion of arbitrary
  shape in a plane or half-plane.
\newblock {\em J. Appl. Mech. ASME}, {\bf 66}:315--322, 1999.

\bibitem{Ru-96}
{\sc C.Q. Ru and P.~Schiavone}.
\newblock On the elliptic inclusion in anti-plane shear.
\newblock {\em Math.\ Mech.\ Solids}, {\bf 1}:327--333, 1996.

\bibitem{Ru-05}
{\sc C.Q. Ru, P.~Schiavone, L.J. Sudak, and A.~Mioduchowski}.
\newblock Uniformity of stresses inside an elliptic inclusion in finite plane
  elastostatics.
\newblock {\em Int.\ J.\ Non-Linear Mech.}, {\bf 40}:281--287, 2005.

\bibitem{Sad-47}
{\sc M.A.\ Sadowsky and E.~Sternberg}.
\newblock Stress concentration around an ellipsoidal cavity in an infinite body
  under arbitrary plane stress perpendicular to the axis of revolution of
  cavity.
\newblock {\em J.\ Appl.\ Mech.\ ASME}, {\bf 14}:1947, 1947.

\bibitem{Sad-49}
{\sc M.A. Sadowsky and E.~Sternberg}.
\newblock Stress concentration around a triaxial ellipsoidal cavity.
\newblock {\em J.\ Appl.\ Mech.\ ASME}, {\bf 16}(2):149--157, 1949.

\bibitem{Sev-05}
{\sc I.~Sevostianov, N.~Yilmaz, V.~Kushch, and V.~Levin}.
\newblock Effective elastic properties of matrix composites with
  transversely-isotropic phases.
\newblock {\em Int. J. Solids Struct.}, {\bf 42}:455--476, 2005.

\bibitem{Sha-04}
{\sc P.~Sharma and S.~Ganti}.
\newblock Size-dependent {E}shelby's tensor for embedded nano-inclusions
  incorporating surface/interface.
\newblock {\em J. Appl. Mech. ASME}, {\bf 71}:663--671, 2004.

\bibitem{Sou-26}
{\sc R.V. Southwell and H.J. Gough}.
\newblock {VI}. {O}n the concentration of stress in the neighbourhood of a
  small spherical flaw; and on the propagation of fatigue fractures in
  ``{S}tatistically {I}sotropic'' materials.
\newblock {\em Lond.\ Edin.\ Dublin Phil.\ Mag.\ J.\ Science}, {\bf
  1}(1):71--97, 1926.

\bibitem{Suv-02}
{\sc A.P.\ Suvarov and G.J. Dvorak}.
\newblock Rate form of the {E}shelby and {H}ill tensors.
\newblock {\em Int. J. Solids Struct.}, {\bf 39}:5659--5678, 2002.

\bibitem{Wal-67}
{\sc L.J. Walpole}.
\newblock The elastic field of an inclusion in an anisotropic medium.
\newblock {\em Proc. R. Soc. A}, {\bf 300}:270--289, 1967.

\bibitem{Wal-77}
{\sc L.J. Walpole}.
\newblock The determination of the elastic field of an ellipsoidal inclusion in
  an anisotropic medium.
\newblock {\em Math.\ Proc.\ Camb.\ Phil.\ Soc.}, {\bf 81}:283--289, 1977.

\bibitem{Wal-81}
{\sc L.J. Walpole}.
\newblock Elastic behaviour of composite materials: Theoretical foundations.
\newblock {\em Advances in Appl. Mech.}, {\bf 21}:169--242, 1981.

\bibitem{Wal-84}
{\sc L.J. Walpole}.
\newblock Fourth-rank tensor of the thirty-two crystal classes: multiplication
  tables.
\newblock {\em Proc. R. Soc. A}, {\bf 391}:149--179, 1984.

\bibitem{Wan-04}
{\sc M.Z. Wang and B.X. Xu}.
\newblock The arithmetic mean theorem of {E}shelby tensor for a rotational
  symmetrical inclusion.
\newblock {\em J. Elasticity}, {\bf 77}:13--23, 2004.

\bibitem{Wan-92}
{\sc Y.M. Wang and G.J. Weng}.
\newblock The influence of inclusion shape on the overall viscoelastic behavior
  of composites.
\newblock {\em J. Appl. Mech. ASME}, {\bf 59}:510--518, 1992.

\bibitem{Wen-84}
{\sc G.J. Weng}.
\newblock Some elastic properties of reinforced solids, with special reference
  to isotropic ones containing spherical inclusions.
\newblock {\em Int. J. Solids Struct.}, {\bf 22}:845--856, 1984.

\bibitem{Wil-64}
{\sc J.R. Willis}.
\newblock Anisotropic elastic inclusion problems.
\newblock {\em Q. J. Mech. Appl. Math.}, {\bf 17}:157--174, 1964.

\bibitem{Wil-77}
{\sc J.R. Willis}.
\newblock Bounds and self-consistent estimates for the overall moduli of
  anisotropic composites.
\newblock {\em J. Mech. Phys. Solids}, {\bf 25}:185--202, 1977.

\bibitem{Wil-80a}
{\sc J.R. Willis}.
\newblock A polarization approach to the scattering of elastic waves - {I}.
  {S}cattering by a single inclusion.
\newblock {\em J. Mech. Phys. Solids}, {\bf 28}:287--305, 1980.

\bibitem{Wil-81}
{\sc J.R. Willis}.
\newblock Variational and related methods for the overall properties of
  composites.
\newblock {\em Advances in Appl. Mech.}, {\bf 21}:1--78, 1981.

\bibitem{Wit-89}
{\sc P.J. Withers}.
\newblock The determination of the elastic field of an ellipsoidal inclusion in
  a transversely isotropic medium, and its relevance to composite materials.
\newblock {\em Phil.\ Mag.\ A.}, {\bf 59}:759--781, 1989.

\bibitem{Wu-66}
{\sc T.T. Wu}.
\newblock On the effect of inclusion shape on the elastic moduli of a two-phase
  material.
\newblock {\em Int. J. Solids Struct.}, {\bf 2}:1--8, 1966.

\bibitem{Yav-13}
{\sc A.~Yavari and A.~Goriely}.
\newblock Nonlinear elastic inclusions in isotropic solids.
\newblock {\em Proc.\ Roy.\ Soc.\ A}, {\bf 469}(2160):20130415, 2013.

\bibitem{Zha-05}
{\sc Z.K. Zhang and A.K. Soh}.
\newblock Micromechanics predictions of the effective moduli of
  magnetoelectroelastic composite materials.
\newblock {\em Eur. J. Mech. A}, {\bf 24}:1054--1067, 2005.

\bibitem{Zhe-06}
{\sc Q.-S. Zheng, Z.-H. Zhao, and D.X. Du}.
\newblock Irreducible structure, symmetry and average of eshelby's tensor
  fields in isotropic elasticity.
\newblock {\em J. Mech. Phys. Solids}, {\bf 54}:368--383, 2006.

\bibitem{Zho-13}
{\sc K.~Zhou, H.~Jen~Hoh, X.~Wang, L.M. Keer, J.H.L. Pang, B.~Song, and Q.J.
  Wang}.
\newblock A review of recent works on inclusions.
\newblock {\em Mech. Mater.}, {\bf 60}:144--158, 2013.

\bibitem{Zho-11}
{\sc K.~Zhou, L.M. Keer, and Q.J. Wang}.
\newblock Semi-analytic solution for multiple interacting three-dimensional
  inhomogeneous inclusions of arbitrary shape in an infinite space.
\newblock {\em Int.\ J.\ Numer.\ Meth.\ Engng}, {\bf 87}:617--638, 2011.

\bibitem{Zou-10}
{\sc W.~Zou, Q.~He, M.~Huang, and Q.~Zheng}.
\newblock Eshelby's problem of non-elliptical inclusions.
\newblock {\em J. Mech. Phys. Solids}, {\bf 58}:346--372, 2010.

\bibitem{Zou-11}
{\sc W.-N. Zou, Q.-S. Zheng, and Q.-C. He}.
\newblock Solutions to {E}shelby's problems of non-elliptical thermal
  inclusions and cylindrical elastic inclusions of non-elliptical
  cross-section.
\newblock {\em Proc. R. Soc. A}, {\bf 467}:607--626, 2011.

\end{thebibliography}
\bibliographystyle{gillow}  

\begin{appendix}

\section{Uniform P-tensors for ellipsoidal inhomogeneities} \label{app:ellipsoid}

Fourier transforms can be applied in a straightforward manner to derive forms of the  Green's tensors that are useful in the context of deriving properties of the Hill and Eshelby tensors. For arbitrary anisotropy in the potential problem the Green's function takes the form  \cite{Mur-82}
\begin{align}
G(\mathbf{z}) &= \frac{1}{16\pi^3}\int_{S^2}\int_{-\infty}^{\infty}\frac{1}{C_{ij}\bar{\xi}_i\bar{\xi}_j}\exp(i\xi\bar{\boldsymbol{\xi}}\cdot\mathbf{z})\hspace{0.1cm}d\xi dS(\bar{\boldsymbol{\xi}}) \label{GFt}
\end{align}
where the Fourier transform variable $\boldsymbol{\xi}=\xi\bar{\boldsymbol{\xi}}$ with $\xi=|\boldsymbol{\xi}|$ and where $S^2$ corresponds to $\xi=1$, i.e.\ the surface of the unit sphere. Next since
\begin{align}
\de(\mathbf{x}) &= \frac{1}{8\pi^3}\int_{-\infty}^{\infty}\int_{-\infty}^{\infty}\int_{-\infty}^{\infty}\exp(i\boldsymbol{\xi}\cdot\mathbf{x})\hspace{0.1cm} d\boldsymbol{\xi}
\end{align}
the form \eqref{GFt} becomes
\begin{align}
G(\mathbf{z}) &= \frac{1}{8\pi^2}\int_{S^2}\de(\bar{\boldsymbol{\xi}}\cdot\mathbf{z})\frac{1}{C_{ij}\bar{\xi}_i\bar{\xi}_j} \hspace{0.1cm}dS(\bar{\boldsymbol{\xi}}). \label{GF3Dint4}
\end{align}

In elastostatics an entirely analogous approach shows that
\begin{align}
G_{ij}(\mathbf{z}) &= \frac{1}{8\pi^2}\int_{S^2}\de(\bar{\boldsymbol{\xi}}\cdot\mathbf{z})N_{ij}(\bar{\boldsymbol{\xi}}) \hspace{0.1cm}dS(\bar{\boldsymbol{\xi}}) \label{GFforelast}
\end{align}
where
\begin{align}
\tilde{N}_{ik}(\boldsymbol{\xi})N_{kj}(\boldsymbol{\xi}) &= \de_{ij}, & \tilde{N}_{ij}(\boldsymbol{\xi}) &= C_{ijk\ell}\xi_j\xi_{\ell}.
\end{align}
%

\subsection{The potential problem} \label{app:ellipsoidpot}

Substitute the general form \eqref{GF3Dint4} of the free-space Green's function into \eqref{4:Pijuniform} and it is found that the resulting expression must be integrated over the inclusion region $V_1$ as well as $\mathbf{x}\in V_1$, leading to the form
\begin{align}
P_{ij}(\mathbf{x}) &=
    -\frac{1}{8\pi^2}\derivtwomix{}{x_i}{x_j}\int_{S^2}{\frac{1}{C^0_{k\ell}\overline{\xi}_k\overline{\xi}_{\ell}}}
    J(\overline{\boldsymbol{\xi}}\cdot\mathbf{x})dS
\label{Pkltransdelta}
\end{align}
where
\begin{align}
J(p) &= \int_{\mathbf{y}\in V_1}\de(p-\overline{\boldsymbol{\xi}}\cdot\mathbf{y})\hspace{0.2cm}d\mathbf{y}.
\end{align}
Take the simplest case, where $V_1$ is a sphere of radius $a$. With $\mathbf{x}\in V_1$ then $p\leq a$ since $\overline{\boldsymbol{\xi}}$ is a unit vector. It is then recognized that the value of $J(p)$ is the area of the disc defined by the intersection of the plane $\overline{\boldsymbol{\xi}}\cdot\mathbf{y}=p$ with the sphere $V_1$. Since $|p|\leq a$, $J(p) = \pi(a^2-p^2)$ and carrying out the necessary differentiation gives
\begin{align}
P_{ij}^{\tn{sphere}}  &= \frac{1}{4\pi}\int_{S^2} \frac{\overline{\xi}_i\overline{\xi}_j}{C_{k\ell}^0\overline{\xi}_k\overline{\xi}_{\ell}}\hspace{0.2cm}dS. \label{PklS}
\end{align}
Since this integral is over the unit sphere $S^2$ and involves only $\overline{\xi}_i$ then this shows that the $P$-tensor is independent of $\mathbf{x}$ for a spherical inhomogeneity. This was evident from the fact that $J(p)$ is a quadratic function of its argument. In order to perform the integration over the unit sphere, introduce the parameters $\vartheta\in[0,\pi)$ and $\varphi\in[0,2\pi)$ via
\begin{align}
\overline{\xi}_1 &= \cos\varphi\sin\vartheta, & \overline{\xi}_2 &= \sin\varphi\sin\vartheta, & \overline{\xi}_3 &= \cos\vartheta \label{appKvectors}
\end{align}
and then
\begin{align}
P_{k\ell}^{\tn{sphere}} &= \frac{1}{4\pi}\int_0^{2\pi}\int_0^{\pi} \Phi_{k\ell}(\overline{\boldsymbol{\xi}})\hspace{0.2cm}\sin\vartheta d\vartheta d\varphi \label{Pklsphere}
\end{align}
where
\begin{align}
\Phi_{ij}(\overline{\boldsymbol{\xi}}) &= \frac{\overline{\xi}_i\overline{\xi}_j}{C_{k\ell}\overline{\xi}_k\overline{\xi}_{\ell}}.
\end{align}

It is straightforward to extend this derivation in order to derive the corresponding result for an ellipsoid. Suppose that the ellipsoid is defined by $V=\{\mathbf{y}:\mathbf{y}^T (\mathbf{a}^T\mathbf{a})^{-1}\mathbf{y}<1\}$ where $\mathbf{a}$ is the second order tensor defined in \eqref{analigned} with $a_j$ being the semi-axes of the ellipsoid. With $\mathbf{x}\in V_1$, the only aspect that changes from the calculation for the sphere is that now the function $J(p)$ will be the area of the region defined by the intersection of the plane $\overline{\boldsymbol{\xi}}\cdot\mathbf{y}=p$ with the ellipsoid $V_1$. It transpires that
\begin{align}
J(p) &= \frac{\det{\mathbf{a}}}{(\overline{\xi}_k a_{k\ell}a_{\ell m}\overline{\xi}_m)^{3/2}}\pi(a^2-p^2).
\end{align}
Importantly the integral is still only over the unit sphere and the result is \eqref{5genPtransport}.

\subsection{Elastostatics}

One can proceed entirely analogously to the transport case in order to derive the following representation of the $P$-tensor for an ellipsoid in the elastostatics context. Using \eqref{GFforelast} and following the same procedure as for the potential problem one obtains \eqref{5Pijklellipsoid}.
%

\section{Potential theory} \label{app:potential}

Two important integrals over ellipsoids arise in potential theory, having important applications in micromechanics. Define the two functions
\begin{align}
\Gamma(\mathbf{x}) &= -\frac{1}{4\pi}\int_{V_1}\frac{1}{|\mathbf{x}-\mathbf{y}|}\hspace{0.1cm}d\mathbf{y}, &
\Psi(\mathbf{x}) &= \frac{1}{4\pi}\int_{V_1}|\mathbf{x}-\mathbf{y}|\hspace{0.1cm}d\mathbf{y}
\end{align}
where $V_1$ is the ellipsoidal domain defined by the inequality
\begin{align}
\frac{y_1^2}{a_1^2} + \frac{y_2^2}{a_2^2} + \frac{y_3^2}{a_3^2} &\leq 1
\end{align}
and of interest is the case when $\mathbf{x}\in V_1$.

Introduce local \textit{spherical} polar coordinates via $y_j=x_j+z_j, j=1,2,3$ where
\begin{align}
z_1 &= r\cos\varphi\sin\vartheta, &  z_2 &= r\sin\varphi\sin\vartheta, & z_3 &= r\cos\vartheta,
\end{align}
where $\vartheta\in[0,\pi), \varphi\in[0,2\pi)$ and $r\in[0,\infty)$. The surface of the ellipsoid is given by $r=R_1(\vartheta,\varphi)$ and a differential volume element is $dV=r^2\sin\vartheta drd\vartheta d\varphi$ so that
\begin{align}
\Gamma(\mathbf{x}) 
 &= -\frac{1}{8\pi}\int_{0}^{2\pi}\int_0^{\pi}R_1^2(\vartheta,\varphi)\sin\vartheta drd\vartheta d\varphi
\end{align}
and similarly
\begin{align}
\Psi(\mathbf{x})  &= \frac{1}{16\pi}\int_{0}^{2\pi}\int_0^{\pi}R_1^4(\vartheta,\varphi)\sin\vartheta d\vartheta d\varphi.
\end{align}
Defining the shifted variable $\psi=\vartheta-\pi/2$ and upon defining $r_1(\psi,\varphi)=R_1(\psi+\pi/2,\varphi)$ these become
\begin{align}
\Gamma(\mathbf{x}) &= -\frac{1}{8\pi}\int_{-\pi/2}^{\pi/2}\int_0^{2\pi}r_1^2(\psi,\varphi)\cos\psi d\varphi d\psi, \\
\Psi(\mathbf{x})  &= \frac{1}{16\pi}\int_{-\pi/2}^{\pi/2}\int_0^{2\pi}r_1^4(\psi,\varphi)\cos\psi d\varphi d\psi.
\end{align}

At this point note that
\begin{multline}
\int_{-\pi/2}^{\pi/2}\int_0^{2\pi} f(\psi,\varphi)d\theta d\psi 
= \int_0^{\pi/2}\int_0^{\pi}[f(\psi,\varphi)+f(\psi,\varphi+\pi)]\\
+[f(\varphi,-\psi)+f(\varphi+\pi,\psi)]\hspace{0.1cm} d\varphi d\psi. \label{appsym2a}
\end{multline}
This pairing is useful to argue that certain integrals below are zero. 
Evaluating the local spherical polar coordinates on the surface of the ellipsoid yields
\begin{align}
A r_1^2 + 2B r_1 + C &= 0
\end{align}
where
\begin{align}
A &= \frac{\cos^2\psi\cos^2\varphi}{a_1^2} + \frac{\cos^2\psi\sin^2\varphi}{a_2^2} + \frac{\sin^2\psi}{a_3^2} \label{app:Aff}\\
B &= \frac{x_1\cos\psi\cos\varphi}{a_1^2} + \frac{x_2\cos\psi\sin\varphi}{a_2^2} - \frac{x_3\sin\psi}{a_3^2} \label{app:Bff}\\
C &= \frac{x_1^2}{a_1^2} + \frac{x_2^2}{a_2^2} + \frac{x_3^2}{a_3^2} - 1 \label{app:Cff}
\end{align}
noting that $A>0$ and $C<0$, and thus
\begin{align}
r_1 &= \frac{-B+\sqrt{B^2-AC}}{A}
\end{align}
where the positive root is chosen since $B^2-AC>B^2$ ($AC<0$). Hence
\begin{align}
\Gamma(\mathbf{x}) &= \frac{1}{8\pi}\int_{-\pi/2}^{\pi/2}\int_0^{2\pi}\frac{AC-2B^2+2B\sqrt{B^2-AC}}{A^2}\cos\psi d\varphi d\psi
\end{align}
and
\begin{multline}
\Psi(\mathbf{x}) = \frac{1}{16\pi}\int_{-\pi/2}^{\pi/2}\int_0^{2\pi} \Big(\frac{8B^4}{A^4} - \frac{8B^2 C}{A^3} + \frac{C^2}{A^2} \\
-\frac{8B^3\sqrt{B^2-AC}}{A^4} +\frac{4B C\sqrt{B^2-AC}}{A^3}\Big)\cos\psi d\varphi d\psi.
\end{multline}
The radical contributions to both potentials can be shown to be zero by appealing to \eqref{appsym2a} since it transpires that the relevant integrand $f(\psi,\varphi)$ possesses the symmetry
\begin{align}
f(\psi,\varphi) &= -f(-\psi,\varphi+\pi), & f(\psi,\varphi+\pi) &= -f(-\psi,\varphi).
\end{align}
The functions $\Gamma$ and $\Psi$ thus reduce to
\begin{align}
\Gamma(\mathbf{x}) &= C \Upsilon - \Gamma_1
\end{align}
where
\begin{align}
 \Gamma_1 &= \frac{1}{4\pi}\int_{-\pi/2}^{\pi/2}\int_0^{2\pi}\frac{B^2}{A^2}\cos\psi d\varphi d\psi, & \Upsilon &= \frac{1}{8\pi}\int_{-\pi/2}^{\pi/2}\int_0^{2\pi}\frac{\cos\psi}{A} d\varphi d\psi \label{J}
\end{align}
and
\begin{align}
\Psi(\mathbf{x}) &= \Psi_1 - C \Psi_2 + C^2 \Omega
\end{align}
where
\begin{align}
\Psi_1 &= \frac{1}{2\pi}\int_{-\pi/2}^{\pi/2}\int_0^{2\pi}\frac{B^4}{A^4}\cos\psi d\varphi d\psi, & \Psi_2 &= \frac{1}{2\pi}\int_{-\pi/2}^{\pi/2}\int_0^{2\pi}\frac{B^2}{A^3}\cos\psi d\varphi d\psi, \label{K}
\end{align}
\begin{align}
\Omega &= \frac{1}{16\pi}\int_{-\pi/2}^{\pi/2}\int_0^{2\pi}\frac{1}{A^2}\cos\psi d\varphi d\psi. \label{M}
\end{align}

\subsection{Closed integral form for $\Gamma(\mathbf{x})$} \label{app:closedGamma}

Write $\Gamma_1 = \Gamma_{11}+\Gamma_{12}$ where
\begin{align}
\Gamma_{11} &= \frac{1}{4\pi}\int_{-\pi/2}^{\pi/2}\int_0^{2\pi}\left(\frac{\cos^2\psi\cos^2\varphi}{a_1^2}\frac{x_1^2}{a_1^2}+
\frac{\cos^2\psi\sin^2\varphi}{a_2^2}\frac{x_2^2}{a_2^2}+\frac{\sin^2\psi}{a_3^2}\frac{x_3^2}{a_3^2}\right)
\frac{\cos\psi}{A^2}d\varphi d\psi  \label{appPhi11}
\end{align}
and
\begin{multline}
\Gamma_{12} = \frac{1}{2\pi}\int_{-\pi/2}^{\pi/2}\int_0^{2\pi}\Big(\frac{x_1x_2\cos^2\psi\sin\varphi\cos\varphi}{a_1^2a_2^2} \\
- \frac{x_2x_3\cos\psi\sin\psi\sin\varphi}{a_2^2a_3^2}\\
- \frac{x_3x_1\cos\psi\sin\psi\cos\varphi}{a_3^2a_1^2}\Big)\frac{\cos\psi}{A^2}d\varphi d\psi. \label{appPhi12}
\end{multline}
The contribution from $\Gamma_{12}$ is zero - the first term due to $2\pi$ periodicity of the integrand in $\varphi$ and the second and third terms due to their being odd in $\psi$. Treating $\Upsilon$ as a function of $a_j$, the form of $\Gamma_1$ can be exploited, writing
\begin{align}
\Gamma(\mathbf{x}) &=
 C \Upsilon - \sum_{j=1}^3 \frac{x_j^2}{a_j}\deriv{\Upsilon}{a_j}.  \label{5intysum}
\end{align}
Therefore once $\Upsilon$ is determined, $\Gamma(\mathbf{x})$ straightforwardly follows. As such, introduce $A$ into the form of $\Upsilon$ in \eqref{J} to obtain
\begin{align}
\Upsilon &= \frac{1}{8\pi}\int_{-\pi/2}^{\pi/2}\cos\psi\int_0^{2\pi}
\frac{1}{M(\psi)\cos^2\theta+N(\psi)\sin^2\theta}d\theta d\psi
\end{align}
where
\begin{align}
M(\psi) &= \frac{\cos^2\psi}{a_1^2} + \frac{\sin^2\psi}{a_3^2}, & N(\psi) &= \frac{\cos^2\psi}{a_2^2} + \frac{\sin^2\psi}{a_3^2}.
\end{align}
Next, the evenness of the integrand is exploited in order to write it as
\begin{align}
\Upsilon &= \frac{1}{\pi}\int_0^{\pi/2}\cos\psi\int_0^{\pi/2}\frac{\sec^2\varphi}{M+N\tan^2\varphi}d\varphi d\psi \non \\
  &= \frac{1}{2} a_1 a_2 a_3^2\int_0^{\pi/2}\frac{\cos\psi}{\sqrt{(a_1^2\sin^2\psi+a_3^2\cos^2\psi)(a_2^2\sin^2\psi+a_3^2\cos^2\psi)}}d\psi. \label{appUpsilonphi}
\end{align}
Make the substitution $\sin\psi = a_3/\sqrt{a_3^2+t}$, where $t\in[0,\infty)$ so that
\begin{align}
\Upsilon &= \frac{1}{4} a_1a_2a_3\int_0^{\infty}\frac{dt}{\Delta(t)}
\end{align}
where $\Delta(t)=\sqrt{(a_1^2+t)(a_2^2+t)(a_3^2+t)}$. Finally therefore using \eqref{5intysum}
\begin{align}
\Gamma(\mathbf{x}) &=
\frac{1}{4}a_1a_2a_3\int_0^{\infty}\frac{X(\mathbf{x},t)\hspace{0.2cm}dt}{\sqrt{(a_1^2+t)(a_2^2+t)(a_3^2+t)}} \label{appgenPhirep}
\end{align}
where
\begin{align}
X(\mathbf{x},t) &= \sum_{n=1}^3\frac{x_n^2}{a_n^2+t}-1
\end{align}
is a quadratic polynomial in $\mathbf{x}$. It is then found that
\begin{align}
\derivtwomix{\Gamma}{x_i}{x_j} &= \sum_{n=1}^3 \ga_n\de_{in}\de_{jn} \label{app:Ptransportellipsoid}
\end{align}
where
\begin{align}
\ga_n &= \frac{a_1a_2a_3}{2}\int_{0}^{\infty}\frac{dt}{(a_n^2+t)\sqrt{(a_1^2+t)(a_2^2+t)(a_3^2+t)}}. \label{5:gammana}
\end{align}
Setting $t=a_3^2 s$, $\ga_n=\mathcal{E}(\vareps_n;\vareps_1,\vareps_2)$ which is defined in \eqref{5ellipsoidfunction}, with $\vareps_n=a_3/a_n$. Mura \cite{Mur-82} writes \eqref{appgenPhirep} in the form
\begin{align}
\Gamma(\mathbf{x}) &=
\frac{1}{8\pi}\left(-I+\sum_{n=1}^3 x_n^2 I_n\right) \label{appgenPhirep2}
\end{align}
where
\begin{align}
I &= 2\pi a_1a_2a_3\int_0^{\infty} \frac{ds}{\Delta(s)}, &
I_n &= 2\pi a_1a_2a_3\int_0^{\infty} \frac{ds}{(a_n^2+s)\Delta(s)}.
\end{align}
The link between $I_n$ and $\mathcal{E}$ is then clear:
\begin{align}
I_n &= 4\pi\mathcal{E}(\vareps_n;\vareps_1,\vareps_2). \label{app:InmathcalE}
\end{align}

Finally, note that the integrals $I_n$ (or equivalently $\mathcal{E}(\vareps_n;\vareps_1,\vareps_2)$) can be expressed in terms of elliptic integrals \cite{Mur-82}. In particular assuming that $a_1>a_2>a_3$,
\begin{align}
I_1 &= \frac{4\pi\vareps_2}{(\vareps_2^2/\vareps_1^2-1)(1-\vareps_1^2)^{1/2}}\left\{F(\theta,k)-E(\theta,k)\right\}, \label{I1FE}\\
I_3 &= \frac{4\pi}{(1-\vareps_2^2)(1-\vareps_1^2)^{1/2}}\left\{(1-\vareps_1^2)^{1/2}-\vareps_2 E(\theta,k)\right\},  \label{I3FE}\\
I_2 &= 4\pi-I_1-I_2, \label{I2FE}
\end{align}
where
\begin{align}
F(\theta,k) &= \int_0^{\theta}\frac{dx}{(1-k^2\sin^2 x)^{1/2}}, & E(\theta,k) &= \int_0^{\theta}(1-k^2\sin^2 x)^{1/2} \hspace{0.1cm}dx, \label{FE} \\
\theta &= \sin^{-1}(1-\vareps_1^2)^{1/2}, & k &= \frac{1}{\vareps_2}\left(\frac{\vareps_2^2-\vareps_1^2}{1-\vareps_1^2}\right)^{1/2}. \label{thetak}
\end{align}

\subsection{Closed integral form for $\Psi(\mathbf{x})$}  \label{app:closedPsi}

One can also derive an integral form for $\Psi(\mathbf{x})$ although such a derivation is rather lengthy. Expression (11.38) of Mura \cite{Mur-82} is employed, which establishes that (no sum over $i$ here, with sums being shown explicitly for clarity)
\begin{align}
\deriv{\Psi(\mathbf{x})}{x_i} &= \frac{x_i}{8\pi}\left(\left(I- a_i^2 I_i\right) - \sum_{n=1}^3 \left(I_n-I_{in}\right)x_n^2\right), \label{app:dpsibydxi}
\end{align}
noting the additional factor of $1/(4\pi)$ here from our modified definition of potentials as compared with Mura. The integral $I_{mn}$ is defined as
\begin{align}
I_{mn} &= 2\pi a_m^2 a_1a_2a_3\int_0^{\infty} \frac{ds}{(a_m^2+s)(a_n^2+s)\Delta(s)} \label{app:Imn}
\end{align}
where the slight modification to Mura's notation should be noted: $I_{mn}=a_m^2I_{mn}^M$ where $I_{mn}^M$ is Mura's definition of this integral as defined in Chapter 11 of \cite{Mur-82}. This modification means that expressions are now defined in terms of non-dimensional quantities. In particular  it is possible to write $I_{mn}$ in terms of $I_m, I_n$ and $\eps_m, \eps_n$ as follows,
\begin{align}
I_{mn} &=
\frac{(I_n-I_m)}{(1-(\vareps_m/\vareps_n)^2)}, & m &\neq n. \label{app:Imn1}
\end{align}
Additional relations are noted as
\begin{align}
I_{mn} &= \frac{\eps_n^2}{\eps_m^2}I_{nm} \label{app:Imn4}
\end{align}
and
\begin{align}
I_{11} &= \frac{4\pi}{3}-\frac{1}{3}(I_{12}+I_{13}), & I_{22} &= \frac{4\pi}{3}-\frac{1}{3}(I_{21}+I_{23}), & I_{33} &= \frac{4\pi}{3}-\frac{1}{3}(I_{31}+I_{32}), \label{app:Imn2}\\
3I_1 &= 3I_{11}+I_{21}+I_{31}, & 3I_2 &= 3I_{22} + I_{12} + I_{32}, & 3I_3 &= 3I_{33} + I_{13}+I_{23}. \label{app:Imn3}
\end{align}
Therefore, differentiating \eqref{app:dpsibydxi} with respect to $x_j, x_k$ and then $x_{\ell}$ (in that order) gives
\begin{align}
\frac{\pa^4 \Psi}{\pa x_i\pa x_j\pa x_k\pa x_{\ell}} &= \frac{1}{4\pi}\de_{ij}\de_{k\ell}(I_{ik}-I_k)+\frac{1}{4\pi}(\de_{ik}\de_{j\ell}+\de_{jk}\de_{i\ell})(I_{ij}-I_j). \label{app:Psiresultellipsoida}
\end{align}
Note that \eqref{app:Psiresultellipsoida} is however not a fully symmetric fourth order tensor as it should be since derivatives should be able to be taken in any order. As such one can enforce in turn major, minor then total symmetry \cite{Moe-08} to show that (no sum over repeated coefficients)
\begin{align}
\frac{\pa^4\Psi}{\pa x_i^4} &= \frac{3}{4\pi}(I_{ii}-I_i), & \frac{\pa^4\Psi}{\pa x_i^2\pa x_j^2} &= \frac{1}{8\pi}(I_{ij}+I_{ji}-I_i-I_j), & i &\neq j \label{app:Psiresultellipsoid}
\end{align}
and odd derivatives are zero.

\section{Cartesian coordinates, rotations and tensors} \label{app:carttensors}

Cartesian tensors are used throughout this article. Some of their properties are summarized shortly, in particular those associated with higher order symmetrized are discussed. Before this a brief review of rotations of Cartesian coordinates is given for completeness.

\subsection{Rotations of Cartesian coordinate systems} \label{app:rotcoords}

Consider a fixed Cartesian coordinate system $x_i, i=1,2,3$ and an associated Cartesian coordinate system $x_i', i=1,2,3$ having the same origin, having been rotated arbitrarily in three dimensions. The general three dimensional rotation matrix is constructed as a product of three rotation matrices, each of which corresponds to a rotation of the axes about a given axis in three dimensional space. Begin by rotating anticlockwise about the $x_3$ axis by use of the matrix
\begin{align}
\mathbf{Q}^1(\varphi) &= \left(
\begin{array}{ccc}
\cos\varphi & \sin\varphi & 0 \\
-\sin\varphi & \cos\varphi & 0 \\
 0  & 0 & 1
\end{array}
\right)
\end{align}
which generates the rotated coordinate system $\bar{x}_i = Q_{ij}^1(\varphi)x_j$. This is followed by an anticlockwise rotation about the $\bar{x}_1$ axis by application of the matrix
\begin{align}
\mathbf{Q}^2(\vartheta) &= \left(
\begin{array}{ccc}
1 & 0 & 0 \\
0 & \cos\vartheta & \sin\vartheta \\
 0  & -\sin\vartheta & \cos\vartheta
\end{array}
\right),
\end{align}
which generates the rotated coordinate system $\hat{x}_i = Q_{ij}^2(\vartheta)Q_{jk}^1(\varphi)x_k$. Finally an anticlockwise rotation about the $\hat{x}_3$ axis is performed by application of the matrix
\begin{align}
\mathbf{Q}^1(\psi) &= \left(
\begin{array}{ccc}
\cos\psi & \sin\psi & 0 \\
-\sin\psi & \cos\psi & 0 \\
 0  & 0 & 1
\end{array}
\right)
\end{align}
so that our required fully rotated system is derived as
\begin{align}
x_i' &= Q_{ij}^1(\psi)Q^2_{jk}(\vartheta)Q_{k\ell}^1(\varphi)x_{\ell} \non \\
  &=  \mathcal{Q}_{i\ell}(\varphi,\vartheta,\psi)x_{\ell}.
\end{align}
The domains of the Euler angles are $\vartheta\in[0,\pi], \varphi\in[0,2\pi), \psi\in[0,2\pi)$. The angles $\vartheta$ and $\varphi$ correspond to the standard angles with identical notation as used in the spherical coordinate system defined above.

\subsection{Cartesian tensors in rotated frames}

Employing tensor product notation $\otimes$, a second order Cartesian tensor $\mathbf{A}$ with components $A_{ij}$ can be written
\begin{align}
\mathbf{A} = A_{ij}\mathbf{e}_i\otimes\mathbf{e}_j
\end{align}
where $\mathbf{e}_j$ is the $j$th Cartesian unit basis vector. Similarly a fourth order tensor $\mathbf{A}$ is defined as
\begin{align}
\mathbf{A} = A_{ijk\ell}\mathbf{e}_i\otimes\mathbf{e}_j\otimes\mathbf{e}_k\otimes\mathbf{e}_{\ell}.
\end{align}

The relation between components of second and fourth order tensors in the rotated system $x_i'$, generated by application of the general rotation matrix $\boldsymbol{\mathcal{Q}}$ to the tensors in the original system $x_i$, i.e.\ $A_{ij}'$ and $A_{ijk\ell}'$ are
\begin{align}
A_{ij}' &= Q_{ik}Q_{j\ell}A_{k\ell}, &  A_{ij} &= Q_{ki}Q_{\ell j}A_{k\ell}', \\
A_{ijk\ell}' &= Q_{im}Q_{jn}Q_{kp}Q_{\ell q}A_{mnpq}, & A_{ijk\ell} &= Q_{mi}Q_{nj}Q_{pk}Q_{q\ell}A_{mnpq}'.
\end{align}
Often it is useful to determine the average of a general second order tensor, say $\mathbf{A}$ over all possible rotations of Cartesian axes (uniformly). The natural way to do this is to take a tensor with components $A_{ij}'$, diagonal in some coordinate system $x_i'$. This frame has been rotated from the fixed system $x_i$. However it is natural to work in a ``fixed'' coordinate system $x_i$, which can be considered as being obtained from the system $x_i'$ via a rotation. Indeed we have $x_i= \mathcal{Q}_{ji}x_j'$. The components of $A_{ij}'$ are diagonal and as such the components $A_{ij}$ are dependent on the Euler angles. If one wishes to determine the average of the tensor $A_{ij}$ over all such orientations uniformly, it can be done by carrying out the following integration
\begin{align}
\underline{A}_{ij} &= \frac{1}{8\pi^2}\int_0^{2\pi} \int_0^{2\pi}\int_0^{\pi} A_{ij}(\varphi,\vartheta,\psi)\lit \sin\vartheta \lit d\vartheta d\varphi d\psi \non\\
 &= \frac{1}{8\pi^2}\int_0^{2\pi} \int_0^{2\pi}\int_0^{\pi} \mathcal{Q}_{ki}(\varphi,\vartheta,\psi)\mathcal{Q}_{\ell j}(\varphi,\vartheta,\psi)A_{k\ell}' \lit\sin\vartheta \lit d\vartheta d\varphi d\psi \label{Aavuniform}
\end{align}
where the underline denotes orientation averaging. Alternatively there may be some orientation distribution function, say $p(\varphi,\vartheta,\psi)$ that weights the importance of certain distributions. A weighted orientation average can then be defined as
\begin{align}
\underline{A}_{ij}
 &= \frac{1}{8\pi^2}\int_0^{2\pi} \int_0^{2\pi}\int_0^{\pi}p(\varphi,\vartheta,\psi) \mathcal{Q}_{ki}(\varphi,\vartheta,\psi)\mathcal{Q}_{\ell j}(\varphi,\vartheta,\psi)A_{k\ell}' \lit\sin\vartheta \lit d\vartheta d\varphi d\psi. \label{app:Aijav1}
\end{align}
Note the normalization condition on the weighting distribution
\begin{align}
\frac{1}{8\pi^2}\int_0^{2\pi} \int_0^{2\pi}\int_0^{\pi}p(\varphi,\vartheta,\psi)  \lit\sin\vartheta \lit d\vartheta d\varphi d\psi &= 1.
\end{align}
Analogous expressions to \eqref{Aavuniform} and \eqref{app:Aijav1} hold for the fourth order tensor case of course.

\subsection{Second order Cartesian tensors} \label{app:tensorstructure2}

\subsubsection{Isotropy}

The second order identity tensor is $I_{ij}=\de_{ij}$ and with $\al$ constant, the general second order isotropic tensor is therefore $A_{ij} = \al\de_{ij}$. Its inverse, with components $\tilde{A}_{ij}$ is $\tilde{A}_{ij} = \frac{1}{\al}\de_{ij}$.

\subsubsection{Transverse isotropy}  \label{sec:apptens2TI}

Upon defining the tensor
\begin{align}
\Theta_{ij} &= \de_{ij}-\de_{i3}\de_{j3} \label{Thetaij}
\end{align}
a second order \textit{transversely isotropic} tensor (with symmetry axis $x_3$) has the form
\begin{align}
A_{ij} &= \al_1\Theta_{ij} + \al_3\de_{i3}\de_{j3} \label{app:genTI} \\
\end{align}
The tensor $\Theta_{ij}$ defined in \eqref{Thetaij} possesses the following properties:
\begin{align}
\Theta_{ij}&=\Theta_{ji}, &
\Theta_{ik}\Theta_{kj} &= \Theta_{ij}, &
\Theta_{ij}\Theta_{ij} &= 2,
\end{align}
Using these properties, the inverse of $\mathbf{A}$ has components $\tilde{A}_{ij}$ that can be written
\begin{align}
\tilde{A}_{ij} &= \frac{1}{\al_1}\Theta_{ij} + \frac{1}{\al_3}\de_{i3}\de_{j3}.
\end{align}
The concept can easily be generalized to other symmetry axes either by use of rotations of coordinate axes as in \S \eqref{app:rotcoords} or by use of the notation $n_{ij}=n_i n_j$ where $n_i$ are the components of the direction vector associated with the axis of symmetry, so that for the example described above $n_i=\de_{i3}$. The associated generalization of \eqref{Thetaij} is therefore $\theta_{ij}=\de_{ij}-n_{ij}$.

\subsubsection{Orthotropy}

A second order orthotropic tensor has the form
\begin{align}
A_{ij} &= \al_1\de_{i1}\de_{j1}+\al_2\de_{i2}\de_{j2} + \al_3\de_{i3}\de_{j3} \label{app:ortho1}
\end{align}
which has as its inverse
\begin{align}
\tilde{A}_{ij} &= \frac{1}{\al_1}\de_{i1}\de_{j1}+\frac{1}{\al_2}\de_{i2}\de_{j2} + \frac{1}{\al_3}\de_{i3}\de_{j3}.
\end{align}

\subsubsection{Averaging over orientations} \label{app:avor2}

Uniform orientation averaging of second order tensors can be done mechanically via rotation tensors as was described above in \S \ref{app:rotcoords} and written explicitly in \eqref{app:Aijav1}. Alternatively for simple uniform orientation averaging, a simple aspect of tensor analysis associated with invariants can be exploited. Averaging uniformly would give rise to an isotropic tensor with components of the form
\[
\underline{A}_{ij} = \al \de_{ij}.
\]
where the underline denotes averaging. Performing a contraction in the original tensor gives rise to a quantity that does not change with rotations, i.e. $a=A_{kk}$ is an invariant. Therefore
\[
\underline{A}_{kk} = a = \al \de_{kk} = 3\al
\]
so that $\al=a/3$. The components of the averaged tensor therefore take the form
\begin{align}
\underline{A}_{ij} = \frac{1}{3}a\de_{ij}. \label{2orderaveragedangles}
\end{align}
If averages need to be taken with respect to some weighting function, then the mechanical process of averaging over angles as in \eqref{app:Aijav1} needs to be followed.

In the case of a transversely isotropic second order tensor $A_{ij}$ with $n_{ij}=\de_{i3}\de_{j3}$, it is easily shown that
\begin{align}
\underline{n}_{ij} &= \frac{1}{3}\de_{ij}, & \underline{\Theta}_{ij} &= \frac{2}{3}\de_{ij}, \label{app:av2orderbasis}
\end{align}
and so taking a uniform orientation average of $A_{ij}$ in \eqref{app:genTI} yields
\begin{align}
\underline{A}_{ij} &= \frac{1}{3}\left(2\al_1 + \al_3\right)\de_{ij}.
\end{align}
In the orthotropic case upon taking uniform averages of \eqref{app:ortho1} it is shown that
\begin{align}
\underline{A}_{ij} &= \frac{1}{3}(\al_1+\al_2+\al_3)\de_{ij}.
\end{align}

\subsection{Fourth order Cartesian Tensors}\label{app:tensorstructure4}

The following tensors are used extensively in elasticity applications. See Walpole \cite{Wal-84} for a comprehensive derivation of all associated theory.

\subsubsection{Isotropy}

First define the following tensors
\begin{align}
I_{ijk\ell}^1 &= \frac{1}{3}\de_{ij}\de_{k\ell}, \label{appnot:I14}\\
I_{ijk\ell}^2 &= \frac{1}{2}\left(\de_{ik}\de_{j\ell}+\de_{i\ell}\de_{jk}\right)-\frac{1}{3}\de_{ij}\de_{k\ell}, \label{appnot:I24}\\
I_{ijk\ell} &= \frac{1}{2}\left(\de_{ik}\de_{j\ell}+\de_{i\ell}\de_{jk}\right)= I_{ijk\ell}^1+I_{ijk\ell}^2. \label{4thIDapp}
\end{align}
These have the following properties
\begin{align}
I^1_{ijk\ell} &= I^1_{k\ell ij}, & I^2_{ijk\ell} &= I^2_{k\ell ij}, & I_{ijk\ell} &= I_{k\ell ij}
\end{align}
and
\begin{align}
I^1_{ijmn}I^1_{mnk\ell} &= I^1_{ijk\ell}, & I^2_{ijmn}I^2_{mnk\ell} &= I^2_{ijk\ell}, &
I^1_{ijmn}I^2_{mnk\ell} &=  0, & I^2_{ijmn}I^1_{mnk\ell} &= 0. \label{app:t4con}
\end{align}
If $\boldsymbol{\si}$ is a second order tensor, whose components are written in the deviatoric/scalar form
\begin{align}
\si_{ij} &= \si'_{ij}+\frac{1}{3}\si\de_{ij} \label{strainsplit2}
\end{align}
where $\si=\si_{kk}$, then
\begin{align}
I_{ijk\ell}^1\si_{k\ell} &= \frac{1}{3}\si\de_{ij}, &
I_{ijk\ell}^2\si_{k\ell} &= \si'_{ij}. \label{2split2}
\end{align}
Given a fourth order isotropic tensor with components of the form
\begin{align}
A_{ijk\ell} 
         &= 3\al_1 I_{ijk\ell}^1 + 2\al_2 I_{ijk\ell}^2 \label{2A}
\end{align}
then due to \eqref{2split2}
\begin{align}
A_{ijk\ell}\si_{k\ell} &= 3\al_1\si\de_{ij}+2\al_2\si'_{ij}.
\end{align}
Introduce a second fourth order isotropic tensor $B_{ijk\ell} = 3\be_1 I_{ijk\ell}^1 + 2\be_2 I_{ijk\ell}^2$ and then
\begin{align}
A_{ijmn}B_{mnk\ell}=B_{ijmn}A_{mnk\ell} &= 9\al_1\be_1 I_{ijk\ell}^1 + 4\al_2\be_2 I_{ijk\ell}^2
\end{align}
and the inverse of $A_{ijk\ell}$, denoted by $\tilde{A}_{ijk\ell}$, such that $A_{ijmn}\tilde{A}_{mnk\ell}=I_{ijk\ell}$ is
\begin{align}
\tilde{A}_{ijk\ell} &= \frac{1}{3\al_1}I_{ijk\ell}^1 + \frac{1}{2\al_2}I_{ijk\ell}^2.
\end{align}

\subsubsection{Cubic system} \label{app:app4cubic}

Define the \textit{cubic} tensor
\begin{align}
A_{ijk\ell} &= \al_1 I_{ijk\ell}^1 + \al_2 I^2_{ijk\ell} + \al_3 \de_{ijk\ell}
\end{align}
and this tensor has the property that $\de_{ijk\ell}=1$ only if $i=j=k=\ell$ and is zero otherwise. Furthermore
\begin{align}
I_{ijmn}^1\de_{mnk\ell} &= I_{ijk\ell}^1, & I_{ijmn}^2\de_{mnk\ell} &= \de_{ijk\ell}-I_{ijk\ell}^1. \label{2prods2}
\end{align}
Writing the inverse of $A_{ijk\ell}$ as
\begin{align}
\tilde{A}_{ijk\ell} &= \frac{\al_2^2+\al_1\al_3}{\al_2^2(\al_1+\al_3)} I_{ijk\ell}^1 + \frac{1}{\al_2} I_{ijk\ell}^2 -\frac{\al_3}{\al_2^2} \de_{ijk\ell}.
\end{align}

\subsubsection{Transverse isotropy} \label{app:apptensTI}

We shall use the Hill basis for transversely isotropic (TI) tensors. There are several slight variants on this but the Hill basis is used commonly in the micromechanics literature and so it appears sensible to adopt it here.
%
This basis set enables a fourth order TI tensor $A_{ijk\ell}$ to be written in the form
\begin{align}
A_{ijk\ell} &= \sum_{n=1}^6 \al_n \mathcal{H}_{ijk\ell}^{n} \label{HTIbasis}
\end{align}
where $X_n$ are constants. We note that in general $X_2\neq X_3$ since contraction of a TI tensor with another TI tensor does not result in a tensor with $X_2=X_3$. The basis tensors $\mathcal{H}^{n}_{ijkl}$ are defined by
\begin{align}
\mathcal{H}^{1}_{ijk\ell} &= \frac{1}{2}\Theta_{ij}\Theta_{k\ell}, &
\mathcal{H}^{2}_{ijk\ell} &= \Theta_{ij}\de_{k3}\de_{\ell 3}, &
\mathcal{H}^{3}_{ijk\ell} &= \Theta_{k\ell}\de_{i3}\de_{j3}, \label{H1}
\end{align}
\vspace{-0.7cm}
\begin{align}
\mathcal{H}^{4}_{ijk\ell} &= \de_{i3}\de_{j3}\de_{k3}\de_{\ell 3}, &
\mathcal{H}^{5}_{ijk\ell} &= \frac{1}{2}(\Theta_{ik}\Theta_{\ell j}+\Theta_{i\ell}\Theta_{kj}-\Theta_{ij}\Theta_{k\ell}),
\end{align}
\vspace{-0.7cm}
\begin{align}
\mathcal{H}^{6}_{ijk\ell} &= \frac{1}{2}(\Theta_{ik}\de_{\ell 3}\de_{j3}+\Theta_{i\ell}\de_{k3}\de_{j3}+\Theta_{jk}\de_{\ell 3}\de_{i3}+\Theta_{j\ell}\de_{k3}\de_{i3}), \label{H6}
\end{align}
where $\Theta_{ij} = \de_{ij}-\de_{i3}\de_{j3}$. The notation $\mathcal{H}$ signifies the \textit{Hill} basis.

Let us define the shorthand notation
\begin{align}
\mathcal{H}^m \mathcal{H}^n &= \mathcal{H}_{ijpq}^m \mathcal{H}_{pqk\ell}^n \label{Hcontr}
\end{align}
for contraction between the basis tensors defined in \eqref{H1}-\eqref{H6}. The contractions defined in \eqref{Hcontr} are then summarized in table \ref{tab:Pbasistab}.

\begin{table}[h]
\begin{center}\begin{tabular}{|c|c|c|c|c|c|c|}
\hline
                &  $\mathcal{H}^{(1)}$ & $\mathcal{H}^{(2)}$ & $\mathcal{H}^{(3)}$ & $\mathcal{H}^{(4)}$ & $\mathcal{H}^{(5)}$ & $\mathcal{H}^{(6)}$\\
\hline
$\mathcal{H}^{(1)}$ & $\mathcal{H}^{(1)}$ & $\mathcal{H}^{(2)}$ & 0 & 0 & 0 & 0 \\
$\mathcal{H}^{(2)}$ & 0              & 0 & 2$\mathcal{H}^{(1)}$ & $\mathcal{H}^{(2)}$ & 0 & 0 \\
$\mathcal{H}^{(3)}$ & $\mathcal{H}^{(3)}$ &  2$\mathcal{H}^{(4)}$ & 0 & 0 & 0 &0 \\
$\mathcal{H}^{(4)}$ & 0 & 0 & $\mathcal{H}^{(3)}$ &  $\mathcal{H}^{(4)}$ & 0 & 0 \\
$\mathcal{H}^{(5)}$ & 0              & 0             & 0             & 0 & $\mathcal{H}^{(5)}$ & 0\\
$\mathcal{H}^{(6)}$ & 0              & 0             & 0             & 0 & 0 & $\mathcal{H}^{(6)}$\\
\hline
\end{tabular}\end{center}
\caption{The contractions of the basis tensors $\mathcal{H}_{ijkl}^{(n)}$.}\label{tab:Pbasistab}
\end{table}

Note that it is often useful to write the fourth order isotropic identity tensor and basis tensors in the Hill TI basis form, i.e.\
\begin{align}
I_{ijk\ell} &= \mathcal{H}_{ijk\ell}^1+\mathcal{H}_{ijk\ell}^4+\mathcal{H}_{ijk\ell}^5+\mathcal{H}_{ijk\ell}^6, \label{5ITI}\\
I_{ijk\ell}^1 &= \frac{1}{3}(2\mathcal{H}_{ijk\ell}^1+\mathcal{H}_{ijk\ell}^2+\mathcal{H}_{ijk\ell}^3+\mathcal{H}_{ijk\ell}^4), \label{5I1TI}\\
I_{ijk\ell}^2 &= \frac{1}{3}\left(\mathcal{H}_{ijk\ell}^1-\mathcal{H}_{ijk\ell}^2-\mathcal{H}_{ijk\ell}^3+2\mathcal{H}_{ijk\ell}^4
+3\mathcal{H}_{ijk\ell}^5+3\mathcal{H}_{ijk\ell}^6\right). \label{5I2TI}
\end{align}

This allows us to define the inverse of the tensor $A_{ijk\ell}$, $\tilde{A}_{ijk\ell}$ in a straightforward manner. It is
\begin{align}
\tilde{A}_{ijk\ell} &= \sum_{n=1}^6 \tilde{\al}_n \mathcal{H}_{ijk\ell}^n
\end{align}
where
\begin{align}
\tilde{\al}_1 &= \frac{\al_4}{2\De}, & \tilde{\al}_2 &= -\frac{\al_2}{2\Delta}, & \tilde{\al}_3 &= -\frac{\al_3}{2\De},\\
\tilde{\al}_4 &= \frac{\al_1}{2\De}, & \tilde{\al}_5 &= \frac{1}{\al_5}, & \tilde{\al}_6 &= \frac{1}{\al_6}.
\end{align}
where $\De = \al_1\al_4/2-\al_2\al_3$.

\subsubsection{Hill's shorthand notation} \label{appHilltensornotation}

Hill introduced a convenient short-hand notation regarding fourth order tensors. For a fourth order isotropic tensor $A_{ijk\ell}$ defined via
\begin{align}
A_{ijk\ell} = \sum_{n=1}^2 \al_n I^n_{ijk\ell}
\end{align}
Hill denoted it in shorthand notation as $A = \left(\al_1,\al_2\right)$, i.e.\ the tensor basis is assumed from the outset so only the coefficients are to be prescribed. Commonly for linear isotropic elasticity, the case when $\al_1=3\ka, \al_2=2\mu$ is assumed where $\ka$ and $\mu$ are as usual the bulk and shear moduli.
For a fourth order TI tensor $A_{ijk\ell}$ defined via
\begin{align}
A_{ijk\ell} = \sum_{n=1}^6 \al_n \mathcal{H}^n_{ijk\ell}
\end{align}
Hill denoted it in shorthand notation as $A = \left(\al_1,\al_2,\al_3,\al_4,\al_5,\al_6\right)$. Commonly for linear TI elasticity, the case when $\al_1=2k, \al_2=\al_3=\ell, \al_4=n, \al_5=2m, \al_6=2p$ is assumed.

\subsubsection{Orthotropy} \label{app:apptensortho}

Orthotropic basis tensors are described in the paper by Walpole \cite{Wal-67} for example. For practical purposes the matrix formulation of fourth order tensors is extremely useful for higher order tensors, beginning with orthotropy for example. This formulation is described in \ref{app:matrix4tensor} below. First averaging over orientations is considered.

\subsubsection{Averaging over orientations} \label{avor4}

As with the second order case, averages of fourth order tensors can also be taken over all rotations. Uniform averaging would give rise to an isotropic tensor with components of the form
\begin{align}
\underline{A}_{ijk\ell} = \al I_{ijk\ell}^1 + \be I_{ijk\ell}^2. \label{app:Aijklisoav}
\end{align}
Here $A_{iikk}=a$ and $A_{ikik}=b$ are invariants of the original tensor. Therefore $\underline{A}_{iikk} = a = 3\al$ so that $\al=a/3$. Furthermore $\underline{A}_{ikik} = b = \al + 5\be$ so that $\be=b/5-a/15$. The components of the averaged tensor take the form
\begin{align}
\underline{A}_{ijk\ell} = \frac{1}{3}\underline{A}_{ppqq} I_{ijk\ell}^1 + \frac{1}{15}\left(3\underline{A}_{pqpq}- \underline{A}_{ppqq}\right) I_{ijk\ell}^2. \label{2orderaveragedangles2}
\end{align}
The process of averaging with respect to a weighting function takes the form
\begin{align}
\underline{A}_{ijk\ell}
 &= \frac{1}{8\pi^2}\int_0^{2\pi} \int_0^{2\pi}\int_0^{\pi}p(\phi,\theta,\psi) Q_{mi}Q_{nj}Q_{pk}Q_{q\ell}A_{klmn}' \lit\sin\theta \lit d\theta d\phi d\psi \label{app:Aijav2}
\end{align}
where arguments on the rotation matrices have been omitted for conciseness.

A similar approach can be adopted for the uniform orientation averaging of a general fourth order TI tensor. Given the general TI form \eqref{HTIbasis}, the averaged tensor will take the isotropic form \eqref{app:Aijklisoav} and therefore it must be the case that
\begin{align}
\underline{A}_{ppqq} &= 3\al = 2\al_1+2\al_2+2\al_3 + \al_4, \\
\underline{A}_{pqpq} &= \al+5\be = \al_1 +\al_4+2\al_5+2\al_6
\end{align}
which defines $\al$ and $\be$ in terms of the coefficients $\al_n$, i.e.\
\begin{align}
\underline{A}_{ijk\ell} = \frac{1}{3}(2\al_1+2\al_2+2\al_3 + \al_4)I^1_{ijk\ell}
+ \frac{1}{15}(\al_1-2\al_2-2\al_3+2\al_4+6\al_5+6\al_6)I^2_{ijk\ell}. \label{app:4thordertensorisoav}
\end{align}

\subsection{Matrix formulation of fourth order tensors} \label{app:matrix4tensor}

Matrix representation and manipulation of second order Cartesian tensors is of course trivial. It is also often of great utility to represent \textit{fourth} order tensors in the form of a six by six matrices. In particular a general fourth order tensor $\mathbf{T}$ with components $T_{ijk\ell}$ with respect to a Cartesian basis can be usefully written in matrix form $[T]$ as
\begin{align}
[T] &=
\left( \begin{array}{cccccc}
T_{1111} & T_{1122} & T_{1133} & T_{1123} & T_{1131} & T_{1112}  \\
T_{2211} & T_{2222} & T_{2233} & T_{2223} & T_{2213} & T_{2212}  \\
T_{3311} & T_{3322} & T_{3333} & T_{3323} & T_{3313} & T_{3312}  \\
T_{2311} & T_{2322} & T_{2333} & T_{2323} & T_{2313} & T_{2312}  \\
T_{1311} & T_{1322} & T_{1333} & T_{1323} & T_{1313} & T_{1312}  \\
T_{1211} & T_{1222} & T_{1233} & T_{1223} & T_{1213}  & T_{1212}
\end{array} \right).
\label{app:Ttensmat}
\end{align}
Of particular importance is the ability to carry out the operations of tensor contraction and inversion with the matrix form stated here.  Defining the matrix $[W]$ as
\begin{align}
[W] &=
\left( \begin{array}{cccccc}
1 & 0 & 0 & 0 & 0 & 0 \\
0 & 1 & 0 & 0 & 0 & 0 \\
0 & 0 & 1 & 0 & 0 & 0 \\
0 & 0 & 0 & 2 & 0 & 0 \\
0 & 0 & 0 & 0 & 2 & 0 \\
0 & 0 & 0 & 0 & 0 & 2
\end{array} \right),
\label{app:Wmat}
\end{align}
the contraction $T_{ijmn}T'_{nmk\ell}$ in matrix form is $[T][W][T']$. Furthermore, defining $[T^{-1}]$ as the matrix associated with the inverse of the tensor $\mathbf{T}$ it is straightforward to show that $[T^{-1}] = [W]^{-1}[T]^{-1}[W]^{-1}$.

The matrix $[W]$ can be used in the formulation of matrix forms of the linear elastic constitutive relations, i.e.\ given the tensor forms
\begin{align}
\si_{ij} &= C_{ijk\ell}e_{k\ell}, & e_{ij} &= D_{ijk\ell}\si_{k\ell},
\end{align}
the equivalent matrix forms as
\begin{align}
[\si]&=[C][W][e], & [e] &= [D][W][\si] \label{app:linelastmat}
\end{align}
where $[\si]=(\si_{11}\hspace{0.1cm} \si_{22} \hspace{0.1cm}\si_{33} \hspace{0.1cm}\si_{23}\hspace{0.1cm}\si_{13}\hspace{0.1cm}\si_{12})^T$ is the $1\times 6$ column vector of stresses and similarly $[e]=(e_{11}\hspace{0.1cm} e_{22} \hspace{0.1cm}e_{33} \hspace{0.1cm}e_{23}\hspace{0.1cm}e_{13}\hspace{0.1cm}e_{12})^T$ . Multiplying the equations in \eqref{app:linelastmat} from the left by $[W]$ we find that
\begin{align}
[\si]&=[\mathcal{C}][\ga], & [\ga] &= [\mathcal{D}][\si] \label{app:linelastmat2}
\end{align}
where
\begin{align}
[\mathcal{C}] &= [C], & [\mathcal{D}]=[W][D][W] = [C]^{-1}
\end{align}
and $[\ga]=[W][e]$ is the engineering strain.

\end{appendix}

\end{document}